\documentclass[
reprint,
superscriptaddress,
amsmath,amssymb,
aps,
prb,
longbibliography]{revtex4-2}

\usepackage{xspace}
\usepackage{graphicx}% include figure files
\usepackage{bm}% bold math
\usepackage[colorlinks=true,urlcolor=blue,linkcolor=blue,citecolor=blue,bookmarks=false]{hyperref}% add hypertext capabilities
\usepackage{amsmath}
\usepackage{amssymb}
\usepackage{bbold}
\usepackage{soul}
\usepackage{dsfont}
\usepackage{braket}
\usepackage{textcomp}

\graphicspath{{./}}

\newcommand{\customref}[2]{\hyperref[#1]{\ref*{#1}#2}}

\usepackage{xcolor}

\definecolor{Ured}{HTML}{cc0000}
\definecolor{Ublue}{HTML}{1f65cf}
\definecolor{Ugreen}{HTML}{198a11}

\DeclareMathOperator{\Tr}{Tr}

\newcommand{\ie}[0]{i.e.\@\xspace}
\newcommand{\eg}[0]{e.g.\@\xspace}

\newcommand{\cf}[0]{cf.\@\xspace}

\newfont{\tensy}{cmsy10}

\renewcommand{\S}[0]{\hat{\mathcal{H}}}

\newcommand{\ban}[1]{\hat{a}^{\vphantom\dagger}_{#1}}
\newcommand{\bcr}[1]{\hat{a}^{\dagger}_{#1}}
\newcommand{\bant}[1]{\hat{A}^{\vphantom\dagger}_{#1}}
\newcommand{\bcrt}[1]{\hat{A}^{\dagger}_{#1}}

\newcommand{\Q}[1]{\hat{Q}_{#1}}
\renewcommand{\P}[1]{\hat{P}_{#1}}

\newcommand{\im}{\mathrm{i}}

\newcommand{\Hc}{\mathrm{H.c.}}
\newcommand{\expvtext}[1]{\langle #1 \rangle}

\newcommand{\spin}[1]{\hat{\mathbf{S}}_{#1}}
\newcommand{\spinz}[1]{\hat{S}^{z}_{#1}}
\newcommand{\spinx}[1]{\hat{S}^{x}_{#1}}
\newcommand{\spinp}[1]{\hat{S}^{+}_{#1}}
\newcommand{\spinm}[1]{\hat{S}^{-}_{#1}}

\newcommand{\proj}[1]{\hat{\Pi}_{#1}}

\newcommand{\omegac}{\omega_\mathrm{c}}

\newcommand{\VBS}{\mathrm{VBS}}

\begin{document}
\title{Quantum spin chains with bond dissipation}

\author{Manuel Weber}
\affiliation{Institut f\"ur Theoretische Physik and W\"urzburg-Dresden Cluster of Excellence ct.qmat, Technische Universit\"at Dresden, 01062 Dresden, Germany}
\affiliation{Max-Planck-Institut f\"ur Physik komplexer Systeme, N\"othnitzer Str.~38, 01187 Dresden, Germany}

\date{\today}

\begin{abstract}
We study the effect of bond dissipation on the one-dimensional antiferromagnetic spin-$1/2$ Heisenberg model. In analogy to the spin-Peierls problem, the dissipative bath is described by local harmonic oscillators that modulate the spin exchange coupling, but instead of a single boson frequency we consider a continuous bath spectrum $\propto \omega^s$. Using an exact quantum Monte Carlo method for retarded interactions, we show that for $s<1$ any finite coupling to the bath induces valence-bond-solid order, whereas for $s>1$ the critical phase of the isolated chain remains stable up to a finite critical coupling. We find that, even in the presence of the gapless bosonic spectrum, the spin-triplet gap remains well defined for any system size, from which we extract a dynamical critical exponent of $z=1$. We provide evidence for a Berezinskii-Kosterlitz-Thouless quantum phase transition that is governed by the SU(2)$_1$ Wess-Zumino-Witten model. Our results suggest that the critical properties of the dissipative system are the same as for the spin-Peierls model, irrespective of the different interaction range, \ie, power-law vs.~exponential decay, of the retarded dimer-dimer interaction, indicating that the spin-Peierls criticality is robust with respect to the bosonic density of states.
\end{abstract}

\maketitle

\section{Introduction}

Quantum critical phases of matter
are a central theme in quantum magnetism
and often emerge from the interplay between strong correlations and quantum fluctuations \cite{Sachdev:2008aa}.
One of the most prominent examples
can be found in the one-dimensional (1D) antiferromagnetic spin-1/2
Heisenberg chain. Within the last years, it has become possible to realize this simple model in solid-state experiments
and, \eg, probe the
spectral signatures of fractionalized spinons
down to very low energy scales
\cite{PhysRevB.44.12361, Lake05, Mourigal:2013aa, Toskovic16, gao2023spinon}.
However, the gapless nature of their excitations
makes critical states highly susceptible to perturbations, raising the important
question of the stability of such phases, 
in particular since experimental setups can never be fully isolated
from their environment. In practical realizations of 1D spin chains, perturbations can arise from weak interchain coupling,
the coupling to a substrate, or interactions with other degrees of freedom like phonons.
Hence, it is important to understand the effects of generic dissipation mechanisms on
quantum spin systems.

The theoretical study of quantum dissipative systems was pioneered by Caldeira and Leggett \cite{PhysRevLett.46.211}, who modeled the effect of the environment on the system via an infinite number of harmonic oscillators. This description has the advantage that
the bath can be integrated out
exactly using the path integral to obtain a retarded interaction in the system's degrees
of freedom. 
A power-law bath spectrum $\propto \omega^s$ then leads to a long-range interaction in imaginary time  with an algebraic decay $\propto 1/|\tau-\tau'|^{1+s}$.
In this way, already a single spin can exhibit nontrivial quantum phase transitions, quantum critical phases, or exotic renormalization-group (RG) phenomena, as the bath exponent $s$ is tuned
\cite{RevModPhys.59.1, doi:10.1080/14786430500070396, PhysRevLett.102.030601,
Smith_1999, PhysRevB.61.4041, Sachdev2479, PhysRevB.61.15152, PhysRevLett.91.096401,
PhysRevLett.108.160401, PhysRevB.90.245130, PhysRevB.106.L081109, PhysRevLett.130.186701}.
In extended systems, bath-induced phase transitions have been explored primarily
in quantum Ising/clock/rotor models
\cite{%
% old work on dissipative Josephson junctions
PhysRevLett.56.2303, PhysRevB.36.1917, PANYUKOV1987325, PhysRevB.37.3283, 1989EL......9..107K, PhysRevB.41.4009, PhysRevB.45.2294, PhysRevLett.78.1779, PhysRevB.75.014522, FEIGELMAN1998107, 
PhysRevLett.94.047201, 2005JPSJ...74S..67W, 
PhysRevB.81.104302, PhysRevB.83.115134, PhysRevB.84.180503, PhysRevB.85.214302, PhysRevB.85.224531,
PhysRevB.72.060505, PhysRevB.73.064503, PhysRevB.73.094516,
PhysRevB.79.184501, PhysRevB.82.174501, PhysRevB.91.205129, PhysRevB.94.235156, 
PhysRevLett.96.227201, PhysRevLett.99.230601, PhysRevLett.100.240601,  PhysRevB.86.075119, Vojta_2011,
PhysRevA.77.051601, PhysRevB.97.155427,
PhysRevB.104.L060410, PhysRevB.105.L180407, PhysRevB.107.L100302}
whereas only a few studies have examined quantum dissipative spin systems
using analytic
\cite{PhysRevLett.79.4629, PhysRevLett.97.076401, PhysRevB.86.035455, friedman2019dissipative, PhysRevB.107.165113, martin2023stable, kuklov2023transverse}
or quantum Monte Carlo (QMC) \cite{PhysRevLett.113.260403, PhysRevB.97.035148, PhysRevLett.129.056402, PhysRevLett.125.206602, PhysRevB.106.L161103}
approaches.
Moreover, previous work has mainly focused on ohmic dissipation with a bath exponent of $s=1$,
although the retardation range can have a significant impact on the dynamical properties of the system, \eg, tuning $s$ can alter the dynamical critical exponent $z$ at the quantum phase transition \cite{PhysRevB.85.214302}.
To which extent the interaction range can affect the ground-state properties of quantum many-particle systems
is not only relevant for dissipative systems, but also in the context of long-range interactions in space
\cite{Laflorencie_2005, PhysRevLett.104.137204, yang2020deconfined, PhysRevLett.119.023001, PhysRevResearch.5.033038, daliao2022caution, PhysRevB.108.045105, PhysRevResearch.5.033046},
which can be engineered on modern quantum simulators \cite{RevModPhys.95.035002}.

The ground-state properties of the antiferromagnetic spin-1/2 Heisenberg chain are
governed by the SU(2)$_k$ Wess-Zumino-Witten nonlinear sigma model at level $k=1$ \cite{tsvelik_2003}.
This field theory contains a topological $\theta$ term originating from the spin-Berry phase which is relevant for half-integer spin chains and renders them critical \cite{PhysRevLett.50.1153}, but which is absent for quantum rotor models. Consequently, dissipation effects on the spin-$1/2$ chain are expected to be different from previous results on rotor models. The SO(4)
symmetry of the SU(2)$_1$ Wess-Zumino-Witten model manifests in the quantum spin chain by being
critical in both spin and dimer channels. It has been shown that the spin channel can develop long-range antiferromagnetic order if the spin chain is coupled to ohmic site dissipation that conserves a global SO(3) symmetry between system and bath \cite{PhysRevLett.129.056402}. While short-range interactions cannot break the continuous spin-rotational symmetry in 1D due to the Mermin-Wagner theorem  \cite{PhysRevLett.17.1133, PhysRev.158.383}, its requirements are not fulfilled in the presence of a long-range retarded interaction. Eventually, ohmic site dissipation is a marginally relevant perturbation that leads to long-range order for any finite coupling to the bath \cite{PhysRevLett.129.056402}.

It is natural to ask if dissipation can also lead to long-range order in the dimer channel, as anticipated with a valence-bond-solid (VBS) ground state.
The VBS state breaks a $\mathds{Z}_2$ translational symmetry by forming dimers between neighboring spins; therefore, it can already be induced by short-range interactions, \eg, by a second-nearest-neighbor spin exchange in the $J_1$-$J_2$ model \cite{PhysRevB.25.4925}
or a dimer-dimer interaction in the $J$-$Q$ model \cite{PhysRevB.84.235129, PhysRevLett.107.157201, PhysRevB.98.014414}.
Another way to obtain dimerization is via the
spin-Peierls instability \cite{PhysRevB.10.4637, PhysRevB.19.402}, where a coupling to phonons induces a periodic lattice distortion accompanied by VBS order. Theoretical studies of the spin-Peierls
problem usually consider the coupling to a single phonon frequency which drives a Berezinskii-Kosterlitz-Thouless (BKT) quantum phase
transition from the critical Luttinger-liquid (LL) phase to a VBS state
\cite{%
PhysRevB.57.R14004,
PhysRevLett.81.3956,
PhysRevLett.83.408,
PhysRevLett.83.195,
PhysRevB.60.6566,
PhysRevB.72.024434,
PhysRevB.74.214426,
2007arXiv0705.0799M,
PhysRevLett.115.080601}.
In all of these problems,
the critical properties are governed by the SU(2)$_1$ Wess-Zumino-Witten model.
It is an open problem whether this picture remains valid if we replace the single phonon frequency by a continuous dissipative bath spectrum, by which the interaction range of the corresponding retarded dimer-dimer interaction changes from an exponential to a power-law decay, respectively.
Relatedly, we can ask how sensitive the properties of the spin-Peierls transition are to the phonon density of states, as real materials might couple to multiple phonon modes \cite{PhysRevB.108.045147}.

In this paper, we study the 1D spin-Peierls model with a gapless bosonic spectrum $\propto \omega^s$
as a function of the bath exponent $s$. For our simulations, we used a recently developed
QMC method for retarded interactions \cite{PhysRevLett.119.097401}, which is based on the directed-loop algorithm \cite{PhysRevE.66.046701} and allows for an efficient sampling of the continuous bath spectrum \cite{PhysRevB.105.165129}.
For $s<1$, we provide evidence that any finite coupling to the bath induces VBS order, whereas for $s>1$ the critical LL phase remains stable up to a critical dissipation strength.
To characterize this dissipation-induced quantum phase transition, we analyze the finite-size dependence of the excitation gaps using
QMC level spectroscopy techniques \cite{PhysRevLett.115.080601}.
Although the spin chain is coupled to a gapless bath,
the spin triplet gap remains well defined and we use it to confirm that the dynamical critical exponent remains $z=1$ within the critical phase. Furthermore, we perform a finite-size-scaling analysis of the total energy and the triplet gap to estimate the central charge and the scaling dimension at the critical point, which are in excellent agreement with the SU(2)$_1$ Wess-Zumino-Witten model. We conclude
that our dissipative model undergoes a BKT transition from the critical to the VBS phase that has the same properties as the transition in the $J_1$-$J_2$ model, the $J$-$Q$ model, or the spin-Peierls model.
In particular, our results suggest that the long-range nature of the dissipative dimer-dimer interaction does not change the critical properties compared to the exponentially-decaying retardation range in the spin-Peierls model.
Therefore,
we do not expect the phonon density of states to have a significant effect on the nature of the 1D spin-Peierls quantum phase transition.

The paper is organized as follows. In Sec~\ref{Sec:Model}, we define the dissipative spin-Peierls model, in Sec.~\ref{Sec:Method}, we describe our QMC method, in Sec.~\ref{Sec:Results}, we present our results, and in Sec.~\ref{Sec:Conclusions} we conclude.

\section{Model\label{Sec:Model}}

We consider the 1D Heisenberg chain coupled to a dissipative bosonic bath,
\begin{align}
\label{Eq:Hamiltonian}
\hat{H}
	=
	- \sum_{b} \Big[J + \sum_k \lambda_k \big( \bcr{bk} + \ban{bk} \big) \Big] \proj{b}
	+ \sum_{bk} \omega_k \bcr{bk} \ban{bk} \, ,
\end{align}
which we have written in terms of the
spin-singlet projector $\proj{b} = \big[\frac{1}{4} - \spin{i(b)} \cdot \spin{i(b)+1} \big]$.
The first term in Eq.~\eqref{Eq:Hamiltonian} describes a nearest-neighbor exchange interaction
between spin-1/2 operators $\spin{i}$ defined on sites $i\in\{1, \dots, L\}$ of a 1D lattice. The antiferromagnetic Heisenberg exchange $J>0$ is modulated via a mode-dependent coupling $\lambda_k$ to harmonic oscillators
sitting on the links $b$ between neighboring sites $i(b)$ and $i(b)+1$; here, $\bcr{bk}$ ($\ban{bk}$) creates (annihilates) a
boson at bond $b$ and in mode $k$ with frequency $\omega_k$. For a single mode of frequency $\omega_0$,
Eq.~\eqref{Eq:Hamiltonian} is the well-known spin-Peierls model, which is coupled to optical bond phonons. In this paper, we consider the coupling to a continuous
dissipative spectrum
\begin{align}
\label{Eq:spec}
J(\omega)
	&=
	\pi \sum_{k} \lambda_{k}^2 \, \delta(\omega - \omega_{k}) \\
	&=
	2\pi \alpha \, \omegac^{1-s} \omega^s 
	\quad \mathrm{for}
	\quad 0< \omega < \omegac \, .
\label{Eq:spec_cont}
\end{align}
In the last step, we have taken the continuum limit and introduced a power-law spectrum with bath exponent $s$ as well as the dimensionless spin-boson coupling $\alpha$;
beyond the cutoff frequency $\omegac$, $J(\omega)$ is zero. Our parameterization of $J(\omega)$ in Eq.~\eqref{Eq:spec_cont} follows the convention for dissipative impurity models \cite{doi:10.1080/14786430500070396}.

Hamiltonian \eqref{Eq:Hamiltonian} is quadratic in the bosonic operators; therefore,
the trace over the bosonic Hilbert space can be calculated exactly and
the partition function becomes
\begin{align}
\label{eq:Zint}
Z =
Z_0
\Tr_\mathrm{s} \hat{\mathcal{T}}_\tau \, e^{-\S} \, .
\end{align}
Here, $Z_0$ is the contribution of the free-boson part and $\hat{\mathcal{T}}_\tau$
the time-ordering operator; note that the time-ordered exponential is defined in the interaction representation
via its Dyson expansion and the imaginary-time labels $\tau$, which appear in the following, are mainly required
to establish time ordering at each expansion order \cite{PhysRevB.105.165129}.
This  representation is convenient for the operator-based formulation of our QMC method discussed in Sec.~\ref{Sec:Method}.
We obtain $\S = \S_{\mathrm{s}} +\S_{\mathrm{ret}}$ with
\begin{gather}
\label{eq:SJ}
\S_{\mathrm{s}}
	=
	 - J \int_0^\beta d\tau \sum_{b} \proj{b}(\tau) \, , \\
\S_\mathrm{ret}
	=
	- \iint_0^\beta d\tau d\tau' \sum_{b}
	\proj{b}(\tau) \, K(\tau-\tau') \, \proj{b}(\tau') \, .
\label{eq:Slam}
\end{gather}
The coupling to the bosons has generated a retarded interaction mediated by the bath propagator 
\begin{align}
\label{Eq:prop}
K(\tau) 
	= 
	 \int_0^{\infty} d\omega \,
	\frac{J(\omega)}{\pi}
	\frac{\cosh[\omega (\beta/2 - \tau)]}{2\sinh[\omega \beta /2]} \, ,
\end{align}
where
$0 \leq \tau < \beta$ and $K(\tau+\beta) = K(\tau)$. 
Here, $\beta=1/T$ is the inverse temperature.
The power-law spectrum in Eq.~(\ref{Eq:spec_cont}) yields a long-range interaction in imaginary time with $K(\tau) \propto 1/\tau^{1+s}$ for $\omegac \tau \gg 1$,
whereas a single mode $\omega_0$ leads to an exponential decay, $K(\tau) \propto \exp(-\omega_0 \tau)$.

To avoid a sign problem in our QMC simulations,
we have defined our Hamiltonian in Eq.~\eqref{Eq:Hamiltonian} in
terms of the spin-singlet projectors $\proj{b}$, which include an additional shift of $1/4$
to the Heisenberg exchange interaction.
Because of these shifts,
the retarded interaction in Eq.~\eqref{eq:Slam} includes a term that is linear in the
Heisenberg exchange interaction and therefore leads to a renormalized coupling
$J' = J + \alpha \, \omegac /s$, where $J'$ is the exchange coupling of
the Hamiltonian defined without the shifts in $\proj{b}$
(for details see App.~\ref{App:Rep}).
In addition, the coupling to bond phonons leads to a finite expectation value
of the bosonic displacements \cite{costa2023comparative}, \ie, $\expvtext{\bcr{bk} + \ban{bk}} \neq 0$,
which further renormalizes the effective exchange coupling \cite{PhysRevLett.83.195}.
None of these effects change the physical properties of our system, but they
only renormalize the energy scales and parameters which we use to describe them.
In the following, we use $J=1$ as the unit of energy because it is most convenient
for our QMC simulations.
Moreover, we set $\omegac / J = 10.0$ for the cutoff frequency as well as $\hbar, k_\mathrm{B}=1$,
and use periodic boundary conditions.
Note that the high-energy cutoff $\omega_\mathrm{c}$ will affect the absolute values of critical couplings, but not the critical properties or the RG (ir)relevance of the bath as a perturbation to the spin chain, as these properties are determined by the long-range decay of the retarded interaction.

At zero dissipation ($\alpha = 0$), the ground state of the isolated spin chain is
critical in the spin and dimer channels, which becomes visible in the corresponding correlation functions.
At equal times and large distances, their asymptotic behavior is given by
\cite{1989JPhA...22..511A, PhysRevB.39.2562, PhysRevB.39.4620}
\begin{align}
\label{eq:spin_r_decay}
C_\mathrm{s}(r) &= (-1)^r \frac{\gamma}{r} \ln^{1/2}(r) \, ,
\\
C_\mathrm{d}(r) &= (-1)^r \frac{\gamma}{r} \ln^{-3/2}(r) \, ,
\label{eq:dimer_r_decay}
\end{align}
where $\gamma=1/(2\pi)^{3/2}$ \cite{affleck_exact_1998, PhysRevB.96.134429};
for microscopic definitions of the correlation functions, see Eqs.~\eqref{eq:spin_r} and \eqref{eq:dimer_r} below.
Note that spin and dimer correlations have different logarithmic corrections to their $1/r$ decay, which stem from a marginally irrelevant operator that is present in the quantum spin chain.
Because the spin chain fulfills conformal invariance, we just need to replace
$r \to \sqrt{r^2 + (v_\mathrm{s} \tau)^2}$ in Eqs.~\eqref{eq:spin_r_decay} and \eqref{eq:dimer_r_decay}
to obtain the corresponding time-dependent correlation functions ($v_\mathrm{s}$ is the spin velocity).
If we ignore the marginally-irrelevant operator, an RG treatment of the long-range interaction
in Eq.~\eqref{eq:Slam} allows us to estimate the relevance of this perturbation \cite{cardy_1996, Laflorencie_2005}: for $s<1$ ($s>1$) the coupling to the bath is always relevant (irrelevant), whereas for $s=1$ it is marginal. Because both spin and dimer operators have a scaling dimension of 1/2, the same analysis holds in the case of site dissipation considered in Ref.~\cite{PhysRevLett.129.056402}.

\section{Method\label{Sec:Method}}

For our simulations, we used an exact QMC method for
retarded interactions \cite{PhysRevLett.119.097401} that is based on a diagrammatic
expansion of the partition function \eqref{eq:Zint} in the full exponent $\S$, \ie,
\begin{align}
\label{Eq:expansion}
\frac{Z}{Z_0}
	=
	\sum_\alpha \sum_{n=0}^\infty \frac{(-1)^n}{n!}
	\braket{\alpha |
	\hat{\mathcal{T}}_\tau \, \S^n  | \alpha} \, .
\end{align}
Here, $n$ is the expansion order and we have rewritten $\Tr_\mathrm{s}$ as a sum
over all spin states $\ket{\alpha}=\ket{s^z_1, \dots, s^z_L}$ in the local $\spinz{}$ eigenbasis.
Beyond that, we introduce the superindex $\nu=\{t_\mathrm{vert}, \tilde\nu\}$ which contains the variable $t_\mathrm{vert}$ which distinguishes
between the two types of vertices $\S_\mathrm{s}$ and $\S_\mathrm{ret}$
as well as another set of variables $\tilde\nu$ which correspond to the sums and integrals
within each vertex. Then,  $\S = - \sum_\nu \S_\nu$ and we can write
Eq.~\eqref{Eq:expansion} as
\begin{align}
\frac{Z}{Z_0}
	=
	\sum_\alpha \sum_{n=0}^\infty \sum_{\nu_1 , \dots , \nu_n} 
	\frac{1}{n!}
	\braket{\alpha |
	\hat{\mathcal{T}}_\tau \, \S_{\nu_1} \dots \S_{\nu_n}  | \alpha} \, .
\label{Eq:expansion2}
\end{align}
In a final step, we apply the time-ordering operator $\hat{\mathcal{T}}_\tau$
to sort all operators within the product $\S_{\nu_1} \dots \S_{\nu_n}$
with respect to their time variables. Then, by subsequent application of the operators $\S_\nu$ on the initial state $\ket{\alpha} \equiv \ket{\alpha_0}$ we obtain the propagated state $\ket{\alpha_l}$, such that the expectation value factorizes into products of vertex weights
$W_\nu$, so that we finally obtain (we will define $W_\nu$ for each vertex further below)
\begin{align}
\frac{Z}{Z_0}
	=
	\sum_\alpha \sum_{n=0}^\infty \sum_{\nu_1 , \dots , \nu_n} 
	\frac{1}{n!} \prod_{p=1}^n W_{\nu_p}
	 \, .
\end{align}
Because our expansion is based on an interaction representation around the free-spin part $\S_0 = 0$ \cite{PhysRevB.105.165129}, the time evolution of each operator is trivial and
does not lead to any additional factors in the weights, as it is, \eg, the case in the worm algorithm \cite{Prokofev:1998aa}. Thus, we have arrived at a representation that is equivalent
to the stochastic series expansion \cite{PhysRevB.43.5950}, which allows us
to apply the efficient directed-loop updating scheme developed within this framework \cite{PhysRevB.59.R14157, PhysRevE.66.046701}.
A generalization of the directed-loop updates to retarded interactions has been
discussed in Ref.~\cite{PhysRevLett.119.097401} and previously been applied
to the 2D spin-Peierls model \cite{PhysRevB.103.L041105}. In the following, we will not repeat
the details of this algorithm, but only define the vertex weights for the dissipative spin-Peierls model.

First, we consider the Heisenberg vertex with 
$t_\mathrm{vert} = \mathrm{s}$, which contains the variables
$\tilde\nu =\{ a, b, \tau \}$. Here, $b$ and $\tau$ are the bond and
imaginary-time variables of the vertex, as apparent from Eq.~\eqref{eq:SJ},
and $a$ is an additional index that distinguishes between the diagonal ($a=1$) and
off-diagonal ($a=2$) parts of the singlet projector, so that 
$\S_{\mathrm{s}, \tilde{\nu}} = J \, \proj{a,b}(\tau)$.
We have
\begin{align}
\label{Eq:proj_diag}
\proj{a=1,b}(\tau) &= \frac{1}{4} - \spinz{i(b)}(\tau) \, \spinz{i(b)+1}(\tau) \, , \\
\proj{a=2,b}(\tau) &= - \frac{1}{2} \left[\spinp{i(b)}(\tau) \, \spinm{i(b)+1}(\tau) + \Hc \right] \, ,
\label{Eq:proj_off}
\end{align}
where $\hat{S}^{\pm}_i = \hat{S}^{x}_i \pm \im \hat{S}^{y}_i$ are the local spin-flip operators.
From this, we obtain the well-known vertex weights of the Heisenberg model,
\begin{align}
W_{\mathrm{s},\tilde{\nu}}
= J \braket{\alpha_{l} | \proj{a,b}| \alpha_{l-1}} \, .
\end{align}
Note that we have dropped the time label $\tau$ which is implicitly contained
in the label $l$ of the propagated state.
As usual, the constant shift of $1/4$ (or larger) in Eq.~\eqref{Eq:proj_diag}
and the cancellation of the minus sign in Eq.~\eqref{Eq:proj_off} on bipartite lattices
(via a sublattice rotation) lead to positive Monte Carlo weights. 
The only difference
to the original formulation of the method \cite{PhysRevB.59.R14157, PhysRevE.66.046701} is the presence of the imaginary-time variable $\tau$, which
requires us to formulate the diagonal updates in a different way. As suggested
in Ref.~\cite{PhysRevLett.119.097401}, diagonal updates are performed using a Metropolis scheme, in which we propose to add or remove diagonal vertices with variables $\{b, \tau\}$ chosen randomly within their range of definition. The construction of the directed-loop updates remains unchanged.

For the retarded interaction vertex with $t_\mathrm{vert} = \mathrm{ret}$,
the vertex variables are $\tilde\nu=\{ \omega, a, b, \tau, a', \tau' \}$ and the vertex reads
$\S_{\mathrm{ret}, \tilde{\nu}} = \proj{a,b}(\tau)\, \tilde{K}(\omega,\tau-\tau') \, \proj{a',b}(\tau')$. In our case, both projectors act on the same bond $b$, therefore we include the bond variable only once in $\tilde\nu$. As a result, the vertex weight becomes
\begin{align}
\nonumber
W_{\mathrm{ret}, \tilde{\nu}} =
\tilde{K}(\omega, \tau - \tau')
& \braket{\alpha_l |  \proj{a, b} |\alpha_{l-1}} \\
\times & \braket{\alpha_{l'} | \proj{a', b} |\alpha_{l'-1}} \, .
\label{eq:weight_ret}
\end{align}
As before, the time labels can be dropped from the spin operators, but they need to be kept for $\tilde{K}$ because here the specific time values affect the total weight of the vertex.
The weight of the retarded vertex, $W_{\mathrm{ret}, \tilde{\nu}}$, factorizes into three independent parts. In particular, the nonlocal interaction splits into two subvertices which can be updated independently of each other during the directed-loop updates, as if they were local Heisenberg vertices; with the shift of $1/4$ in Eq.~\eqref{Eq:proj_diag} the loops can even be constructed deterministically \cite{PhysRevB.59.R14157}. During the diagonal updates, we use a Metropolis scheme to add/remove the product of the two diagonal subvertices at different positions of the world-line configuration. While the first time variable is chosen from a uniform distribution, the second one is chosen according to $\tilde{K}$, for which we also sample the frequency dependence according to the bath spectrum $J(\omega)$ contained in $\tilde{K}$ (for details see Ref.~\cite{PhysRevB.105.165129}).
Because $\tilde{K}$ is a global prefactor in Eq.~\eqref{eq:weight_ret}, once sampled during the diagonal updates, it does not appear during the directed-loop updates anymore,
so that the remaining parts of the algorithm stay unaffected.

The calculation of observables follows the standard procedure in the
interaction representation \cite{PhysRevB.56.14510}.
We consider the equal-time correlation functions
\begin{align}
\label{eq:spin_r}
C_\mathrm{s}(r)
	&=
	\big\langle {\spinx{r} \spinx{0}} \big\rangle \, ,
	\\
C_\mathrm{d}(r)
	&=
	\big\langle {\big[\spinz{r} \spinz{r+1} - D \big] \big[ \spinz{0} \spinz{1} - D \big] } \big\rangle
\label{eq:dimer_r}
\end{align}
as a function of distance $r$ as well as their Fourier transforms
$C_\mathrm{s/d}(q) = \frac{1}{L} \sum_r e^{\im q r} C_\mathrm{s/d}(r)$
with momentum transfer $q$. 
The spin correlations $C_\mathrm{s}(r)$ along the $x$ orientation can be calculated efficiently
during the construction of the directed loop, whereas the dimer correlations
$C_\mathrm{d}(r)$ can be accessed from the propagated state; for the latter,
we subtract the expectation value of the dimer operator,
$D=\frac{1}{L} \sum_i \expvtext{\spinz{i} \spinz{i+1}}$.

We also calculate the dynamical correlation functions
\begin{align}
\label{Eq:sus_spin}
\chi_\mathrm{s}(q,\im\Omega_m)
	&=
	\int_0^\beta d\tau  \sum_{r} e^{\im (\Omega_m \tau - q r)}
	\expvtext{\spinx{r}(\tau) \spinx{0}(0)}	\, ,
\\
\label{Eq:sus_bond}
\chi_\mathrm{d}(q,\im \Omega_m)
	&=
	\int_0^\beta d\tau \sum_{r} e^{\im ( \Omega_m \tau - q r)}
	\expvtext{\proj{r}(\tau)  \proj{0}(0)} \, ,
\end{align}
which can be accessed directly in Matsubara
 frequencies $\Omega_m = 2\pi m / \beta$ with $m \in \mathds{Z}$.
Again, the former is obtained by tracking the propagation of the directed loop,
whereas the latter is recovered from the distribution of
vertices \eqref{eq:SJ} in the perturbation expansion \cite{PhysRevB.56.14510}.

Furthermore, we calculate the total energy of the dissipative spin-Peierls chain.
Because the bosons have been integrated out, the corresponding observables
cannot be accessed directly from the Monte Carlo configurations.
However, it has been shown that bosonic observables can be recovered
from higher-order spin correlation functions with the help of generating functionals
\cite{PhysRevB.94.245138}. The latter can be obtained efficiently from
the distribution of vertices \cite{PhysRevB.94.245138}.

To distinguish between the critical LL phase and the VBS phase, we
also calculate the spin stiffness
\begin{align}
\rho_\mathrm{s} = \left. \frac{1}{L} \frac{\partial^2 F(\phi)}{\partial\phi^2} \right\vert_{\phi = 0} \, ,
\end{align}
which is defined via the second derivative of the free energy with respect
to a twist $\phi$ in the spin orientation \cite{doi:10.1063/1.3518900}.
In our QMC simulations, it can be calculated efficiently from the winding-number fluctuations of the world-line configurations, which remains valid even in the
presence of the retarded dimer-dimer interaction \cite{PhysRevResearch.2.023013}.

\section{Results\label{Sec:Results}}

In this section, we present our QMC results for the dissipative spin-Peierls
model. In Sec.~\ref{Sec:Gap}, we determine the dynamical critical exponent
from a finite-size analysis of excitation gaps,  in Sec.~\ref{Sec:subohmic}, we
study the formation of VBS order in the sub-ohmic regime, in Sec.~\ref{Sec:superohmic}, we characterize the LL--VBS transition in the super-ohmic regime, and in Sec.~\ref{Sec:ohmic},
we approach the ohmic case.

\subsection{Finite-size gaps and dynamical critical exponent\label{Sec:Gap}}

The analysis of finite-size gaps
is a powerful tool to get precise information on the properties
of quantum phases and their phase transitions.
Their finite-size scaling gives direct access to the dynamical critical exponent $z$ or
the spin velocity $v_\mathrm{s}$, but this approach has also been particularly successful in
determining the critical coupling
of the LL--VBS transition in the frustrated $J_1$-$J_2$ Heisenberg chain
\cite{OKAMOTO1992433, PhysRevB.54.R9612}, even for the small system sizes
accessible to exact-diagonalization studies.
For the quantum spin chain, we consider
the lowest-energy excitations from the ground state in the spin-singlet and spin-triplet sectors. In
the thermodynamic limit, both excitation gaps are zero in the critical LL phase,
whereas the triplet gap remains finite in the VBS phase. For finite systems,
the lowest excitation of the LL (VBS) phase lies within the triplet (singlet) sector,
so that the gaps show a crossing as a function of the coupling parameter that drives
the transition. For the $J_1$-$J_2$ model, these crossings can be extrapolated
with high precision, because the functional form of the leading-order correction term is known
\cite{OKAMOTO1992433}.
This methodology has been of great advantage to determine the LL--VBS transition point, as many other estimators suffer from large finite-size corrections at the BKT transition, which are often hard to control.
Recently, Ref.~\cite{PhysRevLett.115.080601} has introduced an unbiased QMC gap estimator and
applied the gap-crossing technique to pinpoint the LL--VBS transition in the 1D spin-Peierls model.
In the following, we want to test this novel estimator for the dissipative spin-Peierls model.

\subsubsection{Triplet and singlet gap estimators}

The triplet and singlet gaps can be obtained from the dynamical spin and dimer
correlation functions defined in Eqs.~\eqref{Eq:sus_spin} and \eqref{Eq:sus_bond}, respectively, using the generalized gap estimator derived in Ref.~\cite{PhysRevLett.115.080601},
\begin{align}
\label{Eq:gap_est}
\Delta^{(n,\beta)}_{\mathrm{s/d}}(q) = \Omega_1 \sqrt{\frac{-\sum_{m=0}^n m^2 \, x_{nm} \, \chi_\mathrm{s/d}(q,\im\Omega_m)}
				{\sum_{m=0}^n x_{nm} \, \chi_\mathrm{s/d}(q,\im\Omega_m)}} \, ,
\end{align}
where $x_{nm} =[\prod_{j=0,j\neq m}^n (m+j)(m-j)]^{-1}$. This estimator
makes use of the analytical structure of the correlation functions, in particular that
a finite-size gap leads to an exponential decay at long imaginary times \cite{PhysRevLett.115.080601}.
For $n=1$, Eq.~\eqref{Eq:gap_est} reduces to the well-known estimator for the inverse correlation length \cite{doi:10.1063/1.3518900} defined along imaginary time.
To obtain an unbiased gap estimate,
$\Delta^{(n,\beta)}$ needs to be converged in the control parameter $n$ and in inverse temperature $\beta$.
A detailed analysis performed in Ref.~\cite{PhysRevLett.115.080601} and its Supplemental Material revealed that $\Delta^{(n,\beta)}$ usually converges quickly with $n$, so that $n\approx 5$ was sufficient for the spin-Peierls model with a single boson frequency.
For convergence in temperature, Ref.~\cite{PhysRevLett.115.080601} suggests to run simulations at
a temperature of the order of the gap, \ie,
$\beta \, \Delta^{(n,\beta)}/2\pi \approx 1$, because for larger $\beta$ statistical fluctuations  are
strongly enhanced with increasing $n$.
Note that this scheme does not require
fitting the long-range decay of the imaginary-time correlation functions, for which the choice of an interval to fit the numerical
data may lead to an unnecessary bias.
Further details on the gap estimators can be found in Ref.~\cite{PhysRevLett.115.080601}.

\subsubsection{Temperature convergence of the gap estimates}

\begin{figure}
\includegraphics[width=\linewidth]{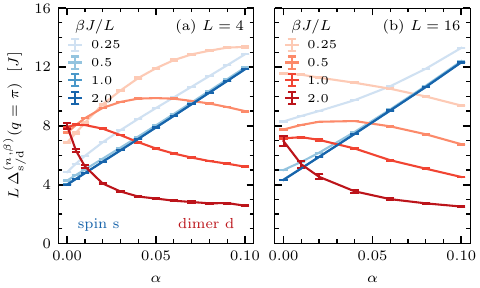}
\caption{%
Triplet/singlet gap estimates $\Delta_\mathrm{s/d}^{(n,\beta)}(q=\pi)$ (rescaled with system size $L$) as a function of the
spin-boson coupling $\alpha$ for different inverse temperatures $\beta$
and system sizes (a) $L=4$ and (b) $L=16$. Here, $s=1.0$ and $n=5$.
}
\label{fig:gap_convergence_s1.0}
\end{figure}
We first study the temperature dependence of the gap estimates for the dissipative spin-Peierls model which, in contrast to the case with a single boson frequency studied in Ref.~\cite{PhysRevLett.115.080601}, contains a gapless continuum of bath modes. The triplet/singlet gaps are obtained from $\Delta_{\mathrm{s/d}}^{(n,\beta)}(q=\pi)$, respectively. Results are shown in Fig.~\ref{fig:gap_convergence_s1.0}
for different inverse temperatures $\beta$ as well as for system sizes $L=4$ and $L=16$ using an ohmic bath
with a bath exponent of $s=1.0$; here we used a fixed projection parameter of $n=5$.
For both system sizes and all couplings $\alpha$, our estimates of the triplet gap $\Delta_\mathrm{s}$ converge quickly with $\beta$ and lie on top of each other for $\beta J /L =1.0$ and $2.0$;
this is in agreement with the condition
$\beta \, \Delta^{(n,\beta)}/2\pi \gtrsim 1$ to get converged results in $\beta$ \cite{PhysRevLett.115.080601}. We have also convinced ourselves that $n=5$ is sufficient to get converged results for $\Delta_\mathrm{s}$ at fixed $\beta$, as it is the case in the presence of a single boson frequency \cite{PhysRevLett.115.080601}. As a result, the triplet gap $\Delta_\mathrm{s}$ remains finite in the presence of the dissipative bath and can be calculated reliably.
By contrast, our estimates for the singlet gap $\Delta_\mathrm{d}$ converge to a finite value only at $\alpha=0$, but steadily decrease with increasing $\beta$ for any $\alpha> 0$. These results suggest that the singlet gap is zero for any finite system size when coupled to an ohmic bath. Note that the convergence of the singlet-gap estimator with the control parameter $n$ is slower than for $\Delta_\mathrm{s}$ and still shows a small decrease beyond $n=5$ (which does not change our conclusions). A slower convergence with $n$ has been explained in the Supplemental Material of Ref.~\cite{PhysRevLett.115.080601} by the presence of a continuous excitation spectrum.

We can understand the difference in the finite-size gaps from the fact that the bosons do not couple to the individual spin operators but to the
spin-singlet projector $\proj{b}$, which enters the dynamical dimer structure factor $\chi_\mathrm{d}(q,\im \Omega_m)$. Since
$\proj{b}$ is coupled linearly to the displacements of the harmonic oscillators, there is an exact relation between $\chi_\mathrm{d}(q,\im \Omega_m)$ and the
dynamical structure factor of the bosons.
As a result, the corresponding spectral functions
contain the same spectral
information, only reweighted differently \cite{PhysRevB.91.235150}. If a system is coupled to a single bosonic mode $\omega_0$,
the boson spectrum shows, besides other features, a renormalized boson frequency $\tilde{\omega}_q$ which deviates from the bare frequency $\omega_0$.
Vice versa, the system's structure factor, which is derived from the operator that is coupled to the bosons
(in our case, this is $\proj{b}$), also includes the bosonic signal at $\tilde{\omega}_q$,
as observed for 1D electron systems coupled to site or bond phonons \cite{PhysRevB.83.115105, PhysRevB.91.245147}, whereas the
spectra of other operators do not show features at $\tilde{\omega}_q$. For the spin-Peierls model with a single boson frequency,
the phonon spectrum has been studied in Ref.~\cite{2007arXiv0705.0799M}, and it was observed that $\tilde{\omega}_q$ does not appear in $\chi_\mathrm{s}(q,\omega)$.
From this we can conclude that if a single bosonic mode enters the dimer structure factor at $\omega \approx \omega_0$, then the gapless spectrum of a
continuous bosonic bath leads to low-energy excitations in $\chi_\mathrm{d}(q,\omega)$, irrespective of the finite system size.
On the other side, if a single mode does not leave a clear signature at $\omega \approx \omega_0$ in $\chi_\mathrm{s}(q,\omega)$, then a continuum of modes will not, either.

Our analysis of finite-size gaps on small system sizes reveals that the spin-triplet gap remains well defined in the presence of a continuous bath, whereas the singlet sector seems to be strongly affected by the bosonic modes.
We have convinced ourselves that this remains true for $s>1$ at system sizes of $L=4$. Since the low-energy contribution of the bath spectrum $J(\omega)\propto \omega^s$ is reduced with increasing $s$, it remains open how this affects the estimate of the singlet gap at larger $L$ and $\beta$. Because our QMC estimator for $\chi_\mathrm{d}(q,\im\Omega_m)$ has substantially larger statistical fluctuations than the one for $\chi_\mathrm{s}(q,\im\Omega_m)$,
our subsequent analysis focuses on the finite-size dependence of the spin-triplet gaps, which can be estimated with good accuracy. For all of these reasons, we do not apply the gap-crossing technique to determine the critical couplings.

\subsubsection{Dynamical critical exponent\label{Sec:Dynamical_z}}

\begin{figure}
\includegraphics[width=\linewidth]{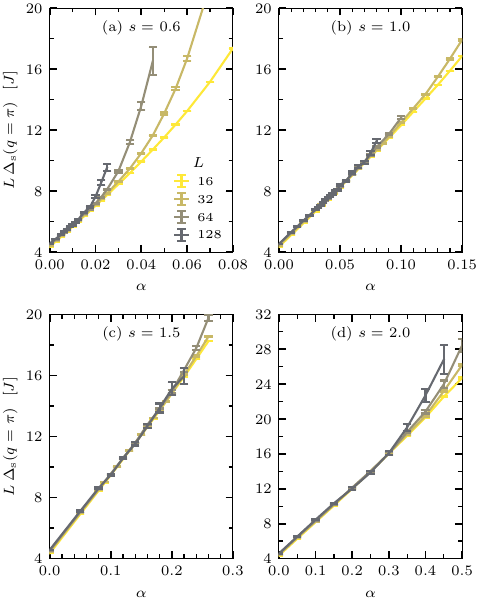}
\caption{%
Finite-size dependence of the rescaled triplet gap $L \, \Delta_\mathrm{s}(q=\pi)$ as a function of the spin-boson coupling $\alpha$ and
for different bath exponents $s$. Here, $\beta J = L$ and $n=5$.
}
\label{fig:gap_different_s}
\end{figure}

The existence of the triplet gap
allows us to get access to the dynamical critical exponent $z$, which is well defined at a quantum phase transition as well as in a critical phase like the LL phase. We have
\begin{align}
\Delta_\mathrm{s}(q=\pi) \propto L^{-z} \, ,
\end{align}
up to logarithmic corrections which are expected to vanish only at the quantum critical point \cite{JulienHaldane83, 1989JPhA...22..511A}.
Because we have $z=1$ for the isolated Heisenberg chain at $\alpha=0$, we assume that
this is also valid if the LL phase remains stable at finite $\alpha$. Therefore, we plot the rescaled gap
$L \, \Delta_\mathrm{s}(q=\pi)$ in Fig.~\ref{fig:gap_different_s} for different system sizes $L$.
For all bath exponents $s$ considered in Fig.~\ref{fig:gap_different_s}, we find a weak-coupling regime
in which all data sets seem to fall on top of each other for different $L$,
before they start to deviate again at stronger couplings. For $s=0.6$ in Fig.~\ref{fig:gap_different_s}(a),
we observe that the values of $\alpha$ for which a curve starts to deviate from the others shift towards lower
$\alpha$ with increasing $L$. This is consistent with the RG prediction that the bath is a relevant perturbation,
such that the LL phase will not be stable at finite $\alpha$. At the marginal point of $s=1.0$ shown in Fig.~\ref{fig:gap_different_s}(b), the regime in which different curves fall on top of each other appears more extended
than at $s=0.6$; because the marginal case is the most delicate to analyze, we will come back to it in Sec.~\ref{Sec:ohmic}. For $s=1.5$ and $s=2.0$
shown in Figs.~\ref{fig:gap_different_s}(c) and \ref{fig:gap_different_s}(d), $L \, \Delta_\mathrm{s}(q=\pi)$ has converged over
a wide parameter range, from which we conclude that $z=1$ is valid in this regime. Note that the weak
finite-size dependence observed within the LL regime is consistent with logarithmic corrections that are expected to be present.

Within the critical phase, the rescaled spin-triplet gap $L \, \Delta_\mathrm{s}(q=\pi)$ depends almost linearly on $\alpha$, which means that the effective exchange coupling is strongly renormalized by the
coupling to the bath. To first order, this is an artifact of our model definition, because we included shifts of $1/4$ in the singlet projectors to avoid a negative-sign problem in our QMC simulations.
As discussed in Sec.~\ref{Sec:Model}, these shifts lead to an effective exchange
coupling $J'=J+\alpha \, \omegac / s$ that varies linearly with $\alpha$, but the spin-boson interaction also induces an additional renormalization that is not captured in this simple redefinition of parameters. Because in our simulations we keep $\beta J = L$ constant for all couplings $\alpha$,
we do not fulfill the condition $\beta \, \Delta^{(n,\beta)}/2\pi \approx 1$ for optimal statistics of our gap estimates;
therefore, we observe increasing error bars with increasing $\alpha$.

Having established that the dynamical critical exponent is $z=1$ greatly simplifies our finite-size-scaling analysis
in the subsequent sections, as we can choose $\beta J = L$ for all simulations. We will return to the gap estimates further below when we extract the central
charge at the LL--VBS transition.

\subsection{VBS order in the sub-ohmic regime\label{Sec:subohmic}}

In the sub-ohmic regime, where the bath exponent fulfills $0<s<1$, the coupling
to the bosonic bath is a relevant perturbation in the RG sense
and we expect that any finite coupling $\alpha$ destabilizes the critical phase of the isolated spin chain.
In the spin-Peierls model, the interaction with a single bosonic mode
eventually induces VBS order; therefore, we also expect VBS order to appear in
the dissipative system.

\begin{figure}[t]
\includegraphics[width=\linewidth]{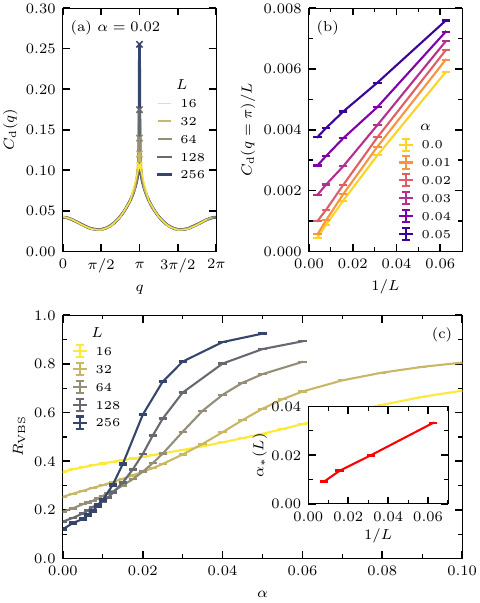}
\caption{%
(a) Finite-size dependence of the dimer structure factor $C_\mathrm{d}(q)$
at $\alpha=0.02$.
(b) Finite-size dependence of the VBS order parameter, $C_\mathrm{d}(q=\pi)/L$,
for different spin-boson couplings $\alpha$.
(c) VBS correlation ratio $R_\mathrm{VBS}$ as a function of $\alpha$ for different system sizes $L$.
The inset shows the crossings $\alpha_\ast(L)$ between
data pairs $(L,2L)$
as a function of $1/L$.
Here, $s=0.6$ and $\beta J = L$.
}
\label{fig:VBS_ratio_s0.6}
\end{figure}
Figure \ref{fig:VBS_ratio_s0.6}(a) shows the dimer structure factor
$C_\mathrm{d}(q)$ for a dissipative bath with $s=0.6$ at $\alpha=0.02$; $C_\mathrm{d}(q)$ has a peak at the VBS
ordering vector $q=\pi$ that diverges with increasing system size $L$. To detect
long-range order, we study the finite-size dependence of the VBS order parameter, $C_\mathrm{d}(q=\pi)/L$,
in Fig.~\ref{fig:VBS_ratio_s0.6}(b). At $\alpha=0$, we know that the order parameter scales to zero,
because $C_\mathrm{d}(q=\pi)$ only grows logarithmically with $L$ within the LL phase.
For any finite $L$, $C_\mathrm{d}(q=\pi)/L$ increases monotonically with increasing $\alpha$.
We infer from Fig.~\ref{fig:VBS_ratio_s0.6}(b) that already at $\alpha=0.02$ the order parameter
extrapolates to a finite value for $L\to\infty$.

To analyze the onset of VBS order in more detail,
we calculate the VBS correlation ratio
\begin{align}
\label{Eq:VBS_ratio}
R_\VBS = 1 - \frac{C_\mathrm{d}(q = \pi+\delta q)}{C_\mathrm{d}(q = \pi)}
\end{align}
from the correlation function $C_\mathrm{d}(q)$ at the VBS ordering vector $q=\pi$
and the closest momentum $q=\pi+\delta q$ on a finite lattice, where the shift by the
momentum resolution $\delta q = 2\pi/L$
takes into account the long-wavelength fluctuations near the ordering vector. In the ordered
phase, $C_\mathrm{d}(q = \pi) \propto L$ so that $R_\VBS \to 1$ for $L \to \infty$, whereas in a disordered phase
$C_\mathrm{d}(q = \pi) \to C_\mathrm{d}(q = \pi + \delta q)$ so that $R_\VBS \to 0$.
At a critical point, $R_\VBS$ becomes scale invariant; in its vicinity $R_\VBS$ captures
$(\xi/L)^2$ where $\xi$ is the correlation length.

Figure \ref{fig:VBS_ratio_s0.6}(c) displays $R_\VBS$ at $s=0.6$ as a function of $\alpha$
and for different system sizes $L$. For large $\alpha$, we find that $R_\VBS$ scales to one
with increasing $L$, while it decreases for $\alpha \to 0$. In between, data
pairs of system sizes $(L,2L)$ exhibit a crossing at $\alpha_\ast(L)$ which we extract
and plot as a function of $1/L$ in the inset of Fig.~\ref{fig:VBS_ratio_s0.6}(c).
Extrapolation of the pseudocritical coupling $\alpha_\ast(L)$ to $L\to\infty$ will give us an estimate
of the critical coupling at which VBS order occurs. From our finite-size estimate
of $\alpha_\ast(L)$ at the largest available $L$ we infer that at $\alpha=0.01$ the system is already in the ordered phase;
note that, at this coupling, it is not yet possible to identify long-range order from the
order parameter in Fig.~\ref{fig:VBS_ratio_s0.6}(b), which has less favorable behavior under finite-size scaling. The finite-size dependence of $\alpha_\ast(L)$
is consistent with a critical coupling of zero, because $\alpha_\ast(L)$ bends down for the largest system size
available. A quantitative extrapolation is difficult because the precise fit function is not known. Our result is in agreement with the RG prediction, that the bath is a relevant perturbation for $s<1$. For dissipation strengths that are close to the unstable fixed point at $\alpha=0$, we expect that the crossover towards the VBS fixed point leads to an initially slow RG flow.

\subsection{LL--VBS transition in the super-ohmic regime\label{Sec:superohmic}}

In the super-ohmic regime where the bath exponent is $s>1$,
the dissipative bath is an irrelevant RG perturbation to the quantum spin chain.
Therefore, we expect the critical LL phase to remain stable for weak spin-boson
couplings. In the following, we study the dissipation-induced
LL--VBS transition at $s=2.0$.

\begin{figure}
\includegraphics[width=\linewidth]{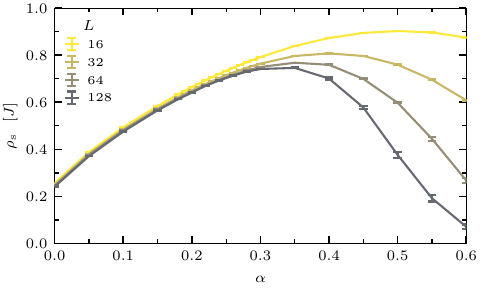}
\caption{%
Spin stiffness $\rho_\mathrm{s}$ as a function of the bath coupling $\alpha$ for different system sizes.
Here, $s=2.0$ and $\beta J = L$.
}
\label{fig:stiffness_s2.0}
\end{figure}
To get a first impression of the ground-state properties as a function
of the dissipation strength $\alpha$, Fig.~\ref{fig:stiffness_s2.0} shows
the spin stiffness $\rho_\mathrm{s}$ for different system sizes at fixed $\beta J = L$.
Because $\rho_\mathrm{s}$ measures the response of the system to a twist
in the spin orientation, we have $\rho_\mathrm{s}>0$ in the critical phase,
whereas $\rho_\mathrm{s} \to 0$ in the VBS phase. Note that the initial increase of $\rho_\mathrm{s}(\alpha)$ in Fig.~\ref{fig:stiffness_s2.0} follows from the renormalization of the effective Heisenberg exchange, which is a consequence of how we chose the unit of energy, as discussed in Sec.~\ref{Sec:Model}.
Our results suggest that the LL--VBS transition in the dissipative spin-Peierls model is of BKT type, for which $\rho_\mathrm{s}$ exhibits a discontinuous jump in the thermodynamic limit, from $\rho_\mathrm{s}>0$ at $\alpha < \alpha_\mathrm{c}$ to $\rho_\mathrm{s}= 0$ at $\alpha > \alpha_\mathrm{c}$. It is characteristic for BKT transitions that this discontinuity gets substantially smeared out at finite lattice sizes, as it is also the case for our results. Finite-size analysis of $\rho_\mathrm{s}$ is further complicated by logarithmic corrections which are expected to be present for all $\alpha < \alpha_\mathrm{c}$ due to a marginally irrelevant operator in the field-theory description; this has also been discussed
for $\rho_\mathrm{s}$ in the spin-Peierls model \cite{PhysRevLett.83.195}. For the isolated Heisenberg chain,
exact results from the Bethe ansatz reveal that the logarithmic corrections of the spin stiffness lead to a significant drop of $\rho_\mathrm{s}$ \cite{2001EPJB...24...77L} which cannot be estimated reliably from the system sizes considered in Fig.~\ref{fig:stiffness_s2.0}. Hence, we will apply different measures to estimate the critical coupling.

Our subsequent analysis of the quantum phase transition relies on what is known from bosonization studies of  the LL--VBS transition
\cite{tsvelik_2003, 10.1093/acprof:oso/9780198525004.001.0001}: The isolated spin chain contains a marginally irrelevant operator that is related to Umklapp scattering. In the $J_1$-$J_2$ model, the nearest-neighbor exchange reduces the prefactor of this operator until it tunes through zero; once its sign has changed, Umklapp scattering becomes relevant and leads to VBS order.
This picture does not only apply to the frustrated spin chain, but also to other systems like the $J$-$Q$ model \cite{PhysRevB.98.014414} or the spin-Peierls chain; for the latter, the RG is supposed to generate the corresponding terms in the bosonized theory from the retarded spin interaction \cite{10.1093/acprof:oso/9780198525004.001.0001}.
A characteristic feature of this description is that the logarithmic corrections of the correlation functions [\cf, Eqs.~\eqref{eq:spin_r_decay} and \eqref{eq:dimer_r_decay}], which stem from the marginally irrelevant operator, disappear exactly at the quantum phase transition \cite{JulienHaldane83, 1989JPhA...22..511A}.
As a result, 
the dynamical spin and dimer correlation functions show the same asymptotic decay, \ie,
$\chi_\mathrm{s/d}(r,\tau) \propto 1/\sqrt{r^2 + (v_\mathrm{s} \tau)^2}$. In the absence of
logarithmic corrections, it follows that $\chi_\mathrm{s/d}(q=\pi) \propto L$ exactly and only at the critical point.
Hence, we can analyze $\chi_\mathrm{s/d}(q=\pi) / L$ for different
$L$ and estimate the critical coupling from a finite-size extrapolation
of the crossings between data pairs $(L,2L)$. For the spin-Peierls model,
such an analysis had also been applied in Refs.~\cite{PhysRevLett.83.195, PhysRevLett.115.080601}.

\begin{figure}
\includegraphics[width=\linewidth]{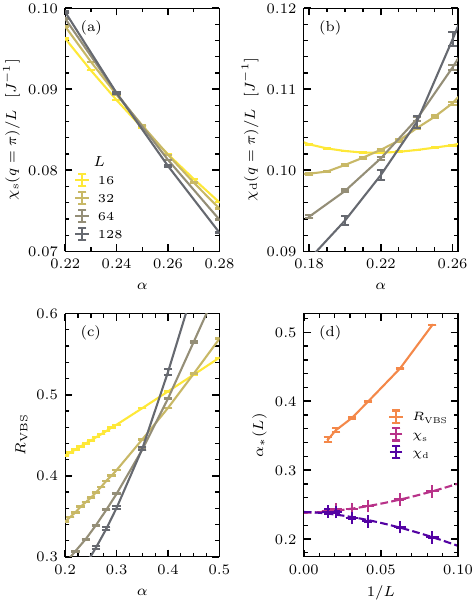}
\caption{%
Finite-size analysis to extract the critical coupling of the dissipative spin chain at bath exponent $s=2.0$.
(a) Spin susceptibility, (b) dimer susceptibility, and (c) VBS correlation ratio
as a function of the bath coupling $\alpha$ for different system sizes $L$.
(d) Finite-size dependence of the crossings $\alpha_\ast(L)$
between data sets $(L,2L)$.
Dashed lines represent fits to the form $\alpha_\ast(L) = \alpha_\mathrm{c} + a_0 \, L^{-a_1}$,
from which we estimate $\alpha_\mathrm{c}(\chi_\mathrm{s}) = 0.238(2)$ and
$\alpha_\mathrm{c}(\chi_\mathrm{d}) = 0.239(5)$.
We use $\beta J = L$.
}
\label{fig:crossings_s2.0}
\end{figure}

Figures \ref{fig:crossings_s2.0}(a) and \ref{fig:crossings_s2.0}(b) show a
finite-size analysis of the rescaled susceptibilities $\chi_\mathrm{s}(q=\pi)/L$ and $\chi_\mathrm{d}(q=\pi)/L$, respectively.
For both susceptibilities, the crossings between data pairs $(L,2L)$ define the pseudocritical
couplings $\alpha_\ast(L)$ which extrapolate to the same critical coupling, \ie,
$\alpha_\mathrm{c}(\chi_\mathrm{s}) = 0.238(2)$ and
$\alpha_\mathrm{c}(\chi_\mathrm{d}) = 0.239(5)$,
as shown in Fig.~\ref{fig:crossings_s2.0}(d). 
We also extract $\alpha_\ast(L)$ for the correlation ratio $R_\mathrm{VBS}$
shown in Fig.~\ref{fig:crossings_s2.0}(c);
$\alpha_\ast(L; R_\VBS)$ exhibits strong finite-size corrections and, for the available system sizes, does not seem to extrapolate to the same value as the crossings of $\chi_\mathrm{s/d}$. 
To understand this discrepancy, we have repeated the same analysis
for the spin-Peierls model, which is presented in App.~\ref{App:SpinPeierls}.
For the spin-Peierls model, the extrapolated crossings of $\chi_\mathrm{s/d}(q=\pi)/L$ are in good agreement with the critical value
that had been determined via gap estimation \cite{PhysRevLett.115.080601}, whereas the correlation ratio again deviates significantly. The correlation ratio also exhibits strong finite-size corrections near the critical point of
the XXZ chain, for which the critical anisotropy is known exactly (not shown).
A possible explanation might be related to the fact that beyond the BKT transition the gap of the ordered phase only opens up exponentially slowly.
From this comparison, we judge that the correlation ratio is less reliable in estimating
the finite critical coupling than the susceptibilities.

\begin{figure}
\includegraphics[width=\linewidth]{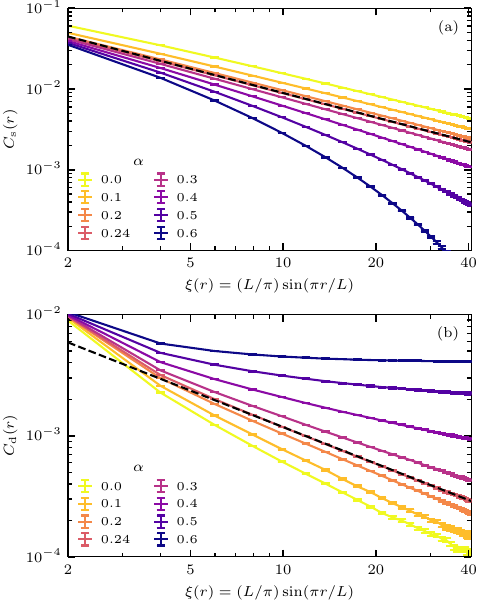}
\caption{%
Real-space spin/dimer correlation functions $C_\mathrm{s/d}(r)$ as a
function of conformal distance
$\xi(r) = (L/\pi) \sin(\pi r / L)$ for different bath couplings $\alpha$.
Results are shown on a log-log scale, for which the dashed lines indicate a $1/r$ decay.
Here, $s=2.0$, $L=128$, and $\beta J = L$
(for $\alpha=0.0$ we use $\beta J =2L$).
}
\label{fig:realspace_s2.0}
\end{figure}
Figure~\ref{fig:realspace_s2.0} shows the equal-time spin and dimer correlation
functions for $L=128$. Our analysis in Sec.~\ref{Sec:Dynamical_z} confirmed a dynamical critical exponent of $z=1$, therefore
we plot $C_\mathrm{s/d}(r)$ as a function of the conformal distance $\xi(r) = (L/\pi) \sin(\pi r / L)$, which
eliminates boundary effects. Within the critical phase, both correlation
functions exhibit a power-law decay, whereas at strong coupling,
$C_\mathrm{d}(r)$ approaches a constant and $C_\mathrm{s}(r)$ decays faster than a power law,
which is the expected behavior within the VBS phase. At the weakest couplings $\alpha$,
the bending of the curves indicates the multiplicative logarithmic corrections which are present within the critical phase.
The logarithmic corrections are expected to disappear at our estimated critical coupling of $\alpha_\mathrm{c} \approx 0.24$, which is confirmed by the excellent agreement with a $1/r$ decay (black dashed line) over almost all available distances $r$. Already at $\alpha = 0.3$, the dimer correlations decay slower than $1/r$, whereas the spin correlations decay slightly faster, indicating that this data point is already in the ordered phase.

\begin{figure}
\includegraphics[width=\linewidth]{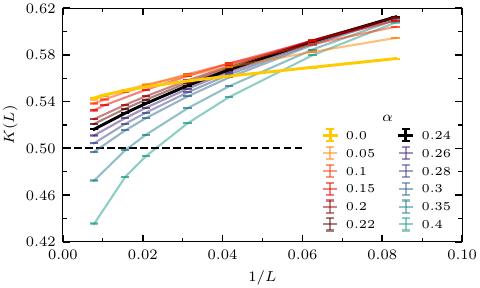}
\caption{%
Finite-size extrapolation of the Luttinger parameter $K(L)$
for different spin-boson couplings $\alpha$. 
The black dashed line indicates the expected Luttinger parameter of the critical phase, \ie,
$K=1/2$.
Here, $s=2.0$ and $\beta J = L$.
}
\label{fig:Krho_s2.0}
\end{figure}
Within the critical phase, the long-distance behavior of the correlation functions is determined by the Luttinger parameter $K$. In the absence of logarithmic corrections,
we have $C_\mathrm{s}(r) = -K/(2\pi^2 r^2) + \tilde{\gamma} \left(-1\right)^r r^{-2K}$. A finite-size estimate of the Luttinger parameter
can be obtained from the spin correlation function at small momenta, \ie,
\begin{align}
K(L) = L \, C_\mathrm{s}(q = 2\pi/L) \, .
\end{align}
Figure \ref{fig:Krho_s2.0} shows $K(L)$ as a function of $1/L$ for different spin-boson couplings $\alpha$. At $\alpha=0$, $K(L\to\infty)= 1/2$ is an exact result, but it is also known that logarithmic corrections make it impossible to perform a naive finite-size extrapolation. However, the absence
of logarithmic corrections at the critical coupling allows for a reliable extrapolation
and we confirm that $K=1/2$ is still valid at $\alpha_\mathrm{c} \approx 0.24$. Therefore, $K(L)$
is expected to scale towards $K=1/2$ for all $\alpha < \alpha_\mathrm{c}$.
By contrast, for any $\alpha > \alpha_\mathrm{c}$ we observe that there is a finite $L$ for which $K(L)$ drops below $1/2$, so that $K(L\to\infty)$ is expected to scale to zero; close to the quantum phase transition, this will only happen very slowly, though.

\begin{figure}
\includegraphics[width=\linewidth]{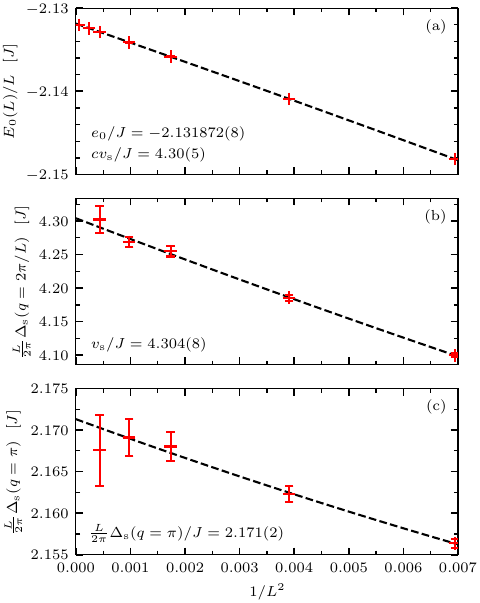}
\caption{%
Finite-size estimation of the central charge $c$ and the scaling dimension $x$ at
$\alpha = 0.24 \approx \alpha_\mathrm{c}$ and $s=2.0$ from (a) the total energy and the rescaled triplet gaps at (b) $q= 2\pi/L$ and (c) $q=\pi$.
(a) We fit the energy density to the form $E_0(L)=e_0 - \pi c v_\mathrm{s}/6 L^{2} + a_3/L^3$
(dashed line), from which we estimate $e_0$ and $c v_\mathrm{s}$, as stated in the panel.
Here, $\beta J = L$. We have subtracted the free-boson contribution to the energy, because it is not uniquely defined for a given spectrum $J(\omega)$. (b) From fitting $(L/2\pi) \Delta_\mathrm{s}(q=2\pi/L) = v_\mathrm{s} +a_2/L^2 + a_4 / L^4$ we estimate the spin velocity $v_\mathrm{s}$. Here, $\beta J = L/4$.
(c) We perform an equivalent fit for $(L/2\pi) \Delta_\mathrm{s}(q=\pi)$. Here, $\beta J = L/2$.
Eventually, we determine $c=0.99(1)$ and $x = 0.504(1)$.
We have checked that our fits do not change beyond error bars if we use different higher-order correction terms. For the high-precision gap estimates, we used a projection parameter of $n=20$.
}
\label{fig:c_s2.0}
\end{figure}
We can determine the properties of the underlying conformal field theory
from the characteristic finite-size dependence of the lowest energy levels at the critical point $\alpha_\mathrm{c} \approx 0.24$; we follow Ref.~\cite{PhysRevLett.115.080601} where the same analysis has been performed for the spin-Peierls model with a single phonon frequency $\omega_0$. At sufficiently large system sizes, the ground-state energy $E_0$ fulfills \cite{PhysRevLett.56.742, PhysRevLett.56.746, 1989JPhA...22..511A}
\begin{align}
\label{eq:E0L}
E_0(L) = e_0 L - \frac{\pi c v_\mathrm{s}}{6 L} + {\scriptstyle \mathcal{O}}(1/L) \, ,
\end{align}
where $e_0$ is the energy density and $v_\mathrm{s}$ the spin velocity of the infinite system.
In particular, the leading corrections to $e_0$ give us access to the central charge $c$.
Figure~\ref{fig:c_s2.0}(a) shows $E_0(L)/L$ as a function of $1/L^2$ as well as a fit to the
scaling form of Eq.~\eqref{eq:E0L}, from which we determine the product $c v_\mathrm{s}/J =4.30(5)$.
Because the spin-triplet gap remains well defined for all momenta $q$ in the presence of the dissipative bath, we
can estimate the spin velocity $v_\mathrm{s}$ from the rescaled spin gap at the smallest momentum transfer, \ie, $v_\mathrm{s} = (L/2\pi) \Delta_\mathrm{s}(q=2\pi/L)$. Extrapolation of the spin gap in Fig.~\ref{fig:c_s2.0}(b)
gives $v_\mathrm{s}/J=4.304(8)$. From the ratio of the two estimates, we obtain a central charge
of $c=0.99(1)$.
Furthermore, we can get access to the scaling dimension $x$ from the ratio of the triplet
gaps \cite{PhysRevLett.115.080601}, \ie, $x=\Delta_\mathrm{s}(q=\pi) / \Delta_\mathrm{s}(q=2\pi/L)$. To this end, we determine
$(L/2\pi) \Delta_\mathrm{s}(q=\pi)/J = 2.171(2)$ in Fig.~\ref{fig:c_s2.0}(c). Hence,
we obtain $x=0.504(1)$, which is in good agreement with a Luttinger exponent of $K=1/2$ (the small deviations beyond error bars might be a result of probing the system slightly away from the critical point).
All in all, our estimates for $c$ and $x$ provide strong evidence that the critical point is described by the SU(2)$_1$ Wess-Zumino-Witten model, which has $c=1$ and $x=1/2$.

\subsection{Approaching the ohmic case\label{Sec:ohmic}}

For ohmic dissipation with a bath exponent of $s=1$,
the coupling to the bath is a marginal RG perturbation to
the isolated spin chain. A marginal coupling usually comes with a logarithmically slow RG flow
and therefore leads to substantial finite-size corrections. To understand the
case of ohmic dissipation, we will approach the limit $s \to 1$ from the sub- and super-ohmic
regimes studied in the previous sections.

\begin{figure}
\includegraphics[width=\linewidth]{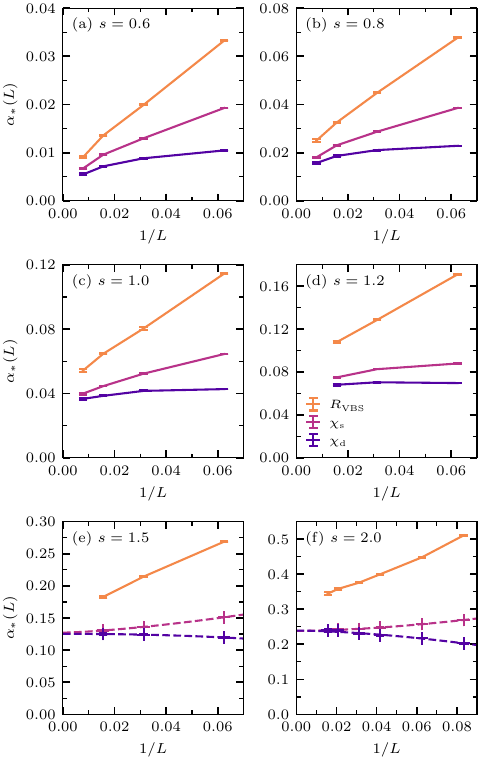}
\caption{%
Finite-size dependence of the pseudocritical coupling $\alpha_\ast(L)$
extracted from $R_\mathrm{VBS}$, $\chi_\mathrm{s}$, and $\chi_\mathrm{d}$
for different bath exponents $s$. For each observable, we determine $\alpha_\ast(L)$
from the crossing between data pairs $(L,2L)$;
raw data are collected in Fig.~\ref{fig:crossing_various_s}.
Dashed lines in panels (e) and (f) represent fits to the form $\alpha_\ast(L) = \alpha_\mathrm{c} + a_0 \, L^{-a_1}$; for $s=1.5$ we determine
$\alpha_\mathrm{c}(\chi_\mathrm{s}) = 0.127(2)$ and
$\alpha_\mathrm{c}(\chi_\mathrm{d}) = 0.125(6)$, whereas for $s=2.0$
results are given in the caption of Fig.~\ref{fig:crossings_s2.0}.
}
\label{fig:crossings_all}
\end{figure}
Figure \ref{fig:crossings_all} shows the finite-size dependence of the pseudocritical coupling $\alpha_\ast(L)$ for $R_\VBS$, $\chi_\mathrm{s}$, and $\chi_\mathrm{d}$
for different bath exponents $s$.
The corresponding raw data can be found in App.~\ref{App:data}, from which the
crossings have been determined as described in Sec.~\ref{Sec:superohmic}.
At $s=0.6$ [Fig.~\ref{fig:crossings_all}(a)], the dissipative spin chain is deep in the
sub-ohmic regime, where the bath is a relevant perturbation. We have already seen
in Sec.~\ref{Sec:subohmic} that the crossings of $R_\VBS$ scale to zero for $L\to\infty$;
the same conclusions can be drawn from the crossings of $\chi_\mathrm{s}$ and $\chi_\mathrm{d}$,
which are also included in Fig.~\ref{fig:crossings_all}(a).
Already at $s=0.8$ in Fig.~\ref{fig:crossings_all}(b), finite-size corrections
have significantly increased and are expected to be largest at $s=1.0$ in
Fig.~\ref{fig:crossings_all}(c). For the latter, a naive extrapolation
of $\alpha_\ast(L)$ would suggest a small but finite critical coupling.
However, the presence of logarithmic corrections makes naive
finite-size extrapolation highly unreliable. This becomes clear
from the finite-size dependence of the Luttinger parameter $K(L)$
for the isolated spin chain in Fig.~\ref{fig:Krho_s2.0};
for the available system sizes,
linear extrapolation would suggest $K(L\to \infty)\approx 0.535$, although
we know from the exact solution of the Heisenberg chain that $K=1/2$.
If we assume that finite-size corrections of similar size occur in Fig.~\ref{fig:crossings_all}(c),
we cannot exclude that $\alpha_\ast(L)$ scales to zero at $s=1$.
Moreover, the logarithmic corrections at the marginal point will also affect the RG flow 
at bath exponents near $s=1$; therefore, we observe
larger finite-size corrections at $s=0.8$ than at $s=0.6$, but
results in Fig.~\ref{fig:crossings_all}(b) are still in agreement with VBS order appearing at any $\alpha>0$.

Our analysis of the LL--VBS transition in the super-ohmic regime in Sec.~\ref{Sec:superohmic}
was based on the observation that logarithmic corrections vanish exactly at the critical point;
therefore, we were able to extrapolate the crossings of $\chi_\mathrm{s/d}$ at $s=2.0$ reliably to $L\to \infty$.
However, as we tune the bath exponent closer towards $s=1$, we expect additional finite-size corrections
to appear because of the proximity to the marginal case. At $s=1.5$
in Fig.~\ref{fig:crossings_all}(e) and $s=2.0$ in Fig.~\ref{fig:crossings_all}(f), the pseudocritical couplings for
$\chi_\mathrm{s}$ and $\chi_\mathrm{d}$ converge with opposite curvature towards $L\to\infty$, so that
we can assume the critical coupling to lie in between the two curves. For $s=2.0$, we observed in Sec.~\ref{Sec:superohmic} that logarithmic corrections disappear at the critical coupling, which, \eg, becomes visible in the finite-size dependence of $K(L)$ in Fig.~\ref{fig:Krho_s2.0}, and obtained
$\alpha_\mathrm{c}(\chi_\mathrm{s}) = 0.238(2)$ and
$\alpha_\mathrm{c}(\chi_\mathrm{d}) = 0.239(5)$.
In the same way, we can extrapolate $\alpha_\ast(L;\chi_\mathrm{s/d})$ at $s=1.5$, as shown in Fig.~\ref{fig:crossings_all}(e), and get
$\alpha_\mathrm{c}(\chi_\mathrm{s}) = 0.127(2)$ and
$\alpha_\mathrm{c}(\chi_\mathrm{d}) = 0.125(6)$; we obtained good fits for the available data points but assumed that no additional corrections appear at larger system sizes. At $s=1.2$ in Fig.~\ref{fig:crossings_all}(d), the system is already close to the marginal case
and the finite-size dependence of $\alpha_\ast(L;\chi_\mathrm{s/d})$ suggests that
additional correction terms are present; as a result, the fitting form which we used for $s=1.5$ and $s=2.0$ does not converge anymore. In the absence of an appropriate fitting function, we roughly extrapolate $\alpha_\ast(L;\chi_\mathrm{s/d})$ to
obtain $\alpha_\mathrm{c} = 0.06(1)$; because $s=1.2$ is close to the marginal point, we likely still overestimate the critical coupling.

\begin{figure}
\includegraphics[width=\linewidth]{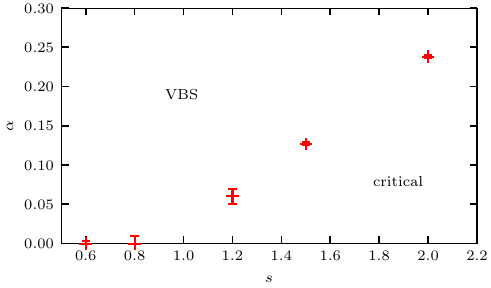}
\caption{%
Phase diagram of the dissipative spin-Peierls model as a function of the bath exponent $s$ and the
spin-boson coupling $\alpha$. Beyond the critical coupling, the critical phase of the 1D Heisenberg chain becomes unstable towards a long-range ordered VBS phase. The critical couplings have been determined as described in the main text. For $s<1$, the error bars indicate the uncertainty of the extrapolation towards $\alpha=0$.
}
\label{fig1:phasediagram}
\end{figure}
All of our estimated critical couplings are collected in Fig.~\ref{fig1:phasediagram},
which shows the phase diagram of the dissipative spin-Peierls model
as a function of  the bath exponent $s$ and the spin-boson coupling $\alpha$.
The evolution of $\alpha_\mathrm{c}$ for $s>1$ is consistent with a critical coupling
of zero at $s=1.0$, but the large finite-size corrections close to the marginal case do not allow for a definite answer based on the available system sizes.

\section{Conclusions\label{Sec:Conclusions}}

We studied the effects of bond dissipation on the 1D antiferromagnetic spin-1/2 Heisenberg model, for which a coupling to a gapless bosonic spectrum $\propto \omega^s$ leads to a retarded dimer-dimer interaction with a power-law decay $\propto 1/|\tau - \tau'|^{1+s}$. For $s<1$ the bath is a relevant perturbation to the isolated spin chain and induces VBS order for any $\alpha>0$, whereas for $s>1$ the critical phase remains stable up to a finite critical coupling. Although the dimer operator couples to a gapless bosonic spectrum, we found that the spin triplet sector retains a well-defined finite-size gap, from which we confirmed that the dynamical critical exponent remains $z=1$ throughout the critical phase.
To characterize the quantum phase transition from the critical to the VBS phase, we applied various measures 
which are
consistent with the prevailing picture of the LL--VBS transition in SU(2)-symmetric quantum spin chains, \ie, that VBS order is induced once the marginally-irrelevant operator changes sign. The resulting absence of logarithmic corrections at the critical point has been observed in different observables; in particular, spin and dimer susceptibilities give consistent estimates for the critical coupling, indicating that conformal invariance holds and that the critical point is described by the SU(2)$_1$ Wess-Zumino-Witten model. The latter has been confirmed via finite-size estimation of the central charge
and the scaling dimension at criticality.

Our results suggest that the critical properties at the LL--VBS transition are the same as for the spin-Peierls model. The latter only couples to a single bosonic mode leading to an exponential decay in the retarded dimer-dimer interaction. As a result, the BKT quantum phase transition in the spin-Peierls model seems to be rather robust with regards to the spectral density of the bosons and to the retardation range (as long as $s>1$ in the dissipative case). This raises the question under which circumstances long-range interactions in space or time will change the ground-state and critical properties compared to their short-range counterparts and when they remain the same.

Bond dissipation represents only one possible channel to induce long-range order in the antiferromagnetic quantum spin chain.
Because of the emergent SO(4) symmetry of the Wess-Zumino-Witten fixed-point theory, coupling a dissipative bath to one component of a local spin operator should lead to similar results as in the dimer channel. By now, it has been confirmed that ohmic dissipation in the $\spinz{i}$ channel is a marginally relevant perturbation to the quantum spin chain \cite{PhysRevLett.113.260403}, whereas a study of interaction-range effects is still missing. Similarly, the coupling to an ohmic bath
is a marginally relevant perturbation for SO(3)-symmetric dissipation in the spin channel
\cite{PhysRevLett.129.056402}. In this case, the finite-size and finite-temperature dependence of the spin susceptibility suggests that dissipation has a more pronounced effect on the dynamical properties at criticality
for $s>1$. An analysis of non-ohmic dissipation with an SO(3) symmetry remains open for future studies, but previous work on dissipative quantum rotor models \cite{PhysRevB.85.214302} as well as on long-range interactions in space \cite{Laflorencie_2005} suggest a quantum phase transition with $z\neq 1$.
Moreover, in the presence of a magnetic field and a spin anisotropy \cite{PhysRevLett.113.260403}, QMC results suggest a more complicated phenomenology beyond the SU(2)-symmetric case,
motivating further studies of quantum dissipative spin chains.

\begin{acknowledgments}
I thank F.~Assaad and R.~Moessner for motivating discussions;
in particular, for pointing out the SO(4) symmetry of the spin-1/2 Heisenberg chain.
Furthermore, I acknowledge discussions with P.~Patil and Z.~Wang.
This work was supported by the Deutsche Forschungsgemeinschaft
through the W\"urzburg-Dresden Cluster of Excellence on Complexity and Topology
in Quantum Matter---\textit{ct.qmat} (EXC 2147, Project No. 390858490).
\end{acknowledgments}

\appendix

\section{Collection of additional data\label{App:data}}

\begin{figure*}
\includegraphics[width=\linewidth]{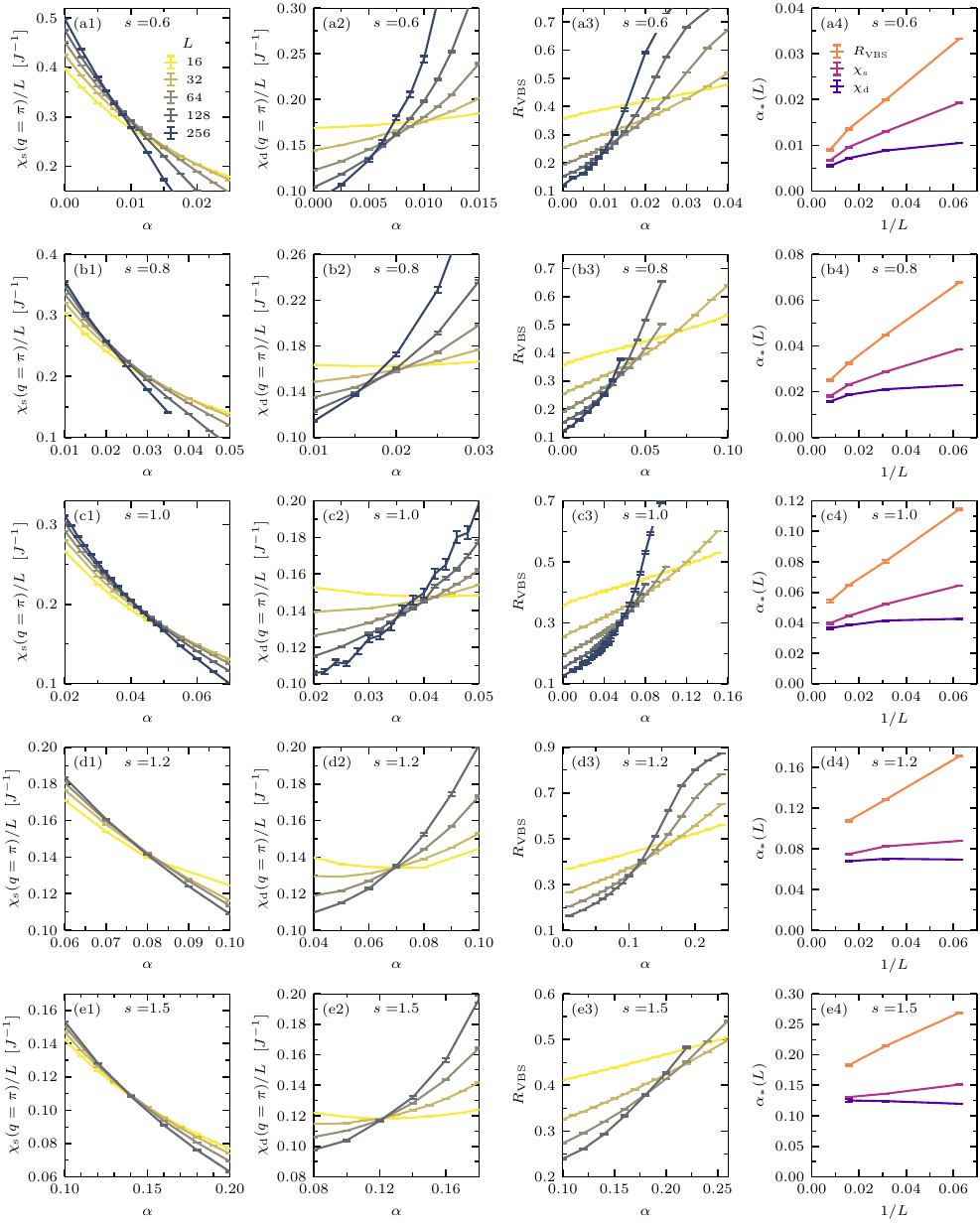}
\caption{%
Collection of the spin susceptibility, the dimer susceptibility, the VBS correlation ratio, and the pseudocritical coupling for bath exponents $s \in \{ 0.6, 0.8, 1.0, 1.2, 1.5\}$, for which we have not shown the crossing analysis in the main text. Here $\beta J = L$.
}
\label{fig:crossing_various_s}
\end{figure*}
Figure \ref{fig:crossing_various_s} shows a collection of all data that have been used
to determine the pseudocritical couplings in Fig.~\ref{fig:crossings_all}, but have not been shown in the main text.

\section{Different definitions of the dissipative spin-Peierls model\label{App:Rep}}

In the following, we will give an overview over different definitions
of the (dissipative) spin-Peierls model. In first quantization, we introduce the model as
\begin{align}
\nonumber
\hat{H}_1
	=
	& \sum_{b} \left[J_1 + \sum_k \gamma_k \, \Q{bk} \right] \, \spin{i(b)} \cdot \spin{j(b)} \\
	&+ \sum_{bk} \left[ \frac{1}{2M_k} \P{bk}^2 +  \frac{K_k}{2} \Q{bk}^2 \right] \, ,
\label{Eq:H1_1}
\end{align}
where $i(b)$ and $j(b)$ are the two sites connected by bond $b$.
For the harmonic oscillators, we use the mode-dependent mass $M_k$
and stiffness constant $K_k$.
The displacement and momentum operators,
\begin{align}
\Q{bk}
	&=
	\frac{1}{\sqrt{2 M_k \omega_k}} \left( \bcrt{bk} + \bant{bk}  \right) \, ,
	\\
\P{bk}
	&=
	\im \sqrt{\frac{M_k\omega_k}{2}} \left( \bcrt{bk} - \bant{bk}  \right) \, ,
\end{align}
can be rewritten in terms of second-quantized creation and annihilation operators,
such that Eq.~\eqref{Eq:H1_1} becomes
\begin{align}
\nonumber
\hat{H}_1
	=
	& \sum_{b} \left[ J_1 + \sum_k \lambda_k \left( \bcrt{bk} + \bant{bk} \right) \right]  \spin{i(b)} \cdot \spin{j(b)} \\
	&+ \sum_{bk} \omega_k \left[ \bcrt{bk} \bant{bk} + \frac{1}{2} \right] \, .
\label{Eq:H1_2}
\end{align}
Here, we have defined the mode-dependent coupling $\lambda_k = \gamma_k / \sqrt{2 M_k \omega_k}$
and the boson frequency $\omega_k = \sqrt{K_k / M_k}$.

\begin{figure}[t]
\includegraphics[width=\linewidth]{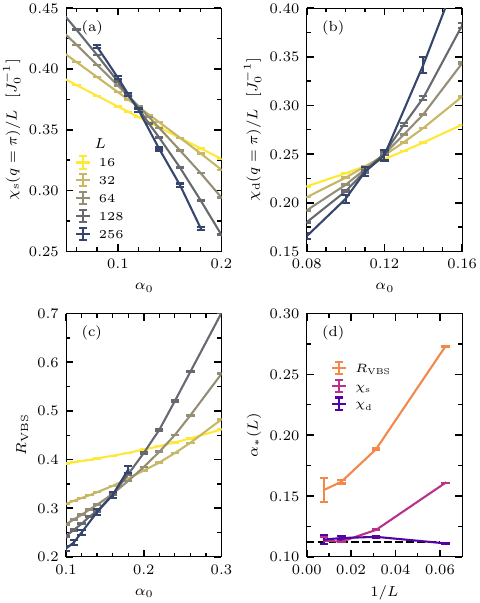}
\caption{%
Finite-size analysis to extract the critical coupling of the spin-Peierls model.
(a) Spin susceptibility, (b) dimer susceptibility, and (c) VBS correlation ratio
as a function of the spin-phonon coupling $\alpha_0$.
(d) Pseudocritical coupling $\alpha_{\ast}(L)$ extracted for each observable in (a)--(c)
from the crossings of data pairs $(L,2L)$. The dashed line indicates the
critical coupling $\alpha_{0,\mathrm{c}}^\mathrm{LS} = 0.1123(8)$ obtained from QMC level spectroscopy \cite{PhysRevLett.115.080601}.
Here, $\omega_0/J_0=0.25$ and $\beta J_0 = 2 L$.
}
\label{fig:crossings_spin-Peierls}
\end{figure}

To avoid a sign problem in our QMC simulations, we need to add a constant shift of $C\geq 1/4$ to the spin
exchange interaction. Therefore, we define
\begin{align}
\nonumber
\hat{H}_2
	=
	&\sum_{b} \left[J_2 + \sum_k \lambda_k  \left( \bcr{bk} + \ban{bk} \right) \right]
	\left[ \spin{i(b)} \cdot \spin{j(b)} - C \right] \\
	&+ \sum_{bk} \omega_k \left[ \bcr{bk} \ban{bk} + \frac{1}{2} \right] \, .
\end{align}
As before, $\hat{H}_2$ can also be written in terms of displacement and momentum operators.
The two Hamiltonians $\hat{H}_1$ and $\hat{H}_2$ are related by a mode-dependent shift of the harmonic oscillators, \ie,
$\ban{bk} = \bant{bk} +  C \lambda_k / \omega_k$,
such that
\begin{align}
\hat{H}_1
	&=
\hat{H}_2 + \sum_b \frac{1}{2} \, C \left(J_1+J_2\right) \, ,
\\
J_1 &= J_2 + \frac{2C}{\pi} \int_0^\infty d\omega \, \frac{J(\omega)}{\omega} \, .
\end{align}
In the last step, we have used the definition of $J(\omega)$ in Eq.~\eqref{Eq:spec}
to identify $\int d\omega J(\omega) / (\pi \omega) = \sum_k \lambda_k^2/\omega_k$.
In particular, we find that $\hat{H}_1$ and $\hat{H}_2$ are equivalent up to
a constant  shift in energy; however, the exchange coupling gets renormalized by including
the shift $C$.
For our simulations, we used the power-law spectrum defined in Eq.~\eqref{Eq:spec_cont}
and set $C=1/4$; as a result, we obtain $J_1 = J_2 + \alpha \, \omegac / s$.

\section{Comparison to the spin-Peierls model\label{App:SpinPeierls}}

The spin-Peierls model can be recovered from the dissipative model
by restricting the bosonic bath to a single mode, \ie,
starting from Eq.~\eqref{Eq:H1_2} we obtain
\begin{align}
\nonumber
\hat{H}_\mathrm{sp}
	=
	&\sum_{b} \Big[J_0 + \lambda_0  \big( \bcrt{b} + \bant{b} \big) \Big] \spin{i(b)} \cdot \spin{j(b)} \\
	&+ \omega_0 \sum_{b}  \bcrt{b} \bant{b} \, .
\label{Eq:spin-Peierls}
\end{align}
Here, $\omega_0$ is the frequency of an optical phonon and we
use the dimensionless coupling constant $\alpha_\mathrm{0}=\lambda_0^2/(\omega_0 J_0)$.
In contrast to the dissipative model in Eq.~\eqref{Eq:Hamiltonian}, we omit the shift of $1/4$ in our definition of the exchange interaction, so that our coupling constants are consistent with previous studies \cite{PhysRevLett.115.080601, PhysRevLett.83.195}.  Our QMC simulations follow
the procedure described in Sec.~\ref{Sec:Method}, for which we need to include shifts as discussed
in App.~\ref{App:Rep}.

For $\omega_0/J_0 =0.25$, the quantum critical coupling has been determined precisely using
level spectroscopy \cite{PhysRevLett.115.080601}.
Here,
we want to test the quality of other estimators 
to extract the critical coupling of the spin-Peierls model. In Fig.~\ref{fig:crossings_spin-Peierls}
we perform a finite-size analysis of the susceptibilities $\chi_\mathrm{s/d}(q=\pi)/L$
as well as of the correlation ratio $R_\VBS$. We find that the susceptibilities are
in good agreement with the extrapolated critical coupling obtained from level
spectroscopy, whereas the crossing analysis of $R_\VBS$ still shows major
finite-size deviations, which might be a consequence of the exponentially-slow gap opening in the VBS phase. Further results on the spin stiffness and other observables
can be found in Ref.~\cite{PhysRevLett.83.195}.

%apsrev4-2.bst 2019-01-14 (MD) hand-edited version of apsrev4-1.bst
%Control: key (0)
%Control: author (8) initials jnrlst
%Control: editor formatted (1) identically to author
%Control: production of article title (0) allowed
%Control: page (0) single
%Control: year (1) truncated
%Control: production of eprint (0) enabled
%

%\bibliography{/Users/mweber/Documents/Papers/Repositories/QMC_for_spin_Peierls/database_spin_boson}

\begin{thebibliography}{122}%
\makeatletter
\providecommand \@ifxundefined [1]{%
 \@ifx{#1\undefined}
}%
\providecommand \@ifnum [1]{%
 \ifnum #1\expandafter \@firstoftwo
 \else \expandafter \@secondoftwo
 \fi
}%
\providecommand \@ifx [1]{%
 \ifx #1\expandafter \@firstoftwo
 \else \expandafter \@secondoftwo
 \fi
}%
\providecommand \natexlab [1]{#1}%
\providecommand \enquote  [1]{``#1''}%
\providecommand \bibnamefont  [1]{#1}%
\providecommand \bibfnamefont [1]{#1}%
\providecommand \citenamefont [1]{#1}%
\providecommand \href@noop [0]{\@secondoftwo}%
\providecommand \href [0]{\begingroup \@sanitize@url \@href}%
\providecommand \@href[1]{\@@startlink{#1}\@@href}%
\providecommand \@@href[1]{\endgroup#1\@@endlink}%
\providecommand \@sanitize@url [0]{\catcode `\\12\catcode `\$12\catcode
  `\&12\catcode `\#12\catcode `\^12\catcode `\_12\catcode `\%12\relax}%
\providecommand \@@startlink[1]{}%
\providecommand \@@endlink[0]{}%
\providecommand \url  [0]{\begingroup\@sanitize@url \@url }%
\providecommand \@url [1]{\endgroup\@href {#1}{\urlprefix }}%
\providecommand \urlprefix  [0]{URL }%
\providecommand \Eprint [0]{\href }%
\providecommand \doibase [0]{https://doi.org/}%
\providecommand \selectlanguage [0]{\@gobble}%
\providecommand \bibinfo  [0]{\@secondoftwo}%
\providecommand \bibfield  [0]{\@secondoftwo}%
\providecommand \translation [1]{[#1]}%
\providecommand \BibitemOpen [0]{}%
\providecommand \bibitemStop [0]{}%
\providecommand \bibitemNoStop [0]{.\EOS\space}%
\providecommand \EOS [0]{\spacefactor3000\relax}%
\providecommand \BibitemShut  [1]{\csname bibitem#1\endcsname}%
\let\auto@bib@innerbib\@empty
%</preamble>
\bibitem [{\citenamefont {Sachdev}(2008)}]{Sachdev:2008aa}%
  \BibitemOpen
  \bibfield  {author} {\bibinfo {author} {\bibfnamefont {S.}~\bibnamefont
  {Sachdev}},\ }\bibfield  {title} {\bibinfo {title} {Quantum magnetism and
  criticality},\ }\href {https://doi.org/10.1038/nphys894} {\bibfield
  {journal} {\bibinfo  {journal} {Nature Physics}\ }\textbf {\bibinfo {volume}
  {4}},\ \bibinfo {pages} {173} (\bibinfo {year} {2008})}\BibitemShut {NoStop}%
\bibitem [{\citenamefont {Nagler}\ \emph {et~al.}(1991)\citenamefont {Nagler},
  \citenamefont {Tennant}, \citenamefont {Cowley}, \citenamefont {Perring},\
  and\ \citenamefont {Satija}}]{PhysRevB.44.12361}%
  \BibitemOpen
  \bibfield  {author} {\bibinfo {author} {\bibfnamefont {S.~E.}\ \bibnamefont
  {Nagler}}, \bibinfo {author} {\bibfnamefont {D.~A.}\ \bibnamefont {Tennant}},
  \bibinfo {author} {\bibfnamefont {R.~A.}\ \bibnamefont {Cowley}}, \bibinfo
  {author} {\bibfnamefont {T.~G.}\ \bibnamefont {Perring}},\ and\ \bibinfo
  {author} {\bibfnamefont {S.~K.}\ \bibnamefont {Satija}},\ }\bibfield  {title}
  {\bibinfo {title} {{Spin dynamics in the quantum antiferromagnetic chain
  compound ${\mathrm{KCuF}}_{3}$}},\ }\href
  {https://doi.org/10.1103/PhysRevB.44.12361} {\bibfield  {journal} {\bibinfo
  {journal} {Phys. Rev. B}\ }\textbf {\bibinfo {volume} {44}},\ \bibinfo
  {pages} {12361} (\bibinfo {year} {1991})}\BibitemShut {NoStop}%
\bibitem [{\citenamefont {Lake}\ \emph {et~al.}(2005)\citenamefont {Lake},
  \citenamefont {Tennant}, \citenamefont {Frost},\ and\ \citenamefont
  {Nagler}}]{Lake05}%
  \BibitemOpen
  \bibfield  {author} {\bibinfo {author} {\bibfnamefont {B.}~\bibnamefont
  {Lake}}, \bibinfo {author} {\bibfnamefont {D.~A.}\ \bibnamefont {Tennant}},
  \bibinfo {author} {\bibfnamefont {C.~D.}\ \bibnamefont {Frost}},\ and\
  \bibinfo {author} {\bibfnamefont {S.~E.}\ \bibnamefont {Nagler}},\ }\bibfield
   {title} {\bibinfo {title} {Quantum criticality and universal scaling of a
  quantum antiferromagnet},\ }\href {https://doi.org/10.1038/nmat1327}
  {\bibfield  {journal} {\bibinfo  {journal} {Nat Mater}\ }\textbf {\bibinfo
  {volume} {4}},\ \bibinfo {pages} {329} (\bibinfo {year} {2005})}\BibitemShut
  {NoStop}%
\bibitem [{\citenamefont {Mourigal}\ \emph {et~al.}(2013)\citenamefont
  {Mourigal}, \citenamefont {Enderle}, \citenamefont {Kl{\"o}pperpieper},
  \citenamefont {Caux}, \citenamefont {Stunault},\ and\ \citenamefont
  {R{\o}nnow}}]{Mourigal:2013aa}%
  \BibitemOpen
  \bibfield  {author} {\bibinfo {author} {\bibfnamefont {M.}~\bibnamefont
  {Mourigal}}, \bibinfo {author} {\bibfnamefont {M.}~\bibnamefont {Enderle}},
  \bibinfo {author} {\bibfnamefont {A.}~\bibnamefont {Kl{\"o}pperpieper}},
  \bibinfo {author} {\bibfnamefont {J.-S.}\ \bibnamefont {Caux}}, \bibinfo
  {author} {\bibfnamefont {A.}~\bibnamefont {Stunault}},\ and\ \bibinfo
  {author} {\bibfnamefont {H.~M.}\ \bibnamefont {R{\o}nnow}},\ }\bibfield
  {title} {\bibinfo {title} {{Fractional spinon excitations in the quantum
  Heisenberg antiferromagnetic chain}},\ }\href
  {https://doi.org/10.1038/nphys2652} {\bibfield  {journal} {\bibinfo
  {journal} {Nature Physics}\ }\textbf {\bibinfo {volume} {9}},\ \bibinfo
  {pages} {435} (\bibinfo {year} {2013})}\BibitemShut {NoStop}%
\bibitem [{\citenamefont {Toskovic}\ \emph {et~al.}(2016)\citenamefont
  {Toskovic}, \citenamefont {van~den Berg}, \citenamefont {Spinelli},
  \citenamefont {Eliens}, \citenamefont {van~den Toorn}, \citenamefont
  {Bryant}, \citenamefont {Caux},\ and\ \citenamefont {Otte}}]{Toskovic16}%
  \BibitemOpen
  \bibfield  {author} {\bibinfo {author} {\bibfnamefont {R.}~\bibnamefont
  {Toskovic}}, \bibinfo {author} {\bibfnamefont {R.}~\bibnamefont {van~den
  Berg}}, \bibinfo {author} {\bibfnamefont {A.}~\bibnamefont {Spinelli}},
  \bibinfo {author} {\bibfnamefont {I.~S.}\ \bibnamefont {Eliens}}, \bibinfo
  {author} {\bibfnamefont {B.}~\bibnamefont {van~den Toorn}}, \bibinfo {author}
  {\bibfnamefont {B.}~\bibnamefont {Bryant}}, \bibinfo {author} {\bibfnamefont
  {J.~S.}\ \bibnamefont {Caux}},\ and\ \bibinfo {author} {\bibfnamefont
  {A.~F.}\ \bibnamefont {Otte}},\ }\bibfield  {title} {\bibinfo {title} {Atomic
  spin-chain realization of a model for quantum criticality},\ }\href
  {https://doi.org/10.1038/nphys3722} {\bibfield  {journal} {\bibinfo
  {journal} {Nature Physics}\ }\textbf {\bibinfo {volume} {12}},\ \bibinfo
  {pages} {656 EP } (\bibinfo {year} {2016})}\BibitemShut {NoStop}%
\bibitem [{\citenamefont {Gao}\ \emph {et~al.}(2023)\citenamefont {Gao},
  \citenamefont {Lin}, \citenamefont {Laurell}, \citenamefont {Chen},
  \citenamefont {Huang}, \citenamefont {dela Cruz}, \citenamefont {Vemuru},
  \citenamefont {Lumsden}, \citenamefont {Nagler}, \citenamefont {Alvarez},
  \citenamefont {Dagotto}, \citenamefont {Zhou}, \citenamefont {Christianson},\
  and\ \citenamefont {Stone}}]{gao2023spinon}%
  \BibitemOpen
  \bibfield  {author} {\bibinfo {author} {\bibfnamefont {S.}~\bibnamefont
  {Gao}}, \bibinfo {author} {\bibfnamefont {L.-F.}\ \bibnamefont {Lin}},
  \bibinfo {author} {\bibfnamefont {P.}~\bibnamefont {Laurell}}, \bibinfo
  {author} {\bibfnamefont {Q.}~\bibnamefont {Chen}}, \bibinfo {author}
  {\bibfnamefont {Q.}~\bibnamefont {Huang}}, \bibinfo {author} {\bibfnamefont
  {C.}~\bibnamefont {dela Cruz}}, \bibinfo {author} {\bibfnamefont {K.~V.}\
  \bibnamefont {Vemuru}}, \bibinfo {author} {\bibfnamefont {M.~D.}\
  \bibnamefont {Lumsden}}, \bibinfo {author} {\bibfnamefont {S.~E.}\
  \bibnamefont {Nagler}}, \bibinfo {author} {\bibfnamefont {G.}~\bibnamefont
  {Alvarez}}, \bibinfo {author} {\bibfnamefont {E.}~\bibnamefont {Dagotto}},
  \bibinfo {author} {\bibfnamefont {H.}~\bibnamefont {Zhou}}, \bibinfo {author}
  {\bibfnamefont {A.~D.}\ \bibnamefont {Christianson}},\ and\ \bibinfo {author}
  {\bibfnamefont {M.~B.}\ \bibnamefont {Stone}},\ }\href@noop {} {\bibinfo
  {title} {{Spinon continuum in the Heisenberg quantum chain compound
  Sr$_2$V$_3$O$_9$}}} (\bibinfo {year} {2023}),\ \Eprint
  {https://arxiv.org/abs/2307.12093} {arXiv:2307.12093 [cond-mat.str-el]}
  \BibitemShut {NoStop}%
\bibitem [{\citenamefont {Caldeira}\ and\ \citenamefont
  {Leggett}(1981)}]{PhysRevLett.46.211}%
  \BibitemOpen
  \bibfield  {author} {\bibinfo {author} {\bibfnamefont {A.~O.}\ \bibnamefont
  {Caldeira}}\ and\ \bibinfo {author} {\bibfnamefont {A.~J.}\ \bibnamefont
  {Leggett}},\ }\bibfield  {title} {\bibinfo {title} {{Influence of Dissipation
  on Quantum Tunneling in Macroscopic Systems}},\ }\href
  {https://doi.org/10.1103/PhysRevLett.46.211} {\bibfield  {journal} {\bibinfo
  {journal} {Phys. Rev. Lett.}\ }\textbf {\bibinfo {volume} {46}},\ \bibinfo
  {pages} {211} (\bibinfo {year} {1981})}\BibitemShut {NoStop}%
\bibitem [{\citenamefont {Leggett}\ \emph {et~al.}(1987)\citenamefont
  {Leggett}, \citenamefont {Chakravarty}, \citenamefont {Dorsey}, \citenamefont
  {Fisher}, \citenamefont {Garg},\ and\ \citenamefont
  {Zwerger}}]{RevModPhys.59.1}%
  \BibitemOpen
  \bibfield  {author} {\bibinfo {author} {\bibfnamefont {A.~J.}\ \bibnamefont
  {Leggett}}, \bibinfo {author} {\bibfnamefont {S.}~\bibnamefont
  {Chakravarty}}, \bibinfo {author} {\bibfnamefont {A.~T.}\ \bibnamefont
  {Dorsey}}, \bibinfo {author} {\bibfnamefont {M.~P.~A.}\ \bibnamefont
  {Fisher}}, \bibinfo {author} {\bibfnamefont {A.}~\bibnamefont {Garg}},\ and\
  \bibinfo {author} {\bibfnamefont {W.}~\bibnamefont {Zwerger}},\ }\bibfield
  {title} {\bibinfo {title} {Dynamics of the dissipative two-state system},\
  }\href {https://doi.org/10.1103/RevModPhys.59.1} {\bibfield  {journal}
  {\bibinfo  {journal} {Rev. Mod. Phys.}\ }\textbf {\bibinfo {volume} {59}},\
  \bibinfo {pages} {1} (\bibinfo {year} {1987})}\BibitemShut {NoStop}%
\bibitem [{\citenamefont {Vojta}(2006)}]{doi:10.1080/14786430500070396}%
  \BibitemOpen
  \bibfield  {author} {\bibinfo {author} {\bibfnamefont {M.}~\bibnamefont
  {Vojta}},\ }\bibfield  {title} {\bibinfo {title} {Impurity quantum phase
  transitions},\ }\href {https://doi.org/10.1080/14786430500070396} {\bibfield
  {journal} {\bibinfo  {journal} {Philosophical Magazine}\ }\textbf {\bibinfo
  {volume} {86}},\ \bibinfo {pages} {1807} (\bibinfo {year}
  {2006})}\BibitemShut {NoStop}%
\bibitem [{\citenamefont {Winter}\ \emph {et~al.}(2009)\citenamefont {Winter},
  \citenamefont {Rieger}, \citenamefont {Vojta},\ and\ \citenamefont
  {Bulla}}]{PhysRevLett.102.030601}%
  \BibitemOpen
  \bibfield  {author} {\bibinfo {author} {\bibfnamefont {A.}~\bibnamefont
  {Winter}}, \bibinfo {author} {\bibfnamefont {H.}~\bibnamefont {Rieger}},
  \bibinfo {author} {\bibfnamefont {M.}~\bibnamefont {Vojta}},\ and\ \bibinfo
  {author} {\bibfnamefont {R.}~\bibnamefont {Bulla}},\ }\bibfield  {title}
  {\bibinfo {title} {{Quantum Phase Transition in the Sub-Ohmic Spin-Boson
  Model: Quantum Monte Carlo Study with a Continuous Imaginary Time Cluster
  Algorithm}},\ }\href {https://doi.org/10.1103/PhysRevLett.102.030601}
  {\bibfield  {journal} {\bibinfo  {journal} {Phys. Rev. Lett.}\ }\textbf
  {\bibinfo {volume} {102}},\ \bibinfo {pages} {030601} (\bibinfo {year}
  {2009})}\BibitemShut {NoStop}%
\bibitem [{\citenamefont {Smith}\ and\ \citenamefont {Si}(1999)}]{Smith_1999}%
  \BibitemOpen
  \bibfield  {author} {\bibinfo {author} {\bibfnamefont {J.~L.}\ \bibnamefont
  {Smith}}\ and\ \bibinfo {author} {\bibfnamefont {Q.}~\bibnamefont {Si}},\
  }\bibfield  {title} {\bibinfo {title} {{Non-Fermi liquids in the two-band
  extended Hubbard model}},\ }\href {https://doi.org/10.1209/epl/i1999-00151-4}
  {\bibfield  {journal} {\bibinfo  {journal} {Europhysics Letters}\ }\textbf
  {\bibinfo {volume} {45}},\ \bibinfo {pages} {228} (\bibinfo {year}
  {1999})}\BibitemShut {NoStop}%
\bibitem [{\citenamefont {Sengupta}(2000)}]{PhysRevB.61.4041}%
  \BibitemOpen
  \bibfield  {author} {\bibinfo {author} {\bibfnamefont {A.~M.}\ \bibnamefont
  {Sengupta}},\ }\bibfield  {title} {\bibinfo {title} {{Spin in a fluctuating
  field: The Bose(+Fermi) Kondo models}},\ }\href
  {https://doi.org/10.1103/PhysRevB.61.4041} {\bibfield  {journal} {\bibinfo
  {journal} {Phys. Rev. B}\ }\textbf {\bibinfo {volume} {61}},\ \bibinfo
  {pages} {4041} (\bibinfo {year} {2000})}\BibitemShut {NoStop}%
\bibitem [{\citenamefont {Sachdev}\ \emph {et~al.}(1999)\citenamefont
  {Sachdev}, \citenamefont {Buragohain},\ and\ \citenamefont
  {Vojta}}]{Sachdev2479}%
  \BibitemOpen
  \bibfield  {author} {\bibinfo {author} {\bibfnamefont {S.}~\bibnamefont
  {Sachdev}}, \bibinfo {author} {\bibfnamefont {C.}~\bibnamefont
  {Buragohain}},\ and\ \bibinfo {author} {\bibfnamefont {M.}~\bibnamefont
  {Vojta}},\ }\bibfield  {title} {\bibinfo {title} {{Quantum Impurity in a
  Nearly Critical Two-Dimensional Antiferromagnet}},\ }\href
  {https://doi.org/10.1126/science.286.5449.2479} {\bibfield  {journal}
  {\bibinfo  {journal} {Science}\ }\textbf {\bibinfo {volume} {286}},\ \bibinfo
  {pages} {2479} (\bibinfo {year} {1999})}\BibitemShut {NoStop}%
\bibitem [{\citenamefont {Vojta}\ \emph {et~al.}(2000)\citenamefont {Vojta},
  \citenamefont {Buragohain},\ and\ \citenamefont
  {Sachdev}}]{PhysRevB.61.15152}%
  \BibitemOpen
  \bibfield  {author} {\bibinfo {author} {\bibfnamefont {M.}~\bibnamefont
  {Vojta}}, \bibinfo {author} {\bibfnamefont {C.}~\bibnamefont {Buragohain}},\
  and\ \bibinfo {author} {\bibfnamefont {S.}~\bibnamefont {Sachdev}},\
  }\bibfield  {title} {\bibinfo {title} {{Quantum impurity dynamics in
  two-dimensional antiferromagnets and superconductors}},\ }\href
  {https://doi.org/10.1103/PhysRevB.61.15152} {\bibfield  {journal} {\bibinfo
  {journal} {Phys. Rev. B}\ }\textbf {\bibinfo {volume} {61}},\ \bibinfo
  {pages} {15152} (\bibinfo {year} {2000})}\BibitemShut {NoStop}%
\bibitem [{\citenamefont {Castro~Neto}\ \emph {et~al.}(2003)\citenamefont
  {Castro~Neto}, \citenamefont {Novais}, \citenamefont {Borda}, \citenamefont
  {Zar\'and},\ and\ \citenamefont {Affleck}}]{PhysRevLett.91.096401}%
  \BibitemOpen
  \bibfield  {author} {\bibinfo {author} {\bibfnamefont {A.~H.}\ \bibnamefont
  {Castro~Neto}}, \bibinfo {author} {\bibfnamefont {E.}~\bibnamefont {Novais}},
  \bibinfo {author} {\bibfnamefont {L.}~\bibnamefont {Borda}}, \bibinfo
  {author} {\bibfnamefont {G.}~\bibnamefont {Zar\'and}},\ and\ \bibinfo
  {author} {\bibfnamefont {I.}~\bibnamefont {Affleck}},\ }\bibfield  {title}
  {\bibinfo {title} {{Quantum Magnetic Impurities in Magnetically Ordered
  Systems}},\ }\href {https://doi.org/10.1103/PhysRevLett.91.096401} {\bibfield
   {journal} {\bibinfo  {journal} {Phys. Rev. Lett.}\ }\textbf {\bibinfo
  {volume} {91}},\ \bibinfo {pages} {096401} (\bibinfo {year}
  {2003})}\BibitemShut {NoStop}%
\bibitem [{\citenamefont {Guo}\ \emph {et~al.}(2012)\citenamefont {Guo},
  \citenamefont {Weichselbaum}, \citenamefont {von Delft},\ and\ \citenamefont
  {Vojta}}]{PhysRevLett.108.160401}%
  \BibitemOpen
  \bibfield  {author} {\bibinfo {author} {\bibfnamefont {C.}~\bibnamefont
  {Guo}}, \bibinfo {author} {\bibfnamefont {A.}~\bibnamefont {Weichselbaum}},
  \bibinfo {author} {\bibfnamefont {J.}~\bibnamefont {von Delft}},\ and\
  \bibinfo {author} {\bibfnamefont {M.}~\bibnamefont {Vojta}},\ }\bibfield
  {title} {\bibinfo {title} {{Critical and Strong-Coupling Phases in One- and
  Two-Bath Spin-Boson Models}},\ }\href
  {https://doi.org/10.1103/PhysRevLett.108.160401} {\bibfield  {journal}
  {\bibinfo  {journal} {Phys. Rev. Lett.}\ }\textbf {\bibinfo {volume} {108}},\
  \bibinfo {pages} {160401} (\bibinfo {year} {2012})}\BibitemShut {NoStop}%
\bibitem [{\citenamefont {Bruognolo}\ \emph {et~al.}(2014)\citenamefont
  {Bruognolo}, \citenamefont {Weichselbaum}, \citenamefont {Guo}, \citenamefont
  {von Delft}, \citenamefont {Schneider},\ and\ \citenamefont
  {Vojta}}]{PhysRevB.90.245130}%
  \BibitemOpen
  \bibfield  {author} {\bibinfo {author} {\bibfnamefont {B.}~\bibnamefont
  {Bruognolo}}, \bibinfo {author} {\bibfnamefont {A.}~\bibnamefont
  {Weichselbaum}}, \bibinfo {author} {\bibfnamefont {C.}~\bibnamefont {Guo}},
  \bibinfo {author} {\bibfnamefont {J.}~\bibnamefont {von Delft}}, \bibinfo
  {author} {\bibfnamefont {I.}~\bibnamefont {Schneider}},\ and\ \bibinfo
  {author} {\bibfnamefont {M.}~\bibnamefont {Vojta}},\ }\bibfield  {title}
  {\bibinfo {title} {{Two-bath spin-boson model: Phase diagram and critical
  properties}},\ }\href {https://doi.org/10.1103/PhysRevB.90.245130} {\bibfield
   {journal} {\bibinfo  {journal} {Phys. Rev. B}\ }\textbf {\bibinfo {volume}
  {90}},\ \bibinfo {pages} {245130} (\bibinfo {year} {2014})}\BibitemShut
  {NoStop}%
\bibitem [{\citenamefont {Nahum}(2022)}]{PhysRevB.106.L081109}%
  \BibitemOpen
  \bibfield  {author} {\bibinfo {author} {\bibfnamefont {A.}~\bibnamefont
  {Nahum}},\ }\bibfield  {title} {\bibinfo {title} {Fixed point annihilation
  for a spin in a fluctuating field},\ }\href
  {https://doi.org/10.1103/PhysRevB.106.L081109} {\bibfield  {journal}
  {\bibinfo  {journal} {Phys. Rev. B}\ }\textbf {\bibinfo {volume} {106}},\
  \bibinfo {pages} {L081109} (\bibinfo {year} {2022})}\BibitemShut {NoStop}%
\bibitem [{\citenamefont {Weber}\ and\ \citenamefont
  {Vojta}(2023)}]{PhysRevLett.130.186701}%
  \BibitemOpen
  \bibfield  {author} {\bibinfo {author} {\bibfnamefont {M.}~\bibnamefont
  {Weber}}\ and\ \bibinfo {author} {\bibfnamefont {M.}~\bibnamefont {Vojta}},\
  }\bibfield  {title} {\bibinfo {title} {{SU(2)-Symmetric Spin-Boson Model:
  Quantum Criticality, Fixed-Point Annihilation, and Duality}},\ }\href
  {https://doi.org/10.1103/PhysRevLett.130.186701} {\bibfield  {journal}
  {\bibinfo  {journal} {Phys. Rev. Lett.}\ }\textbf {\bibinfo {volume} {130}},\
  \bibinfo {pages} {186701} (\bibinfo {year} {2023})}\BibitemShut {NoStop}%
\bibitem [{\citenamefont {Chakravarty}\ \emph {et~al.}(1986)\citenamefont
  {Chakravarty}, \citenamefont {Ingold}, \citenamefont {Kivelson},\ and\
  \citenamefont {Luther}}]{PhysRevLett.56.2303}%
  \BibitemOpen
  \bibfield  {author} {\bibinfo {author} {\bibfnamefont {S.}~\bibnamefont
  {Chakravarty}}, \bibinfo {author} {\bibfnamefont {G.-L.}\ \bibnamefont
  {Ingold}}, \bibinfo {author} {\bibfnamefont {S.}~\bibnamefont {Kivelson}},\
  and\ \bibinfo {author} {\bibfnamefont {A.}~\bibnamefont {Luther}},\
  }\bibfield  {title} {\bibinfo {title} {{Onset of Global Phase Coherence in
  Josephson-Junction Arrays: A Dissipative Phase Transition}},\ }\href
  {https://doi.org/10.1103/PhysRevLett.56.2303} {\bibfield  {journal} {\bibinfo
   {journal} {Phys. Rev. Lett.}\ }\textbf {\bibinfo {volume} {56}},\ \bibinfo
  {pages} {2303} (\bibinfo {year} {1986})}\BibitemShut {NoStop}%
\bibitem [{\citenamefont {Fisher}(1987)}]{PhysRevB.36.1917}%
  \BibitemOpen
  \bibfield  {author} {\bibinfo {author} {\bibfnamefont {M.~P.~A.}\
  \bibnamefont {Fisher}},\ }\bibfield  {title} {\bibinfo {title} {{Dissipation
  and quantum fluctuations in granular superconductivity}},\ }\href
  {https://doi.org/10.1103/PhysRevB.36.1917} {\bibfield  {journal} {\bibinfo
  {journal} {Phys. Rev. B}\ }\textbf {\bibinfo {volume} {36}},\ \bibinfo
  {pages} {1917} (\bibinfo {year} {1987})}\BibitemShut {NoStop}%
\bibitem [{\citenamefont {Panyukov}\ and\ \citenamefont
  {Zaikin}(1987)}]{PANYUKOV1987325}%
  \BibitemOpen
  \bibfield  {author} {\bibinfo {author} {\bibfnamefont {S.}~\bibnamefont
  {Panyukov}}\ and\ \bibinfo {author} {\bibfnamefont {A.}~\bibnamefont
  {Zaikin}},\ }\bibfield  {title} {\bibinfo {title} {{Quantum fluctuations and
  dissipative phase transition in granular superconductors}},\ }\href
  {https://doi.org/https://doi.org/10.1016/0375-9601(87)90020-X} {\bibfield
  {journal} {\bibinfo  {journal} {Physics Letters A}\ }\textbf {\bibinfo
  {volume} {124}},\ \bibinfo {pages} {325} (\bibinfo {year}
  {1987})}\BibitemShut {NoStop}%
\bibitem [{\citenamefont {Chakravarty}\ \emph {et~al.}(1988)\citenamefont
  {Chakravarty}, \citenamefont {Ingold}, \citenamefont {Kivelson},\ and\
  \citenamefont {Zimanyi}}]{PhysRevB.37.3283}%
  \BibitemOpen
  \bibfield  {author} {\bibinfo {author} {\bibfnamefont {S.}~\bibnamefont
  {Chakravarty}}, \bibinfo {author} {\bibfnamefont {G.-L.}\ \bibnamefont
  {Ingold}}, \bibinfo {author} {\bibfnamefont {S.}~\bibnamefont {Kivelson}},\
  and\ \bibinfo {author} {\bibfnamefont {G.}~\bibnamefont {Zimanyi}},\
  }\bibfield  {title} {\bibinfo {title} {{Quantum statistical mechanics of an
  array of resistively shunted Josephson junctions}},\ }\href
  {https://doi.org/10.1103/PhysRevB.37.3283} {\bibfield  {journal} {\bibinfo
  {journal} {Phys. Rev. B}\ }\textbf {\bibinfo {volume} {37}},\ \bibinfo
  {pages} {3283} (\bibinfo {year} {1988})}\BibitemShut {NoStop}%
\bibitem [{\citenamefont {{Korshunov}}(1989)}]{1989EL......9..107K}%
  \BibitemOpen
  \bibfield  {author} {\bibinfo {author} {\bibfnamefont {S.~E.}\ \bibnamefont
  {{Korshunov}}},\ }\bibfield  {title} {\bibinfo {title} {{Phase diagram of a
  chain of dissipative Josephson junctions}},\ }\href
  {https://doi.org/10.1209/0295-5075/9/2/003} {\bibfield  {journal} {\bibinfo
  {journal} {EPL (Europhysics Letters)}\ }\textbf {\bibinfo {volume} {9}},\
  \bibinfo {pages} {107} (\bibinfo {year} {1989})}\BibitemShut {NoStop}%
\bibitem [{\citenamefont {Bobbert}\ \emph {et~al.}(1990)\citenamefont
  {Bobbert}, \citenamefont {Fazio}, \citenamefont {Sch\"on},\ and\
  \citenamefont {Zimanyi}}]{PhysRevB.41.4009}%
  \BibitemOpen
  \bibfield  {author} {\bibinfo {author} {\bibfnamefont {P.~A.}\ \bibnamefont
  {Bobbert}}, \bibinfo {author} {\bibfnamefont {R.}~\bibnamefont {Fazio}},
  \bibinfo {author} {\bibfnamefont {G.}~\bibnamefont {Sch\"on}},\ and\ \bibinfo
  {author} {\bibfnamefont {G.~T.}\ \bibnamefont {Zimanyi}},\ }\bibfield
  {title} {\bibinfo {title} {{Phase transitions in dissipative Josephson
  chains}},\ }\href {https://doi.org/10.1103/PhysRevB.41.4009} {\bibfield
  {journal} {\bibinfo  {journal} {Phys. Rev. B}\ }\textbf {\bibinfo {volume}
  {41}},\ \bibinfo {pages} {4009} (\bibinfo {year} {1990})}\BibitemShut
  {NoStop}%
\bibitem [{\citenamefont {Bobbert}\ \emph {et~al.}(1992)\citenamefont
  {Bobbert}, \citenamefont {Fazio}, \citenamefont {Sch\"on},\ and\
  \citenamefont {Zaikin}}]{PhysRevB.45.2294}%
  \BibitemOpen
  \bibfield  {author} {\bibinfo {author} {\bibfnamefont {P.~A.}\ \bibnamefont
  {Bobbert}}, \bibinfo {author} {\bibfnamefont {R.}~\bibnamefont {Fazio}},
  \bibinfo {author} {\bibfnamefont {G.}~\bibnamefont {Sch\"on}},\ and\ \bibinfo
  {author} {\bibfnamefont {A.~D.}\ \bibnamefont {Zaikin}},\ }\bibfield  {title}
  {\bibinfo {title} {{Phase transitions in dissipative Josephson chains: Monte
  Carlo results and response functions}},\ }\href
  {https://doi.org/10.1103/PhysRevB.45.2294} {\bibfield  {journal} {\bibinfo
  {journal} {Phys. Rev. B}\ }\textbf {\bibinfo {volume} {45}},\ \bibinfo
  {pages} {2294} (\bibinfo {year} {1992})}\BibitemShut {NoStop}%
\bibitem [{\citenamefont {Wagenblast}\ \emph {et~al.}(1997)\citenamefont
  {Wagenblast}, \citenamefont {van Otterlo}, \citenamefont {Sch\"on},\ and\
  \citenamefont {Zim\'anyi}}]{PhysRevLett.78.1779}%
  \BibitemOpen
  \bibfield  {author} {\bibinfo {author} {\bibfnamefont {K.-H.}\ \bibnamefont
  {Wagenblast}}, \bibinfo {author} {\bibfnamefont {A.}~\bibnamefont {van
  Otterlo}}, \bibinfo {author} {\bibfnamefont {G.}~\bibnamefont {Sch\"on}},\
  and\ \bibinfo {author} {\bibfnamefont {G.~T.}\ \bibnamefont {Zim\'anyi}},\
  }\bibfield  {title} {\bibinfo {title} {{New Universality Class at the
  Superconductor-Insulator Transition}},\ }\href
  {https://doi.org/10.1103/PhysRevLett.78.1779} {\bibfield  {journal} {\bibinfo
   {journal} {Phys. Rev. Lett.}\ }\textbf {\bibinfo {volume} {78}},\ \bibinfo
  {pages} {1779} (\bibinfo {year} {1997})}\BibitemShut {NoStop}%
\bibitem [{\citenamefont {Refael}\ \emph {et~al.}(2007)\citenamefont {Refael},
  \citenamefont {Demler}, \citenamefont {Oreg},\ and\ \citenamefont
  {Fisher}}]{PhysRevB.75.014522}%
  \BibitemOpen
  \bibfield  {author} {\bibinfo {author} {\bibfnamefont {G.}~\bibnamefont
  {Refael}}, \bibinfo {author} {\bibfnamefont {E.}~\bibnamefont {Demler}},
  \bibinfo {author} {\bibfnamefont {Y.}~\bibnamefont {Oreg}},\ and\ \bibinfo
  {author} {\bibfnamefont {D.~S.}\ \bibnamefont {Fisher}},\ }\bibfield  {title}
  {\bibinfo {title} {Superconductor-to-normal transitions in dissipative chains
  of mesoscopic grains and nanowires},\ }\href
  {https://doi.org/10.1103/PhysRevB.75.014522} {\bibfield  {journal} {\bibinfo
  {journal} {Phys. Rev. B}\ }\textbf {\bibinfo {volume} {75}},\ \bibinfo
  {pages} {014522} (\bibinfo {year} {2007})}\BibitemShut {NoStop}%
\bibitem [{\citenamefont {Feigel'man}\ and\ \citenamefont
  {Larkin}(1998)}]{FEIGELMAN1998107}%
  \BibitemOpen
  \bibfield  {author} {\bibinfo {author} {\bibfnamefont {M.}~\bibnamefont
  {Feigel'man}}\ and\ \bibinfo {author} {\bibfnamefont {A.}~\bibnamefont
  {Larkin}},\ }\bibfield  {title} {\bibinfo {title} {{Quantum
  superconductor--metal transition in a 2D proximity-coupled array}},\ }\href
  {https://doi.org/https://doi.org/10.1016/S0301-0104(98)00075-5} {\bibfield
  {journal} {\bibinfo  {journal} {Chemical Physics}\ }\textbf {\bibinfo
  {volume} {235}},\ \bibinfo {pages} {107} (\bibinfo {year}
  {1998})}\BibitemShut {NoStop}%
\bibitem [{\citenamefont {Werner}\ \emph {et~al.}(2005)\citenamefont {Werner},
  \citenamefont {V\"olker}, \citenamefont {Troyer},\ and\ \citenamefont
  {Chakravarty}}]{PhysRevLett.94.047201}%
  \BibitemOpen
  \bibfield  {author} {\bibinfo {author} {\bibfnamefont {P.}~\bibnamefont
  {Werner}}, \bibinfo {author} {\bibfnamefont {K.}~\bibnamefont {V\"olker}},
  \bibinfo {author} {\bibfnamefont {M.}~\bibnamefont {Troyer}},\ and\ \bibinfo
  {author} {\bibfnamefont {S.}~\bibnamefont {Chakravarty}},\ }\bibfield
  {title} {\bibinfo {title} {{Phase Diagram and Critical Exponents of a
  Dissipative Ising Spin Chain in a Transverse Magnetic Field}},\ }\href
  {https://doi.org/10.1103/PhysRevLett.94.047201} {\bibfield  {journal}
  {\bibinfo  {journal} {Phys. Rev. Lett.}\ }\textbf {\bibinfo {volume} {94}},\
  \bibinfo {pages} {047201} (\bibinfo {year} {2005})}\BibitemShut {NoStop}%
\bibitem [{\citenamefont {{Werner}}\ \emph {et~al.}(2005)\citenamefont
  {{Werner}}, \citenamefont {{Troyer}},\ and\ \citenamefont
  {{Sachdev}}}]{2005JPSJ...74S..67W}%
  \BibitemOpen
  \bibfield  {author} {\bibinfo {author} {\bibfnamefont {P.}~\bibnamefont
  {{Werner}}}, \bibinfo {author} {\bibfnamefont {M.}~\bibnamefont {{Troyer}}},\
  and\ \bibinfo {author} {\bibfnamefont {S.}~\bibnamefont {{Sachdev}}},\
  }\bibfield  {title} {\bibinfo {title} {{Quantum Spin Chains with Site
  Dissipation}},\ }\href {https://doi.org/10.1143/JPSJS.74S.67} {\bibfield
  {journal} {\bibinfo  {journal} {Journal of the Physical Society of Japan}\
  }\textbf {\bibinfo {volume} {74}},\ \bibinfo {pages} {67} (\bibinfo {year}
  {2005})}\BibitemShut {NoStop}%
\bibitem [{\citenamefont {Sperstad}\ \emph {et~al.}(2010)\citenamefont
  {Sperstad}, \citenamefont {Stiansen},\ and\ \citenamefont
  {Sudb\o{}}}]{PhysRevB.81.104302}%
  \BibitemOpen
  \bibfield  {author} {\bibinfo {author} {\bibfnamefont {I.~B.}\ \bibnamefont
  {Sperstad}}, \bibinfo {author} {\bibfnamefont {E.~B.}\ \bibnamefont
  {Stiansen}},\ and\ \bibinfo {author} {\bibfnamefont {A.}~\bibnamefont
  {Sudb\o{}}},\ }\bibfield  {title} {\bibinfo {title} {{Monte Carlo simulations
  of dissipative quantum Ising models}},\ }\href
  {https://doi.org/10.1103/PhysRevB.81.104302} {\bibfield  {journal} {\bibinfo
  {journal} {Phys. Rev. B}\ }\textbf {\bibinfo {volume} {81}},\ \bibinfo
  {pages} {104302} (\bibinfo {year} {2010})}\BibitemShut {NoStop}%
\bibitem [{\citenamefont {Stiansen}\ \emph {et~al.}(2011)\citenamefont
  {Stiansen}, \citenamefont {Sperstad},\ and\ \citenamefont
  {Sudb\o{}}}]{PhysRevB.83.115134}%
  \BibitemOpen
  \bibfield  {author} {\bibinfo {author} {\bibfnamefont {E.~B.}\ \bibnamefont
  {Stiansen}}, \bibinfo {author} {\bibfnamefont {I.~B.}\ \bibnamefont
  {Sperstad}},\ and\ \bibinfo {author} {\bibfnamefont {A.}~\bibnamefont
  {Sudb\o{}}},\ }\bibfield  {title} {\bibinfo {title} {{Criticality of compact
  and noncompact quantum dissipative ${Z}_{4}$ models in $(1+1)$ dimensions}},\
  }\href {https://doi.org/10.1103/PhysRevB.83.115134} {\bibfield  {journal}
  {\bibinfo  {journal} {Phys. Rev. B}\ }\textbf {\bibinfo {volume} {83}},\
  \bibinfo {pages} {115134} (\bibinfo {year} {2011})}\BibitemShut {NoStop}%
\bibitem [{\citenamefont {Sperstad}\ \emph {et~al.}(2011)\citenamefont
  {Sperstad}, \citenamefont {Stiansen},\ and\ \citenamefont
  {Sudb\o{}}}]{PhysRevB.84.180503}%
  \BibitemOpen
  \bibfield  {author} {\bibinfo {author} {\bibfnamefont {I.~B.}\ \bibnamefont
  {Sperstad}}, \bibinfo {author} {\bibfnamefont {E.~B.}\ \bibnamefont
  {Stiansen}},\ and\ \bibinfo {author} {\bibfnamefont {A.}~\bibnamefont
  {Sudb\o{}}},\ }\bibfield  {title} {\bibinfo {title} {{Quantum criticality in
  a dissipative (2+1)-dimensional $XY$ model of circulating currents in
  high-${T}_{c}$ cuprates}},\ }\href
  {https://doi.org/10.1103/PhysRevB.84.180503} {\bibfield  {journal} {\bibinfo
  {journal} {Phys. Rev. B}\ }\textbf {\bibinfo {volume} {84}},\ \bibinfo
  {pages} {180503} (\bibinfo {year} {2011})}\BibitemShut {NoStop}%
\bibitem [{\citenamefont {Sperstad}\ \emph {et~al.}(2012)\citenamefont
  {Sperstad}, \citenamefont {Stiansen},\ and\ \citenamefont
  {Sudb\o{}}}]{PhysRevB.85.214302}%
  \BibitemOpen
  \bibfield  {author} {\bibinfo {author} {\bibfnamefont {I.~B.}\ \bibnamefont
  {Sperstad}}, \bibinfo {author} {\bibfnamefont {E.~B.}\ \bibnamefont
  {Stiansen}},\ and\ \bibinfo {author} {\bibfnamefont {A.}~\bibnamefont
  {Sudb\o{}}},\ }\bibfield  {title} {\bibinfo {title} {{Quantum criticality in
  spin chains with non-Ohmic dissipation}},\ }\href
  {https://doi.org/10.1103/PhysRevB.85.214302} {\bibfield  {journal} {\bibinfo
  {journal} {Phys. Rev. B}\ }\textbf {\bibinfo {volume} {85}},\ \bibinfo
  {pages} {214302} (\bibinfo {year} {2012})}\BibitemShut {NoStop}%
\bibitem [{\citenamefont {Stiansen}\ \emph {et~al.}(2012)\citenamefont
  {Stiansen}, \citenamefont {Sperstad},\ and\ \citenamefont
  {Sudb\o{}}}]{PhysRevB.85.224531}%
  \BibitemOpen
  \bibfield  {author} {\bibinfo {author} {\bibfnamefont {E.~B.}\ \bibnamefont
  {Stiansen}}, \bibinfo {author} {\bibfnamefont {I.~B.}\ \bibnamefont
  {Sperstad}},\ and\ \bibinfo {author} {\bibfnamefont {A.}~\bibnamefont
  {Sudb\o{}}},\ }\bibfield  {title} {\bibinfo {title} {{Three distinct types of
  quantum phase transitions in a (2+1)-dimensional array of dissipative
  Josephson junctions}},\ }\href {https://doi.org/10.1103/PhysRevB.85.224531}
  {\bibfield  {journal} {\bibinfo  {journal} {Phys. Rev. B}\ }\textbf {\bibinfo
  {volume} {85}},\ \bibinfo {pages} {224531} (\bibinfo {year}
  {2012})}\BibitemShut {NoStop}%
\bibitem [{\citenamefont {Tewari}\ \emph {et~al.}(2005)\citenamefont {Tewari},
  \citenamefont {Toner},\ and\ \citenamefont
  {Chakravarty}}]{PhysRevB.72.060505}%
  \BibitemOpen
  \bibfield  {author} {\bibinfo {author} {\bibfnamefont {S.}~\bibnamefont
  {Tewari}}, \bibinfo {author} {\bibfnamefont {J.}~\bibnamefont {Toner}},\ and\
  \bibinfo {author} {\bibfnamefont {S.}~\bibnamefont {Chakravarty}},\
  }\bibfield  {title} {\bibinfo {title} {{Floating phase in a dissipative
  Josephson junction array}},\ }\href
  {https://doi.org/10.1103/PhysRevB.72.060505} {\bibfield  {journal} {\bibinfo
  {journal} {Phys. Rev. B}\ }\textbf {\bibinfo {volume} {72}},\ \bibinfo
  {pages} {060505} (\bibinfo {year} {2005})}\BibitemShut {NoStop}%
\bibitem [{\citenamefont {Tewari}\ \emph {et~al.}(2006)\citenamefont {Tewari},
  \citenamefont {Toner},\ and\ \citenamefont
  {Chakravarty}}]{PhysRevB.73.064503}%
  \BibitemOpen
  \bibfield  {author} {\bibinfo {author} {\bibfnamefont {S.}~\bibnamefont
  {Tewari}}, \bibinfo {author} {\bibfnamefont {J.}~\bibnamefont {Toner}},\ and\
  \bibinfo {author} {\bibfnamefont {S.}~\bibnamefont {Chakravarty}},\
  }\bibfield  {title} {\bibinfo {title} {{Nature and boundary of the floating
  phase in a dissipative Josephson junction array}},\ }\href
  {https://doi.org/10.1103/PhysRevB.73.064503} {\bibfield  {journal} {\bibinfo
  {journal} {Phys. Rev. B}\ }\textbf {\bibinfo {volume} {73}},\ \bibinfo
  {pages} {064503} (\bibinfo {year} {2006})}\BibitemShut {NoStop}%
\bibitem [{\citenamefont {Goswami}\ and\ \citenamefont
  {Chakravarty}(2006)}]{PhysRevB.73.094516}%
  \BibitemOpen
  \bibfield  {author} {\bibinfo {author} {\bibfnamefont {P.}~\bibnamefont
  {Goswami}}\ and\ \bibinfo {author} {\bibfnamefont {S.}~\bibnamefont
  {Chakravarty}},\ }\bibfield  {title} {\bibinfo {title} {{Dissipation,
  topology, and quantum phase transition in a one-dimensional Josephson
  junction array}},\ }\href {https://doi.org/10.1103/PhysRevB.73.094516}
  {\bibfield  {journal} {\bibinfo  {journal} {Phys. Rev. B}\ }\textbf {\bibinfo
  {volume} {73}},\ \bibinfo {pages} {094516} (\bibinfo {year}
  {2006})}\BibitemShut {NoStop}%
\bibitem [{\citenamefont {Aji}\ and\ \citenamefont
  {Varma}(2009)}]{PhysRevB.79.184501}%
  \BibitemOpen
  \bibfield  {author} {\bibinfo {author} {\bibfnamefont {V.}~\bibnamefont
  {Aji}}\ and\ \bibinfo {author} {\bibfnamefont {C.~M.}\ \bibnamefont
  {Varma}},\ }\bibfield  {title} {\bibinfo {title} {{Quantum criticality in
  dissipative quantum two-dimensional $XY$ and Ashkin-Teller models:
  Application to the cuprates}},\ }\href
  {https://doi.org/10.1103/PhysRevB.79.184501} {\bibfield  {journal} {\bibinfo
  {journal} {Phys. Rev. B}\ }\textbf {\bibinfo {volume} {79}},\ \bibinfo
  {pages} {184501} (\bibinfo {year} {2009})}\BibitemShut {NoStop}%
\bibitem [{\citenamefont {Aji}\ and\ \citenamefont
  {Varma}(2010)}]{PhysRevB.82.174501}%
  \BibitemOpen
  \bibfield  {author} {\bibinfo {author} {\bibfnamefont {V.}~\bibnamefont
  {Aji}}\ and\ \bibinfo {author} {\bibfnamefont {C.~M.}\ \bibnamefont
  {Varma}},\ }\bibfield  {title} {\bibinfo {title} {{Topological excitations
  near the local critical point in the dissipative two-dimensional $XY$
  model}},\ }\href {https://doi.org/10.1103/PhysRevB.82.174501} {\bibfield
  {journal} {\bibinfo  {journal} {Phys. Rev. B}\ }\textbf {\bibinfo {volume}
  {82}},\ \bibinfo {pages} {174501} (\bibinfo {year} {2010})}\BibitemShut
  {NoStop}%
\bibitem [{\citenamefont {Zhu}\ \emph {et~al.}(2015)\citenamefont {Zhu},
  \citenamefont {Chen},\ and\ \citenamefont {Varma}}]{PhysRevB.91.205129}%
  \BibitemOpen
  \bibfield  {author} {\bibinfo {author} {\bibfnamefont {L.}~\bibnamefont
  {Zhu}}, \bibinfo {author} {\bibfnamefont {Y.}~\bibnamefont {Chen}},\ and\
  \bibinfo {author} {\bibfnamefont {C.~M.}\ \bibnamefont {Varma}},\ }\bibfield
  {title} {\bibinfo {title} {{Local quantum criticality in the two-dimensional
  dissipative quantum XY model}},\ }\href
  {https://doi.org/10.1103/PhysRevB.91.205129} {\bibfield  {journal} {\bibinfo
  {journal} {Phys. Rev. B}\ }\textbf {\bibinfo {volume} {91}},\ \bibinfo
  {pages} {205129} (\bibinfo {year} {2015})}\BibitemShut {NoStop}%
\bibitem [{\citenamefont {Zhu}\ \emph {et~al.}(2016)\citenamefont {Zhu},
  \citenamefont {Hou},\ and\ \citenamefont {Varma}}]{PhysRevB.94.235156}%
  \BibitemOpen
  \bibfield  {author} {\bibinfo {author} {\bibfnamefont {L.}~\bibnamefont
  {Zhu}}, \bibinfo {author} {\bibfnamefont {C.}~\bibnamefont {Hou}},\ and\
  \bibinfo {author} {\bibfnamefont {C.~M.}\ \bibnamefont {Varma}},\ }\bibfield
  {title} {\bibinfo {title} {{Quantum criticality in the two-dimensional
  dissipative quantum XY model}},\ }\href
  {https://doi.org/10.1103/PhysRevB.94.235156} {\bibfield  {journal} {\bibinfo
  {journal} {Phys. Rev. B}\ }\textbf {\bibinfo {volume} {94}},\ \bibinfo
  {pages} {235156} (\bibinfo {year} {2016})}\BibitemShut {NoStop}%
\bibitem [{\citenamefont {Schehr}\ and\ \citenamefont
  {Rieger}(2006)}]{PhysRevLett.96.227201}%
  \BibitemOpen
  \bibfield  {author} {\bibinfo {author} {\bibfnamefont {G.}~\bibnamefont
  {Schehr}}\ and\ \bibinfo {author} {\bibfnamefont {H.}~\bibnamefont
  {Rieger}},\ }\bibfield  {title} {\bibinfo {title} {{Strong-Disorder Fixed
  Point in the Dissipative Random Transverse-Field Ising Model}},\ }\href
  {https://doi.org/10.1103/PhysRevLett.96.227201} {\bibfield  {journal}
  {\bibinfo  {journal} {Phys. Rev. Lett.}\ }\textbf {\bibinfo {volume} {96}},\
  \bibinfo {pages} {227201} (\bibinfo {year} {2006})}\BibitemShut {NoStop}%
\bibitem [{\citenamefont {Hoyos}\ \emph {et~al.}(2007)\citenamefont {Hoyos},
  \citenamefont {Kotabage},\ and\ \citenamefont
  {Vojta}}]{PhysRevLett.99.230601}%
  \BibitemOpen
  \bibfield  {author} {\bibinfo {author} {\bibfnamefont {J.~A.}\ \bibnamefont
  {Hoyos}}, \bibinfo {author} {\bibfnamefont {C.}~\bibnamefont {Kotabage}},\
  and\ \bibinfo {author} {\bibfnamefont {T.}~\bibnamefont {Vojta}},\ }\bibfield
   {title} {\bibinfo {title} {{Effects of Dissipation on a Quantum Critical
  Point with Disorder}},\ }\href
  {https://doi.org/10.1103/PhysRevLett.99.230601} {\bibfield  {journal}
  {\bibinfo  {journal} {Phys. Rev. Lett.}\ }\textbf {\bibinfo {volume} {99}},\
  \bibinfo {pages} {230601} (\bibinfo {year} {2007})}\BibitemShut {NoStop}%
\bibitem [{\citenamefont {Hoyos}\ and\ \citenamefont
  {Vojta}(2008)}]{PhysRevLett.100.240601}%
  \BibitemOpen
  \bibfield  {author} {\bibinfo {author} {\bibfnamefont {J.~A.}\ \bibnamefont
  {Hoyos}}\ and\ \bibinfo {author} {\bibfnamefont {T.}~\bibnamefont {Vojta}},\
  }\bibfield  {title} {\bibinfo {title} {{Theory of Smeared Quantum Phase
  Transitions}},\ }\href {https://doi.org/10.1103/PhysRevLett.100.240601}
  {\bibfield  {journal} {\bibinfo  {journal} {Phys. Rev. Lett.}\ }\textbf
  {\bibinfo {volume} {100}},\ \bibinfo {pages} {240601} (\bibinfo {year}
  {2008})}\BibitemShut {NoStop}%
\bibitem [{\citenamefont {Al-Ali}\ \emph {et~al.}(2012)\citenamefont {Al-Ali},
  \citenamefont {Hoyos},\ and\ \citenamefont {Vojta}}]{PhysRevB.86.075119}%
  \BibitemOpen
  \bibfield  {author} {\bibinfo {author} {\bibfnamefont {M.}~\bibnamefont
  {Al-Ali}}, \bibinfo {author} {\bibfnamefont {J.~A.}\ \bibnamefont {Hoyos}},\
  and\ \bibinfo {author} {\bibfnamefont {T.}~\bibnamefont {Vojta}},\ }\bibfield
   {title} {\bibinfo {title} {{Percolation transition in quantum Ising and
  rotor models with sub-Ohmic dissipation}},\ }\href
  {https://doi.org/10.1103/PhysRevB.86.075119} {\bibfield  {journal} {\bibinfo
  {journal} {Phys. Rev. B}\ }\textbf {\bibinfo {volume} {86}},\ \bibinfo
  {pages} {075119} (\bibinfo {year} {2012})}\BibitemShut {NoStop}%
\bibitem [{\citenamefont {Vojta}\ \emph {et~al.}(2011)\citenamefont {Vojta},
  \citenamefont {Hoyos}, \citenamefont {Mohan},\ and\ \citenamefont
  {Narayanan}}]{Vojta_2011}%
  \BibitemOpen
  \bibfield  {author} {\bibinfo {author} {\bibfnamefont {T.}~\bibnamefont
  {Vojta}}, \bibinfo {author} {\bibfnamefont {J.~A.}\ \bibnamefont {Hoyos}},
  \bibinfo {author} {\bibfnamefont {P.}~\bibnamefont {Mohan}},\ and\ \bibinfo
  {author} {\bibfnamefont {R.}~\bibnamefont {Narayanan}},\ }\bibfield  {title}
  {\bibinfo {title} {Influence of super-ohmic dissipation on a disordered
  quantum critical point},\ }\href
  {https://doi.org/10.1088/0953-8984/23/9/094206} {\bibfield  {journal}
  {\bibinfo  {journal} {Journal of Physics: Condensed Matter}\ }\textbf
  {\bibinfo {volume} {23}},\ \bibinfo {pages} {094206} (\bibinfo {year}
  {2011})}\BibitemShut {NoStop}%
\bibitem [{\citenamefont {Orth}\ \emph {et~al.}(2008)\citenamefont {Orth},
  \citenamefont {Stanic},\ and\ \citenamefont {Le~Hur}}]{PhysRevA.77.051601}%
  \BibitemOpen
  \bibfield  {author} {\bibinfo {author} {\bibfnamefont {P.~P.}\ \bibnamefont
  {Orth}}, \bibinfo {author} {\bibfnamefont {I.}~\bibnamefont {Stanic}},\ and\
  \bibinfo {author} {\bibfnamefont {K.}~\bibnamefont {Le~Hur}},\ }\bibfield
  {title} {\bibinfo {title} {{Dissipative quantum Ising model in a cold-atom
  spin-boson mixture}},\ }\href {https://doi.org/10.1103/PhysRevA.77.051601}
  {\bibfield  {journal} {\bibinfo  {journal} {Phys. Rev. A}\ }\textbf {\bibinfo
  {volume} {77}},\ \bibinfo {pages} {051601} (\bibinfo {year}
  {2008})}\BibitemShut {NoStop}%
\bibitem [{\citenamefont {Maile}\ \emph {et~al.}(2018)\citenamefont {Maile},
  \citenamefont {Andergassen}, \citenamefont {Belzig},\ and\ \citenamefont
  {Rastelli}}]{PhysRevB.97.155427}%
  \BibitemOpen
  \bibfield  {author} {\bibinfo {author} {\bibfnamefont {D.}~\bibnamefont
  {Maile}}, \bibinfo {author} {\bibfnamefont {S.}~\bibnamefont {Andergassen}},
  \bibinfo {author} {\bibfnamefont {W.}~\bibnamefont {Belzig}},\ and\ \bibinfo
  {author} {\bibfnamefont {G.}~\bibnamefont {Rastelli}},\ }\bibfield  {title}
  {\bibinfo {title} {{Quantum phase transition with dissipative frustration}},\
  }\href {https://doi.org/10.1103/PhysRevB.97.155427} {\bibfield  {journal}
  {\bibinfo  {journal} {Phys. Rev. B}\ }\textbf {\bibinfo {volume} {97}},\
  \bibinfo {pages} {155427} (\bibinfo {year} {2018})}\BibitemShut {NoStop}%
\bibitem [{\citenamefont {De~Filippis}\ \emph {et~al.}(2021)\citenamefont
  {De~Filippis}, \citenamefont {de~Candia}, \citenamefont {Mishchenko},
  \citenamefont {Cangemi}, \citenamefont {Nocera}, \citenamefont {Mishchenko},
  \citenamefont {Sassetti}, \citenamefont {Fazio}, \citenamefont {Nagaosa},\
  and\ \citenamefont {Cataudella}}]{PhysRevB.104.L060410}%
  \BibitemOpen
  \bibfield  {author} {\bibinfo {author} {\bibfnamefont {G.}~\bibnamefont
  {De~Filippis}}, \bibinfo {author} {\bibfnamefont {A.}~\bibnamefont
  {de~Candia}}, \bibinfo {author} {\bibfnamefont {A.~S.}\ \bibnamefont
  {Mishchenko}}, \bibinfo {author} {\bibfnamefont {L.~M.}\ \bibnamefont
  {Cangemi}}, \bibinfo {author} {\bibfnamefont {A.}~\bibnamefont {Nocera}},
  \bibinfo {author} {\bibfnamefont {P.~A.}\ \bibnamefont {Mishchenko}},
  \bibinfo {author} {\bibfnamefont {M.}~\bibnamefont {Sassetti}}, \bibinfo
  {author} {\bibfnamefont {R.}~\bibnamefont {Fazio}}, \bibinfo {author}
  {\bibfnamefont {N.}~\bibnamefont {Nagaosa}},\ and\ \bibinfo {author}
  {\bibfnamefont {V.}~\bibnamefont {Cataudella}},\ }\bibfield  {title}
  {\bibinfo {title} {{Quantum phase transition of many interacting spins
  coupled to a bosonic bath: Static and dynamical properties}},\ }\href
  {https://doi.org/10.1103/PhysRevB.104.L060410} {\bibfield  {journal}
  {\bibinfo  {journal} {Phys. Rev. B}\ }\textbf {\bibinfo {volume} {104}},\
  \bibinfo {pages} {L060410} (\bibinfo {year} {2021})}\BibitemShut {NoStop}%
\bibitem [{\citenamefont {Butcher}\ \emph {et~al.}(2022)\citenamefont
  {Butcher}, \citenamefont {Pixley},\ and\ \citenamefont
  {Nevidomskyy}}]{PhysRevB.105.L180407}%
  \BibitemOpen
  \bibfield  {author} {\bibinfo {author} {\bibfnamefont {M.~W.}\ \bibnamefont
  {Butcher}}, \bibinfo {author} {\bibfnamefont {J.~H.}\ \bibnamefont
  {Pixley}},\ and\ \bibinfo {author} {\bibfnamefont {A.~H.}\ \bibnamefont
  {Nevidomskyy}},\ }\bibfield  {title} {\bibinfo {title} {{Long-range order and
  quantum criticality in a dissipative spin chain}},\ }\href
  {https://doi.org/10.1103/PhysRevB.105.L180407} {\bibfield  {journal}
  {\bibinfo  {journal} {Phys. Rev. B}\ }\textbf {\bibinfo {volume} {105}},\
  \bibinfo {pages} {L180407} (\bibinfo {year} {2022})}\BibitemShut {NoStop}%
\bibitem [{\citenamefont {Perroni}\ \emph {et~al.}(2023)\citenamefont
  {Perroni}, \citenamefont {De~Candia}, \citenamefont {Cataudella},
  \citenamefont {Fazio},\ and\ \citenamefont
  {De~Filippis}}]{PhysRevB.107.L100302}%
  \BibitemOpen
  \bibfield  {author} {\bibinfo {author} {\bibfnamefont {C.~A.}\ \bibnamefont
  {Perroni}}, \bibinfo {author} {\bibfnamefont {A.}~\bibnamefont {De~Candia}},
  \bibinfo {author} {\bibfnamefont {V.}~\bibnamefont {Cataudella}}, \bibinfo
  {author} {\bibfnamefont {R.}~\bibnamefont {Fazio}},\ and\ \bibinfo {author}
  {\bibfnamefont {G.}~\bibnamefont {De~Filippis}},\ }\bibfield  {title}
  {\bibinfo {title} {First-order transitions in spin chains coupled to quantum
  baths},\ }\href {https://doi.org/10.1103/PhysRevB.107.L100302} {\bibfield
  {journal} {\bibinfo  {journal} {Phys. Rev. B}\ }\textbf {\bibinfo {volume}
  {107}},\ \bibinfo {pages} {L100302} (\bibinfo {year} {2023})}\BibitemShut
  {NoStop}%
\bibitem [{\citenamefont {Castro~Neto}\ \emph {et~al.}(1997)\citenamefont
  {Castro~Neto}, \citenamefont {de~C.~Chamon},\ and\ \citenamefont
  {Nayak}}]{PhysRevLett.79.4629}%
  \BibitemOpen
  \bibfield  {author} {\bibinfo {author} {\bibfnamefont {A.~H.}\ \bibnamefont
  {Castro~Neto}}, \bibinfo {author} {\bibfnamefont {C.}~\bibnamefont
  {de~C.~Chamon}},\ and\ \bibinfo {author} {\bibfnamefont {C.}~\bibnamefont
  {Nayak}},\ }\bibfield  {title} {\bibinfo {title} {{Open Luttinger Liquids}},\
  }\href {https://doi.org/10.1103/PhysRevLett.79.4629} {\bibfield  {journal}
  {\bibinfo  {journal} {Phys. Rev. Lett.}\ }\textbf {\bibinfo {volume} {79}},\
  \bibinfo {pages} {4629} (\bibinfo {year} {1997})}\BibitemShut {NoStop}%
\bibitem [{\citenamefont {Cazalilla}\ \emph {et~al.}(2006)\citenamefont
  {Cazalilla}, \citenamefont {Sols},\ and\ \citenamefont
  {Guinea}}]{PhysRevLett.97.076401}%
  \BibitemOpen
  \bibfield  {author} {\bibinfo {author} {\bibfnamefont {M.~A.}\ \bibnamefont
  {Cazalilla}}, \bibinfo {author} {\bibfnamefont {F.}~\bibnamefont {Sols}},\
  and\ \bibinfo {author} {\bibfnamefont {F.}~\bibnamefont {Guinea}},\
  }\bibfield  {title} {\bibinfo {title} {{Dissipation-Driven Quantum Phase
  Transitions in a Tomonaga-Luttinger Liquid Electrostatically Coupled to a
  Metallic Gate}},\ }\href {https://doi.org/10.1103/PhysRevLett.97.076401}
  {\bibfield  {journal} {\bibinfo  {journal} {Phys. Rev. Lett.}\ }\textbf
  {\bibinfo {volume} {97}},\ \bibinfo {pages} {076401} (\bibinfo {year}
  {2006})}\BibitemShut {NoStop}%
\bibitem [{\citenamefont {Lobos}\ \emph {et~al.}(2012)\citenamefont {Lobos},
  \citenamefont {Cazalilla},\ and\ \citenamefont
  {Chudzinski}}]{PhysRevB.86.035455}%
  \BibitemOpen
  \bibfield  {author} {\bibinfo {author} {\bibfnamefont {A.~M.}\ \bibnamefont
  {Lobos}}, \bibinfo {author} {\bibfnamefont {M.~A.}\ \bibnamefont
  {Cazalilla}},\ and\ \bibinfo {author} {\bibfnamefont {P.}~\bibnamefont
  {Chudzinski}},\ }\bibfield  {title} {\bibinfo {title} {{Magnetic phases in
  the one-dimensional Kondo chain on a metallic surface}},\ }\href
  {https://doi.org/10.1103/PhysRevB.86.035455} {\bibfield  {journal} {\bibinfo
  {journal} {Phys. Rev. B}\ }\textbf {\bibinfo {volume} {86}},\ \bibinfo
  {pages} {035455} (\bibinfo {year} {2012})}\BibitemShut {NoStop}%
\bibitem [{\citenamefont {Friedman}(2019)}]{friedman2019dissipative}%
  \BibitemOpen
  \bibfield  {author} {\bibinfo {author} {\bibfnamefont {A.~J.}\ \bibnamefont
  {Friedman}},\ }\bibfield  {title} {\bibinfo {title} {{Dissipative Luttinger
  liquids}},\ }\href@noop {} {\bibfield  {journal} {\bibinfo  {journal}
  {arXiv:1910.06371}\ } (\bibinfo {year} {2019})},\ \Eprint
  {https://arxiv.org/abs/1910.06371} {arXiv:1910.06371 [cond-mat.quant-gas]}
  \BibitemShut {NoStop}%
\bibitem [{\citenamefont {Majumdar}\ \emph {et~al.}(2023)\citenamefont
  {Majumdar}, \citenamefont {Foini}, \citenamefont {Giamarchi},\ and\
  \citenamefont {Rosso}}]{PhysRevB.107.165113}%
  \BibitemOpen
  \bibfield  {author} {\bibinfo {author} {\bibfnamefont {S.}~\bibnamefont
  {Majumdar}}, \bibinfo {author} {\bibfnamefont {L.}~\bibnamefont {Foini}},
  \bibinfo {author} {\bibfnamefont {T.}~\bibnamefont {Giamarchi}},\ and\
  \bibinfo {author} {\bibfnamefont {A.}~\bibnamefont {Rosso}},\ }\bibfield
  {title} {\bibinfo {title} {{Bath-induced phase transition in a Luttinger
  liquid}},\ }\href {https://doi.org/10.1103/PhysRevB.107.165113} {\bibfield
  {journal} {\bibinfo  {journal} {Phys. Rev. B}\ }\textbf {\bibinfo {volume}
  {107}},\ \bibinfo {pages} {165113} (\bibinfo {year} {2023})}\BibitemShut
  {NoStop}%
\bibitem [{\citenamefont {Martin}\ and\ \citenamefont
  {Grover}(2023)}]{martin2023stable}%
  \BibitemOpen
  \bibfield  {author} {\bibinfo {author} {\bibfnamefont {S.}~\bibnamefont
  {Martin}}\ and\ \bibinfo {author} {\bibfnamefont {T.}~\bibnamefont
  {Grover}},\ }\href@noop {} {\bibinfo {title} {{A stable, critical phase
  induced by Berry phase and dissipation in a spin-chain}}} (\bibinfo {year}
  {2023}),\ \Eprint {https://arxiv.org/abs/2307.13889} {arXiv:2307.13889
  [cond-mat.str-el]} \BibitemShut {NoStop}%
\bibitem [{\citenamefont {Kuklov}\ \emph {et~al.}(2023)\citenamefont {Kuklov},
  \citenamefont {Prokof'ev}, \citenamefont {Radzihovsky},\ and\ \citenamefont
  {Svistunov}}]{kuklov2023transverse}%
  \BibitemOpen
  \bibfield  {author} {\bibinfo {author} {\bibfnamefont {A.}~\bibnamefont
  {Kuklov}}, \bibinfo {author} {\bibfnamefont {N.}~\bibnamefont {Prokof'ev}},
  \bibinfo {author} {\bibfnamefont {L.}~\bibnamefont {Radzihovsky}},\ and\
  \bibinfo {author} {\bibfnamefont {B.}~\bibnamefont {Svistunov}},\ }\href@noop
  {} {\bibinfo {title} {Transverse quantum fluids}} (\bibinfo {year} {2023}),\
  \Eprint {https://arxiv.org/abs/2309.02501} {arXiv:2309.02501
  [cond-mat.other]} \BibitemShut {NoStop}%
\bibitem [{\citenamefont {Cai}\ \emph {et~al.}(2014)\citenamefont {Cai},
  \citenamefont {Schollw\"ock},\ and\ \citenamefont
  {Pollet}}]{PhysRevLett.113.260403}%
  \BibitemOpen
  \bibfield  {author} {\bibinfo {author} {\bibfnamefont {Z.}~\bibnamefont
  {Cai}}, \bibinfo {author} {\bibfnamefont {U.}~\bibnamefont {Schollw\"ock}},\
  and\ \bibinfo {author} {\bibfnamefont {L.}~\bibnamefont {Pollet}},\
  }\bibfield  {title} {\bibinfo {title} {{Identifying a Bath-Induced Bose
  Liquid in Interacting Spin-Boson Models}},\ }\href
  {https://doi.org/10.1103/PhysRevLett.113.260403} {\bibfield  {journal}
  {\bibinfo  {journal} {Phys. Rev. Lett.}\ }\textbf {\bibinfo {volume} {113}},\
  \bibinfo {pages} {260403} (\bibinfo {year} {2014})}\BibitemShut {NoStop}%
\bibitem [{\citenamefont {Yan}\ \emph {et~al.}(2018)\citenamefont {Yan},
  \citenamefont {Pollet}, \citenamefont {Lou}, \citenamefont {Wang},
  \citenamefont {Chen},\ and\ \citenamefont {Cai}}]{PhysRevB.97.035148}%
  \BibitemOpen
  \bibfield  {author} {\bibinfo {author} {\bibfnamefont {Z.}~\bibnamefont
  {Yan}}, \bibinfo {author} {\bibfnamefont {L.}~\bibnamefont {Pollet}},
  \bibinfo {author} {\bibfnamefont {J.}~\bibnamefont {Lou}}, \bibinfo {author}
  {\bibfnamefont {X.}~\bibnamefont {Wang}}, \bibinfo {author} {\bibfnamefont
  {Y.}~\bibnamefont {Chen}},\ and\ \bibinfo {author} {\bibfnamefont
  {Z.}~\bibnamefont {Cai}},\ }\bibfield  {title} {\bibinfo {title}
  {{Interacting lattice systems with quantum dissipation: A quantum Monte Carlo
  study}},\ }\href {https://doi.org/10.1103/PhysRevB.97.035148} {\bibfield
  {journal} {\bibinfo  {journal} {Phys. Rev. B}\ }\textbf {\bibinfo {volume}
  {97}},\ \bibinfo {pages} {035148} (\bibinfo {year} {2018})}\BibitemShut
  {NoStop}%
\bibitem [{\citenamefont {Weber}\ \emph {et~al.}(2022)\citenamefont {Weber},
  \citenamefont {Luitz},\ and\ \citenamefont
  {Assaad}}]{PhysRevLett.129.056402}%
  \BibitemOpen
  \bibfield  {author} {\bibinfo {author} {\bibfnamefont {M.}~\bibnamefont
  {Weber}}, \bibinfo {author} {\bibfnamefont {D.~J.}\ \bibnamefont {Luitz}},\
  and\ \bibinfo {author} {\bibfnamefont {F.~F.}\ \bibnamefont {Assaad}},\
  }\bibfield  {title} {\bibinfo {title} {{Dissipation-Induced Order: The
  $S=1/2$ Quantum Spin Chain Coupled to an Ohmic Bath}},\ }\href
  {https://doi.org/10.1103/PhysRevLett.129.056402} {\bibfield  {journal}
  {\bibinfo  {journal} {Phys. Rev. Lett.}\ }\textbf {\bibinfo {volume} {129}},\
  \bibinfo {pages} {056402} (\bibinfo {year} {2022})}\BibitemShut {NoStop}%
\bibitem [{\citenamefont {Danu}\ \emph {et~al.}(2020)\citenamefont {Danu},
  \citenamefont {Vojta}, \citenamefont {Assaad},\ and\ \citenamefont
  {Grover}}]{PhysRevLett.125.206602}%
  \BibitemOpen
  \bibfield  {author} {\bibinfo {author} {\bibfnamefont {B.}~\bibnamefont
  {Danu}}, \bibinfo {author} {\bibfnamefont {M.}~\bibnamefont {Vojta}},
  \bibinfo {author} {\bibfnamefont {F.~F.}\ \bibnamefont {Assaad}},\ and\
  \bibinfo {author} {\bibfnamefont {T.}~\bibnamefont {Grover}},\ }\bibfield
  {title} {\bibinfo {title} {{Kondo Breakdown in a Spin-$1/2$ Chain of Adatoms
  on a Dirac Semimetal}},\ }\href
  {https://doi.org/10.1103/PhysRevLett.125.206602} {\bibfield  {journal}
  {\bibinfo  {journal} {Phys. Rev. Lett.}\ }\textbf {\bibinfo {volume} {125}},\
  \bibinfo {pages} {206602} (\bibinfo {year} {2020})}\BibitemShut {NoStop}%
\bibitem [{\citenamefont {Danu}\ \emph {et~al.}(2022)\citenamefont {Danu},
  \citenamefont {Vojta}, \citenamefont {Grover},\ and\ \citenamefont
  {Assaad}}]{PhysRevB.106.L161103}%
  \BibitemOpen
  \bibfield  {author} {\bibinfo {author} {\bibfnamefont {B.}~\bibnamefont
  {Danu}}, \bibinfo {author} {\bibfnamefont {M.}~\bibnamefont {Vojta}},
  \bibinfo {author} {\bibfnamefont {T.}~\bibnamefont {Grover}},\ and\ \bibinfo
  {author} {\bibfnamefont {F.~F.}\ \bibnamefont {Assaad}},\ }\bibfield  {title}
  {\bibinfo {title} {{Spin chain on a metallic surface: Dissipation-induced
  order versus Kondo entanglement}},\ }\href
  {https://doi.org/10.1103/PhysRevB.106.L161103} {\bibfield  {journal}
  {\bibinfo  {journal} {Phys. Rev. B}\ }\textbf {\bibinfo {volume} {106}},\
  \bibinfo {pages} {L161103} (\bibinfo {year} {2022})}\BibitemShut {NoStop}%
\bibitem [{\citenamefont {Laflorencie}\ \emph {et~al.}(2005)\citenamefont
  {Laflorencie}, \citenamefont {Affleck},\ and\ \citenamefont
  {Berciu}}]{Laflorencie_2005}%
  \BibitemOpen
  \bibfield  {author} {\bibinfo {author} {\bibfnamefont {N.}~\bibnamefont
  {Laflorencie}}, \bibinfo {author} {\bibfnamefont {I.}~\bibnamefont
  {Affleck}},\ and\ \bibinfo {author} {\bibfnamefont {M.}~\bibnamefont
  {Berciu}},\ }\bibfield  {title} {\bibinfo {title} {{Critical phenomena and
  quantum phase transition in long range Heisenberg antiferromagnetic
  chains}},\ }\href {https://doi.org/10.1088/1742-5468/2005/12/p12001}
  {\bibfield  {journal} {\bibinfo  {journal} {Journal of Statistical Mechanics:
  Theory and Experiment}\ }\textbf {\bibinfo {volume} {2005}},\ \bibinfo
  {pages} {P12001} (\bibinfo {year} {2005})}\BibitemShut {NoStop}%
\bibitem [{\citenamefont
  {Sandvik}(2010{\natexlab{a}})}]{PhysRevLett.104.137204}%
  \BibitemOpen
  \bibfield  {author} {\bibinfo {author} {\bibfnamefont {A.~W.}\ \bibnamefont
  {Sandvik}},\ }\bibfield  {title} {\bibinfo {title} {{Ground States of a
  Frustrated Quantum Spin Chain with Long-Range Interactions}},\ }\href
  {https://doi.org/10.1103/PhysRevLett.104.137204} {\bibfield  {journal}
  {\bibinfo  {journal} {Phys. Rev. Lett.}\ }\textbf {\bibinfo {volume} {104}},\
  \bibinfo {pages} {137204} (\bibinfo {year} {2010}{\natexlab{a}})}\BibitemShut
  {NoStop}%
\bibitem [{\citenamefont {Yang}\ \emph {et~al.}(2020)\citenamefont {Yang},
  \citenamefont {Yao},\ and\ \citenamefont {Sandvik}}]{yang2020deconfined}%
  \BibitemOpen
  \bibfield  {author} {\bibinfo {author} {\bibfnamefont {S.}~\bibnamefont
  {Yang}}, \bibinfo {author} {\bibfnamefont {D.-X.}\ \bibnamefont {Yao}},\ and\
  \bibinfo {author} {\bibfnamefont {A.~W.}\ \bibnamefont {Sandvik}},\
  }\href@noop {} {\bibinfo {title} {Deconfined quantum criticality in spin-1/2
  chains with long-range interactions}} (\bibinfo {year} {2020}),\ \Eprint
  {https://arxiv.org/abs/2001.02821} {arXiv:2001.02821 [physics.comp-ph]}
  \BibitemShut {NoStop}%
\bibitem [{\citenamefont {Maghrebi}\ \emph {et~al.}(2017)\citenamefont
  {Maghrebi}, \citenamefont {Gong},\ and\ \citenamefont
  {Gorshkov}}]{PhysRevLett.119.023001}%
  \BibitemOpen
  \bibfield  {author} {\bibinfo {author} {\bibfnamefont {M.~F.}\ \bibnamefont
  {Maghrebi}}, \bibinfo {author} {\bibfnamefont {Z.-X.}\ \bibnamefont {Gong}},\
  and\ \bibinfo {author} {\bibfnamefont {A.~V.}\ \bibnamefont {Gorshkov}},\
  }\bibfield  {title} {\bibinfo {title} {{Continuous Symmetry Breaking in 1D
  Long-Range Interacting Quantum Systems}},\ }\href
  {https://doi.org/10.1103/PhysRevLett.119.023001} {\bibfield  {journal}
  {\bibinfo  {journal} {Phys. Rev. Lett.}\ }\textbf {\bibinfo {volume} {119}},\
  \bibinfo {pages} {023001} (\bibinfo {year} {2017})}\BibitemShut {NoStop}%
\bibitem [{\citenamefont {Diessel}\ \emph {et~al.}(2023)\citenamefont
  {Diessel}, \citenamefont {Diehl}, \citenamefont {Defenu}, \citenamefont
  {Rosch},\ and\ \citenamefont {Chiocchetta}}]{PhysRevResearch.5.033038}%
  \BibitemOpen
  \bibfield  {author} {\bibinfo {author} {\bibfnamefont {O.~K.}\ \bibnamefont
  {Diessel}}, \bibinfo {author} {\bibfnamefont {S.}~\bibnamefont {Diehl}},
  \bibinfo {author} {\bibfnamefont {N.}~\bibnamefont {Defenu}}, \bibinfo
  {author} {\bibfnamefont {A.}~\bibnamefont {Rosch}},\ and\ \bibinfo {author}
  {\bibfnamefont {A.}~\bibnamefont {Chiocchetta}},\ }\bibfield  {title}
  {\bibinfo {title} {{Generalized Higgs mechanism in long-range-interacting
  quantum systems}},\ }\href {https://doi.org/10.1103/PhysRevResearch.5.033038}
  {\bibfield  {journal} {\bibinfo  {journal} {Phys. Rev. Res.}\ }\textbf
  {\bibinfo {volume} {5}},\ \bibinfo {pages} {033038} (\bibinfo {year}
  {2023})}\BibitemShut {NoStop}%
\bibitem [{\citenamefont {Liao}\ \emph {et~al.}(2022)\citenamefont {Liao},
  \citenamefont {Xu}, \citenamefont {Meng},\ and\ \citenamefont
  {Qi}}]{daliao2022caution}%
  \BibitemOpen
  \bibfield  {author} {\bibinfo {author} {\bibfnamefont {Y.~D.}\ \bibnamefont
  {Liao}}, \bibinfo {author} {\bibfnamefont {X.~Y.}\ \bibnamefont {Xu}},
  \bibinfo {author} {\bibfnamefont {Z.~Y.}\ \bibnamefont {Meng}},\ and\
  \bibinfo {author} {\bibfnamefont {Y.}~\bibnamefont {Qi}},\ }\href@noop {}
  {\bibinfo {title} {{Caution on Gross-Neveu criticality with a single Dirac
  cone: Violation of locality and its consequence of unexpected
  finite-temperature transition}}} (\bibinfo {year} {2022}),\ \Eprint
  {https://arxiv.org/abs/2210.04272} {arXiv:2210.04272 [cond-mat.str-el]}
  \BibitemShut {NoStop}%
\bibitem [{\citenamefont {Wang}\ \emph {et~al.}(2023)\citenamefont {Wang},
  \citenamefont {Assaad},\ and\ \citenamefont
  {Ulybyshev}}]{PhysRevB.108.045105}%
  \BibitemOpen
  \bibfield  {author} {\bibinfo {author} {\bibfnamefont {Z.}~\bibnamefont
  {Wang}}, \bibinfo {author} {\bibfnamefont {F.}~\bibnamefont {Assaad}},\ and\
  \bibinfo {author} {\bibfnamefont {M.}~\bibnamefont {Ulybyshev}},\ }\bibfield
  {title} {\bibinfo {title} {{Validity of SLAC fermions for the
  $(1+1)$-dimensional helical Luttinger liquid}},\ }\href
  {https://doi.org/10.1103/PhysRevB.108.045105} {\bibfield  {journal} {\bibinfo
   {journal} {Phys. Rev. B}\ }\textbf {\bibinfo {volume} {108}},\ \bibinfo
  {pages} {045105} (\bibinfo {year} {2023})}\BibitemShut {NoStop}%
\bibitem [{\citenamefont {Song}\ \emph {et~al.}(2023)\citenamefont {Song},
  \citenamefont {Zhao}, \citenamefont {Zhou},\ and\ \citenamefont
  {Meng}}]{PhysRevResearch.5.033046}%
  \BibitemOpen
  \bibfield  {author} {\bibinfo {author} {\bibfnamefont {M.}~\bibnamefont
  {Song}}, \bibinfo {author} {\bibfnamefont {J.}~\bibnamefont {Zhao}}, \bibinfo
  {author} {\bibfnamefont {C.}~\bibnamefont {Zhou}},\ and\ \bibinfo {author}
  {\bibfnamefont {Z.~Y.}\ \bibnamefont {Meng}},\ }\bibfield  {title} {\bibinfo
  {title} {{Dynamical properties of quantum many-body systems with long-range
  interactions}},\ }\href {https://doi.org/10.1103/PhysRevResearch.5.033046}
  {\bibfield  {journal} {\bibinfo  {journal} {Phys. Rev. Res.}\ }\textbf
  {\bibinfo {volume} {5}},\ \bibinfo {pages} {033046} (\bibinfo {year}
  {2023})}\BibitemShut {NoStop}%
\bibitem [{\citenamefont {Defenu}\ \emph {et~al.}(2023)\citenamefont {Defenu},
  \citenamefont {Donner}, \citenamefont {Macr\`{\i}}, \citenamefont {Pagano},
  \citenamefont {Ruffo},\ and\ \citenamefont
  {Trombettoni}}]{RevModPhys.95.035002}%
  \BibitemOpen
  \bibfield  {author} {\bibinfo {author} {\bibfnamefont {N.}~\bibnamefont
  {Defenu}}, \bibinfo {author} {\bibfnamefont {T.}~\bibnamefont {Donner}},
  \bibinfo {author} {\bibfnamefont {T.}~\bibnamefont {Macr\`{\i}}}, \bibinfo
  {author} {\bibfnamefont {G.}~\bibnamefont {Pagano}}, \bibinfo {author}
  {\bibfnamefont {S.}~\bibnamefont {Ruffo}},\ and\ \bibinfo {author}
  {\bibfnamefont {A.}~\bibnamefont {Trombettoni}},\ }\bibfield  {title}
  {\bibinfo {title} {Long-range interacting quantum systems},\ }\href
  {https://doi.org/10.1103/RevModPhys.95.035002} {\bibfield  {journal}
  {\bibinfo  {journal} {Rev. Mod. Phys.}\ }\textbf {\bibinfo {volume} {95}},\
  \bibinfo {pages} {035002} (\bibinfo {year} {2023})}\BibitemShut {NoStop}%
\bibitem [{\citenamefont {Tsvelik}(2003)}]{tsvelik_2003}%
  \BibitemOpen
  \bibfield  {author} {\bibinfo {author} {\bibfnamefont {A.~M.}\ \bibnamefont
  {Tsvelik}},\ }\href {https://doi.org/10.1017/CBO9780511615832} {\emph
  {\bibinfo {title} {{Quantum Field Theory in Condensed Matter Physics}}}},\
  \bibinfo {edition} {2nd}\ ed.\ (\bibinfo  {publisher} {Cambridge University
  Press},\ \bibinfo {year} {2003})\BibitemShut {NoStop}%
\bibitem [{\citenamefont {Haldane}(1983)}]{PhysRevLett.50.1153}%
  \BibitemOpen
  \bibfield  {author} {\bibinfo {author} {\bibfnamefont {F.~D.~M.}\
  \bibnamefont {Haldane}},\ }\bibfield  {title} {\bibinfo {title} {{Nonlinear
  Field Theory of Large-Spin Heisenberg Antiferromagnets: Semiclassically
  Quantized Solitons of the One-Dimensional Easy-Axis N\'eel State}},\ }\href
  {https://doi.org/10.1103/PhysRevLett.50.1153} {\bibfield  {journal} {\bibinfo
   {journal} {Phys. Rev. Lett.}\ }\textbf {\bibinfo {volume} {50}},\ \bibinfo
  {pages} {1153} (\bibinfo {year} {1983})}\BibitemShut {NoStop}%
\bibitem [{\citenamefont {Mermin}\ and\ \citenamefont
  {Wagner}(1966)}]{PhysRevLett.17.1133}%
  \BibitemOpen
  \bibfield  {author} {\bibinfo {author} {\bibfnamefont {N.~D.}\ \bibnamefont
  {Mermin}}\ and\ \bibinfo {author} {\bibfnamefont {H.}~\bibnamefont
  {Wagner}},\ }\bibfield  {title} {\bibinfo {title} {{Absence of Ferromagnetism
  or Antiferromagnetism in One- or Two-Dimensional Isotropic Heisenberg
  Models}},\ }\href {https://doi.org/10.1103/PhysRevLett.17.1133} {\bibfield
  {journal} {\bibinfo  {journal} {Phys. Rev. Lett.}\ }\textbf {\bibinfo
  {volume} {17}},\ \bibinfo {pages} {1133} (\bibinfo {year}
  {1966})}\BibitemShut {NoStop}%
\bibitem [{\citenamefont {Hohenberg}(1967)}]{PhysRev.158.383}%
  \BibitemOpen
  \bibfield  {author} {\bibinfo {author} {\bibfnamefont {P.~C.}\ \bibnamefont
  {Hohenberg}},\ }\bibfield  {title} {\bibinfo {title} {{Existence of
  Long-Range Order in One and Two Dimensions}},\ }\href
  {https://doi.org/10.1103/PhysRev.158.383} {\bibfield  {journal} {\bibinfo
  {journal} {Phys. Rev.}\ }\textbf {\bibinfo {volume} {158}},\ \bibinfo {pages}
  {383} (\bibinfo {year} {1967})}\BibitemShut {NoStop}%
\bibitem [{\citenamefont {Haldane}(1982)}]{PhysRevB.25.4925}%
  \BibitemOpen
  \bibfield  {author} {\bibinfo {author} {\bibfnamefont {F.~D.~M.}\
  \bibnamefont {Haldane}},\ }\bibfield  {title} {\bibinfo {title} {{Spontaneous
  dimerization in the $S=\frac{1}{2}$ Heisenberg antiferromagnetic chain with
  competing interactions}},\ }\href {https://doi.org/10.1103/PhysRevB.25.4925}
  {\bibfield  {journal} {\bibinfo  {journal} {Phys. Rev. B}\ }\textbf {\bibinfo
  {volume} {25}},\ \bibinfo {pages} {4925} (\bibinfo {year}
  {1982})}\BibitemShut {NoStop}%
\bibitem [{\citenamefont {Sanyal}\ \emph {et~al.}(2011)\citenamefont {Sanyal},
  \citenamefont {Banerjee},\ and\ \citenamefont {Damle}}]{PhysRevB.84.235129}%
  \BibitemOpen
  \bibfield  {author} {\bibinfo {author} {\bibfnamefont {S.}~\bibnamefont
  {Sanyal}}, \bibinfo {author} {\bibfnamefont {A.}~\bibnamefont {Banerjee}},\
  and\ \bibinfo {author} {\bibfnamefont {K.}~\bibnamefont {Damle}},\ }\bibfield
   {title} {\bibinfo {title} {{Vacancy-induced spin texture in a
  one-dimensional $S=\frac{1}{2}$ Heisenberg antiferromagnet}},\ }\href
  {https://doi.org/10.1103/PhysRevB.84.235129} {\bibfield  {journal} {\bibinfo
  {journal} {Phys. Rev. B}\ }\textbf {\bibinfo {volume} {84}},\ \bibinfo
  {pages} {235129} (\bibinfo {year} {2011})}\BibitemShut {NoStop}%
\bibitem [{\citenamefont {Tang}\ and\ \citenamefont
  {Sandvik}(2011)}]{PhysRevLett.107.157201}%
  \BibitemOpen
  \bibfield  {author} {\bibinfo {author} {\bibfnamefont {Y.}~\bibnamefont
  {Tang}}\ and\ \bibinfo {author} {\bibfnamefont {A.~W.}\ \bibnamefont
  {Sandvik}},\ }\bibfield  {title} {\bibinfo {title} {{Method to Characterize
  Spinons as Emergent Elementary Particles}},\ }\href
  {https://doi.org/10.1103/PhysRevLett.107.157201} {\bibfield  {journal}
  {\bibinfo  {journal} {Phys. Rev. Lett.}\ }\textbf {\bibinfo {volume} {107}},\
  \bibinfo {pages} {157201} (\bibinfo {year} {2011})}\BibitemShut {NoStop}%
\bibitem [{\citenamefont {Patil}\ \emph {et~al.}(2018)\citenamefont {Patil},
  \citenamefont {Katz},\ and\ \citenamefont {Sandvik}}]{PhysRevB.98.014414}%
  \BibitemOpen
  \bibfield  {author} {\bibinfo {author} {\bibfnamefont {P.}~\bibnamefont
  {Patil}}, \bibinfo {author} {\bibfnamefont {E.}~\bibnamefont {Katz}},\ and\
  \bibinfo {author} {\bibfnamefont {A.~W.}\ \bibnamefont {Sandvik}},\
  }\bibfield  {title} {\bibinfo {title} {{Numerical investigations of SO(4)
  emergent extended symmetry in spin-$\frac{1}{2}$ Heisenberg antiferromagnetic
  chains}},\ }\href {https://doi.org/10.1103/PhysRevB.98.014414} {\bibfield
  {journal} {\bibinfo  {journal} {Phys. Rev. B}\ }\textbf {\bibinfo {volume}
  {98}},\ \bibinfo {pages} {014414} (\bibinfo {year} {2018})}\BibitemShut
  {NoStop}%
\bibitem [{\citenamefont {Pytte}(1974)}]{PhysRevB.10.4637}%
  \BibitemOpen
  \bibfield  {author} {\bibinfo {author} {\bibfnamefont {E.}~\bibnamefont
  {Pytte}},\ }\bibfield  {title} {\bibinfo {title} {{Peierls instability in
  Heisenberg chains}},\ }\href {https://doi.org/10.1103/PhysRevB.10.4637}
  {\bibfield  {journal} {\bibinfo  {journal} {Phys. Rev. B}\ }\textbf {\bibinfo
  {volume} {10}},\ \bibinfo {pages} {4637} (\bibinfo {year}
  {1974})}\BibitemShut {NoStop}%
\bibitem [{\citenamefont {Cross}\ and\ \citenamefont
  {Fisher}(1979)}]{PhysRevB.19.402}%
  \BibitemOpen
  \bibfield  {author} {\bibinfo {author} {\bibfnamefont {M.~C.}\ \bibnamefont
  {Cross}}\ and\ \bibinfo {author} {\bibfnamefont {D.~S.}\ \bibnamefont
  {Fisher}},\ }\bibfield  {title} {\bibinfo {title} {{A new theory of the
  spin-Peierls transition with special relevance to the experiments on
  TTFCuBDT}},\ }\href {https://doi.org/10.1103/PhysRevB.19.402} {\bibfield
  {journal} {\bibinfo  {journal} {Phys. Rev. B}\ }\textbf {\bibinfo {volume}
  {19}},\ \bibinfo {pages} {402} (\bibinfo {year} {1979})}\BibitemShut
  {NoStop}%
\bibitem [{\citenamefont {Uhrig}(1998)}]{PhysRevB.57.R14004}%
  \BibitemOpen
  \bibfield  {author} {\bibinfo {author} {\bibfnamefont {G.~S.}\ \bibnamefont
  {Uhrig}},\ }\bibfield  {title} {\bibinfo {title} {{Nonadiabatic approach to
  spin-Peierls transitions via flow equations}},\ }\href
  {https://doi.org/10.1103/PhysRevB.57.R14004} {\bibfield  {journal} {\bibinfo
  {journal} {Phys. Rev. B}\ }\textbf {\bibinfo {volume} {57}},\ \bibinfo
  {pages} {R14004} (\bibinfo {year} {1998})}\BibitemShut {NoStop}%
\bibitem [{\citenamefont {Wellein}\ \emph {et~al.}(1998)\citenamefont
  {Wellein}, \citenamefont {Fehske},\ and\ \citenamefont
  {Kampf}}]{PhysRevLett.81.3956}%
  \BibitemOpen
  \bibfield  {author} {\bibinfo {author} {\bibfnamefont {G.}~\bibnamefont
  {Wellein}}, \bibinfo {author} {\bibfnamefont {H.}~\bibnamefont {Fehske}},\
  and\ \bibinfo {author} {\bibfnamefont {A.~P.}\ \bibnamefont {Kampf}},\
  }\bibfield  {title} {\bibinfo {title} {{Peierls Dimerization with
  Nonadiabatic Spin-Phonon Coupling}},\ }\href
  {https://doi.org/10.1103/PhysRevLett.81.3956} {\bibfield  {journal} {\bibinfo
   {journal} {Phys. Rev. Lett.}\ }\textbf {\bibinfo {volume} {81}},\ \bibinfo
  {pages} {3956} (\bibinfo {year} {1998})}\BibitemShut {NoStop}%
\bibitem [{\citenamefont {Bursill}\ \emph {et~al.}(1999)\citenamefont
  {Bursill}, \citenamefont {McKenzie},\ and\ \citenamefont
  {Hamer}}]{PhysRevLett.83.408}%
  \BibitemOpen
  \bibfield  {author} {\bibinfo {author} {\bibfnamefont {R.~J.}\ \bibnamefont
  {Bursill}}, \bibinfo {author} {\bibfnamefont {R.~H.}\ \bibnamefont
  {McKenzie}},\ and\ \bibinfo {author} {\bibfnamefont {C.~J.}\ \bibnamefont
  {Hamer}},\ }\bibfield  {title} {\bibinfo {title} {{Phase Diagram of a
  Heisenberg Spin-Peierls Model with Quantum Phonons}},\ }\href
  {https://doi.org/10.1103/PhysRevLett.83.408} {\bibfield  {journal} {\bibinfo
  {journal} {Phys. Rev. Lett.}\ }\textbf {\bibinfo {volume} {83}},\ \bibinfo
  {pages} {408} (\bibinfo {year} {1999})}\BibitemShut {NoStop}%
\bibitem [{\citenamefont {Sandvik}\ and\ \citenamefont
  {Campbell}(1999)}]{PhysRevLett.83.195}%
  \BibitemOpen
  \bibfield  {author} {\bibinfo {author} {\bibfnamefont {A.~W.}\ \bibnamefont
  {Sandvik}}\ and\ \bibinfo {author} {\bibfnamefont {D.~K.}\ \bibnamefont
  {Campbell}},\ }\bibfield  {title} {\bibinfo {title} {{Spin-Peierls Transition
  in the Heisenberg Chain with Finite-Frequency Phonons}},\ }\href
  {https://doi.org/10.1103/PhysRevLett.83.195} {\bibfield  {journal} {\bibinfo
  {journal} {Phys. Rev. Lett.}\ }\textbf {\bibinfo {volume} {83}},\ \bibinfo
  {pages} {195} (\bibinfo {year} {1999})}\BibitemShut {NoStop}%
\bibitem [{\citenamefont {Wei\ss{}e}\ \emph {et~al.}(1999)\citenamefont
  {Wei\ss{}e}, \citenamefont {Wellein},\ and\ \citenamefont
  {Fehske}}]{PhysRevB.60.6566}%
  \BibitemOpen
  \bibfield  {author} {\bibinfo {author} {\bibfnamefont {A.}~\bibnamefont
  {Wei\ss{}e}}, \bibinfo {author} {\bibfnamefont {G.}~\bibnamefont {Wellein}},\
  and\ \bibinfo {author} {\bibfnamefont {H.}~\bibnamefont {Fehske}},\
  }\bibfield  {title} {\bibinfo {title} {{Quantum lattice fluctuations in a
  frustrated Heisenberg spin-Peierls chain}},\ }\href
  {https://doi.org/10.1103/PhysRevB.60.6566} {\bibfield  {journal} {\bibinfo
  {journal} {Phys. Rev. B}\ }\textbf {\bibinfo {volume} {60}},\ \bibinfo
  {pages} {6566} (\bibinfo {year} {1999})}\BibitemShut {NoStop}%
\bibitem [{\citenamefont {Citro}\ \emph {et~al.}(2005)\citenamefont {Citro},
  \citenamefont {Orignac},\ and\ \citenamefont
  {Giamarchi}}]{PhysRevB.72.024434}%
  \BibitemOpen
  \bibfield  {author} {\bibinfo {author} {\bibfnamefont {R.}~\bibnamefont
  {Citro}}, \bibinfo {author} {\bibfnamefont {E.}~\bibnamefont {Orignac}},\
  and\ \bibinfo {author} {\bibfnamefont {T.}~\bibnamefont {Giamarchi}},\
  }\bibfield  {title} {\bibinfo {title} {{Adiabatic-antiadiabatic crossover in
  a spin-Peierls chain}},\ }\href {https://doi.org/10.1103/PhysRevB.72.024434}
  {\bibfield  {journal} {\bibinfo  {journal} {Phys. Rev. B}\ }\textbf {\bibinfo
  {volume} {72}},\ \bibinfo {pages} {024434} (\bibinfo {year}
  {2005})}\BibitemShut {NoStop}%
\bibitem [{\citenamefont {Wei\ss{}e}\ \emph {et~al.}(2006)\citenamefont
  {Wei\ss{}e}, \citenamefont {Hager}, \citenamefont {Bishop},\ and\
  \citenamefont {Fehske}}]{PhysRevB.74.214426}%
  \BibitemOpen
  \bibfield  {author} {\bibinfo {author} {\bibfnamefont {A.}~\bibnamefont
  {Wei\ss{}e}}, \bibinfo {author} {\bibfnamefont {G.}~\bibnamefont {Hager}},
  \bibinfo {author} {\bibfnamefont {A.~R.}\ \bibnamefont {Bishop}},\ and\
  \bibinfo {author} {\bibfnamefont {H.}~\bibnamefont {Fehske}},\ }\bibfield
  {title} {\bibinfo {title} {{Phase diagram of the spin-Peierls chain with
  local coupling: Density-matrix renormalization-group calculations and unitary
  transformations}},\ }\href {https://doi.org/10.1103/PhysRevB.74.214426}
  {\bibfield  {journal} {\bibinfo  {journal} {Phys. Rev. B}\ }\textbf {\bibinfo
  {volume} {74}},\ \bibinfo {pages} {214426} (\bibinfo {year}
  {2006})}\BibitemShut {NoStop}%
\bibitem [{\citenamefont {Michel}\ and\ \citenamefont
  {Evertz}(2007)}]{2007arXiv0705.0799M}%
  \BibitemOpen
  \bibfield  {author} {\bibinfo {author} {\bibfnamefont {F.}~\bibnamefont
  {Michel}}\ and\ \bibinfo {author} {\bibfnamefont {H.-G.}\ \bibnamefont
  {Evertz}},\ }\href@noop {} {\bibinfo {title} {{Lattice dynamics of the
  Heisenberg chain coupled to finite frequency bond phonons}}} (\bibinfo {year}
  {2007}),\ \Eprint {https://arxiv.org/abs/0705.0799} {arXiv:0705.0799
  [cond-mat.str-el]} \BibitemShut {NoStop}%
\bibitem [{\citenamefont {Suwa}\ and\ \citenamefont
  {Todo}(2015)}]{PhysRevLett.115.080601}%
  \BibitemOpen
  \bibfield  {author} {\bibinfo {author} {\bibfnamefont {H.}~\bibnamefont
  {Suwa}}\ and\ \bibinfo {author} {\bibfnamefont {S.}~\bibnamefont {Todo}},\
  }\bibfield  {title} {\bibinfo {title} {{Generalized Moment Method for Gap
  Estimation and Quantum Monte Carlo Level Spectroscopy}},\ }\href
  {https://doi.org/10.1103/PhysRevLett.115.080601} {\bibfield  {journal}
  {\bibinfo  {journal} {Phys. Rev. Lett.}\ }\textbf {\bibinfo {volume} {115}},\
  \bibinfo {pages} {080601} (\bibinfo {year} {2015})}\BibitemShut {NoStop}%
\bibitem [{\citenamefont {Warawa}\ \emph {et~al.}(2023)\citenamefont {Warawa},
  \citenamefont {Christophel}, \citenamefont {Sobolev}, \citenamefont {Demsar},
  \citenamefont {Roskos},\ and\ \citenamefont {Thomson}}]{PhysRevB.108.045147}%
  \BibitemOpen
  \bibfield  {author} {\bibinfo {author} {\bibfnamefont {K.}~\bibnamefont
  {Warawa}}, \bibinfo {author} {\bibfnamefont {N.}~\bibnamefont {Christophel}},
  \bibinfo {author} {\bibfnamefont {S.}~\bibnamefont {Sobolev}}, \bibinfo
  {author} {\bibfnamefont {J.}~\bibnamefont {Demsar}}, \bibinfo {author}
  {\bibfnamefont {H.~G.}\ \bibnamefont {Roskos}},\ and\ \bibinfo {author}
  {\bibfnamefont {M.~D.}\ \bibnamefont {Thomson}},\ }\bibfield  {title}
  {\bibinfo {title} {{Combined investigation of collective amplitude and phase
  modes in a quasi-one-dimensional charge density wave system over a wide
  spectral range}},\ }\href {https://doi.org/10.1103/PhysRevB.108.045147}
  {\bibfield  {journal} {\bibinfo  {journal} {Phys. Rev. B}\ }\textbf {\bibinfo
  {volume} {108}},\ \bibinfo {pages} {045147} (\bibinfo {year}
  {2023})}\BibitemShut {NoStop}%
\bibitem [{\citenamefont {Weber}\ \emph {et~al.}(2017)\citenamefont {Weber},
  \citenamefont {Assaad},\ and\ \citenamefont
  {Hohenadler}}]{PhysRevLett.119.097401}%
  \BibitemOpen
  \bibfield  {author} {\bibinfo {author} {\bibfnamefont {M.}~\bibnamefont
  {Weber}}, \bibinfo {author} {\bibfnamefont {F.~F.}\ \bibnamefont {Assaad}},\
  and\ \bibinfo {author} {\bibfnamefont {M.}~\bibnamefont {Hohenadler}},\
  }\bibfield  {title} {\bibinfo {title} {{Directed-Loop Quantum Monte Carlo
  Method for Retarded Interactions}},\ }\href
  {https://doi.org/10.1103/PhysRevLett.119.097401} {\bibfield  {journal}
  {\bibinfo  {journal} {Phys. Rev. Lett.}\ }\textbf {\bibinfo {volume} {119}},\
  \bibinfo {pages} {097401} (\bibinfo {year} {2017})}\BibitemShut {NoStop}%
\bibitem [{\citenamefont {Sylju\aa{}sen}\ and\ \citenamefont
  {Sandvik}(2002)}]{PhysRevE.66.046701}%
  \BibitemOpen
  \bibfield  {author} {\bibinfo {author} {\bibfnamefont {O.~F.}\ \bibnamefont
  {Sylju\aa{}sen}}\ and\ \bibinfo {author} {\bibfnamefont {A.~W.}\ \bibnamefont
  {Sandvik}},\ }\bibfield  {title} {\bibinfo {title} {{Quantum Monte Carlo with
  directed loops}},\ }\href {https://doi.org/10.1103/PhysRevE.66.046701}
  {\bibfield  {journal} {\bibinfo  {journal} {Phys. Rev. E}\ }\textbf {\bibinfo
  {volume} {66}},\ \bibinfo {pages} {046701} (\bibinfo {year}
  {2002})}\BibitemShut {NoStop}%
\bibitem [{\citenamefont {Weber}(2022)}]{PhysRevB.105.165129}%
  \BibitemOpen
  \bibfield  {author} {\bibinfo {author} {\bibfnamefont {M.}~\bibnamefont
  {Weber}},\ }\bibfield  {title} {\bibinfo {title} {{Quantum Monte Carlo
  simulation of spin-boson models using wormhole updates}},\ }\href
  {https://doi.org/10.1103/PhysRevB.105.165129} {\bibfield  {journal} {\bibinfo
   {journal} {Phys. Rev. B}\ }\textbf {\bibinfo {volume} {105}},\ \bibinfo
  {pages} {165129} (\bibinfo {year} {2022})}\BibitemShut {NoStop}%
\bibitem [{\citenamefont {Costa}\ \emph {et~al.}(2023)\citenamefont {Costa},
  \citenamefont {Cohen-Stead}, \citenamefont {Ly}, \citenamefont {Neuhaus},\
  and\ \citenamefont {Johnston}}]{costa2023comparative}%
  \BibitemOpen
  \bibfield  {author} {\bibinfo {author} {\bibfnamefont {S.~M.}\ \bibnamefont
  {Costa}}, \bibinfo {author} {\bibfnamefont {B.}~\bibnamefont {Cohen-Stead}},
  \bibinfo {author} {\bibfnamefont {A.~T.}\ \bibnamefont {Ly}}, \bibinfo
  {author} {\bibfnamefont {J.}~\bibnamefont {Neuhaus}},\ and\ \bibinfo {author}
  {\bibfnamefont {S.}~\bibnamefont {Johnston}},\ }\href@noop {} {\bibinfo
  {title} {{A comparative determinant quantum Monte Carlo study of the acoustic
  and optical variants of the Su-Schrieffer-Heeger model}}} (\bibinfo {year}
  {2023}),\ \Eprint {https://arxiv.org/abs/2307.10058} {arXiv:2307.10058
  [cond-mat.str-el]} \BibitemShut {NoStop}%
\bibitem [{\citenamefont {{Affleck}}\ \emph {et~al.}(1989)\citenamefont
  {{Affleck}}, \citenamefont {{Gepner}}, \citenamefont {{Schulz}},\ and\
  \citenamefont {{Ziman}}}]{1989JPhA...22..511A}%
  \BibitemOpen
  \bibfield  {author} {\bibinfo {author} {\bibfnamefont {I.}~\bibnamefont
  {{Affleck}}}, \bibinfo {author} {\bibfnamefont {D.}~\bibnamefont {{Gepner}}},
  \bibinfo {author} {\bibfnamefont {H.~J.}\ \bibnamefont {{Schulz}}},\ and\
  \bibinfo {author} {\bibfnamefont {T.}~\bibnamefont {{Ziman}}},\ }\bibfield
  {title} {\bibinfo {title} {{Critical behaviour of spin-s Heisenberg
  antiferromagnetic chains: analytic and numerical results}},\ }\href
  {https://doi.org/10.1088/0305-4470/22/5/015} {\bibfield  {journal} {\bibinfo
  {journal} {Journal of Physics A Mathematical General}\ }\textbf {\bibinfo
  {volume} {22}},\ \bibinfo {pages} {511} (\bibinfo {year} {1989})}\BibitemShut
  {NoStop}%
\bibitem [{\citenamefont {Singh}\ \emph {et~al.}(1989)\citenamefont {Singh},
  \citenamefont {Fisher},\ and\ \citenamefont {Shankar}}]{PhysRevB.39.2562}%
  \BibitemOpen
  \bibfield  {author} {\bibinfo {author} {\bibfnamefont {R.~R.~P.}\
  \bibnamefont {Singh}}, \bibinfo {author} {\bibfnamefont {M.~E.}\ \bibnamefont
  {Fisher}},\ and\ \bibinfo {author} {\bibfnamefont {R.}~\bibnamefont
  {Shankar}},\ }\bibfield  {title} {\bibinfo {title} {{Spin-(1/2
  antiferromagnetic XXZ chain: New results and insights}},\ }\href
  {https://doi.org/10.1103/PhysRevB.39.2562} {\bibfield  {journal} {\bibinfo
  {journal} {Phys. Rev. B}\ }\textbf {\bibinfo {volume} {39}},\ \bibinfo
  {pages} {2562} (\bibinfo {year} {1989})}\BibitemShut {NoStop}%
\bibitem [{\citenamefont {Giamarchi}\ and\ \citenamefont
  {Schulz}(1989)}]{PhysRevB.39.4620}%
  \BibitemOpen
  \bibfield  {author} {\bibinfo {author} {\bibfnamefont {T.}~\bibnamefont
  {Giamarchi}}\ and\ \bibinfo {author} {\bibfnamefont {H.~J.}\ \bibnamefont
  {Schulz}},\ }\bibfield  {title} {\bibinfo {title} {Correlation functions of
  one-dimensional quantum systems},\ }\href
  {https://doi.org/10.1103/PhysRevB.39.4620} {\bibfield  {journal} {\bibinfo
  {journal} {Phys. Rev. B}\ }\textbf {\bibinfo {volume} {39}},\ \bibinfo
  {pages} {4620} (\bibinfo {year} {1989})}\BibitemShut {NoStop}%
\bibitem [{\citenamefont {Affleck}(1998)}]{affleck_exact_1998}%
  \BibitemOpen
  \bibfield  {author} {\bibinfo {author} {\bibfnamefont {I.}~\bibnamefont
  {Affleck}},\ }\bibfield  {title} {\bibinfo {title} {Exact correlation
  amplitude for the {Heisenberg} antiferromagnetic chain},\ }\href
  {https://doi.org/10.1088/0305-4470/31/20/002} {\bibfield  {journal} {\bibinfo
   {journal} {Journal of Physics A: Mathematical and General}\ }\textbf
  {\bibinfo {volume} {31}},\ \bibinfo {pages} {4573} (\bibinfo {year}
  {1998})}\BibitemShut {NoStop}%
\bibitem [{\citenamefont {Hikihara}\ \emph {et~al.}(2017)\citenamefont
  {Hikihara}, \citenamefont {Furusaki},\ and\ \citenamefont
  {Lukyanov}}]{PhysRevB.96.134429}%
  \BibitemOpen
  \bibfield  {author} {\bibinfo {author} {\bibfnamefont {T.}~\bibnamefont
  {Hikihara}}, \bibinfo {author} {\bibfnamefont {A.}~\bibnamefont {Furusaki}},\
  and\ \bibinfo {author} {\bibfnamefont {S.}~\bibnamefont {Lukyanov}},\
  }\bibfield  {title} {\bibinfo {title} {{Dimer correlation amplitudes and
  dimer excitation gap in spin-$\frac{1}{2}$ XXZ and Heisenberg chains}},\
  }\href {https://doi.org/10.1103/PhysRevB.96.134429} {\bibfield  {journal}
  {\bibinfo  {journal} {Phys. Rev. B}\ }\textbf {\bibinfo {volume} {96}},\
  \bibinfo {pages} {134429} (\bibinfo {year} {2017})}\BibitemShut {NoStop}%
\bibitem [{\citenamefont {Cardy}(1996)}]{cardy_1996}%
  \BibitemOpen
  \bibfield  {author} {\bibinfo {author} {\bibfnamefont {J.}~\bibnamefont
  {Cardy}},\ }\href {https://doi.org/10.1017/CBO9781316036440} {\emph {\bibinfo
  {title} {{Scaling and Renormalization in Statistical Physics}}}},\ Cambridge
  Lecture Notes in Physics\ (\bibinfo  {publisher} {Cambridge University
  Press},\ \bibinfo {year} {1996})\BibitemShut {NoStop}%
\bibitem [{\citenamefont {Prokof'ev}\ \emph {et~al.}(1998)\citenamefont
  {Prokof'ev}, \citenamefont {Svistunov},\ and\ \citenamefont
  {Tupitsyn}}]{Prokofev:1998aa}%
  \BibitemOpen
  \bibfield  {author} {\bibinfo {author} {\bibfnamefont {N.~V.}\ \bibnamefont
  {Prokof'ev}}, \bibinfo {author} {\bibfnamefont {B.~V.}\ \bibnamefont
  {Svistunov}},\ and\ \bibinfo {author} {\bibfnamefont {I.~S.}\ \bibnamefont
  {Tupitsyn}},\ }\bibfield  {title} {\bibinfo {title} {{Exact, complete, and
  universal continuous-time worldline Monte Carlo approach to the statistics of
  discrete quantum systems}},\ }\href {https://doi.org/10.1134/1.558661}
  {\bibfield  {journal} {\bibinfo  {journal} {Journal of Experimental and
  Theoretical Physics}\ }\textbf {\bibinfo {volume} {87}},\ \bibinfo {pages}
  {310} (\bibinfo {year} {1998})}\BibitemShut {NoStop}%
\bibitem [{\citenamefont {Sandvik}\ and\ \citenamefont
  {Kurkij\"arvi}(1991)}]{PhysRevB.43.5950}%
  \BibitemOpen
  \bibfield  {author} {\bibinfo {author} {\bibfnamefont {A.~W.}\ \bibnamefont
  {Sandvik}}\ and\ \bibinfo {author} {\bibfnamefont {J.}~\bibnamefont
  {Kurkij\"arvi}},\ }\bibfield  {title} {\bibinfo {title} {{Quantum Monte Carlo
  simulation method for spin systems}},\ }\href
  {https://doi.org/10.1103/PhysRevB.43.5950} {\bibfield  {journal} {\bibinfo
  {journal} {Phys. Rev. B}\ }\textbf {\bibinfo {volume} {43}},\ \bibinfo
  {pages} {5950} (\bibinfo {year} {1991})}\BibitemShut {NoStop}%
\bibitem [{\citenamefont {Sandvik}(1999)}]{PhysRevB.59.R14157}%
  \BibitemOpen
  \bibfield  {author} {\bibinfo {author} {\bibfnamefont {A.~W.}\ \bibnamefont
  {Sandvik}},\ }\bibfield  {title} {\bibinfo {title} {Stochastic series
  expansion method with operator-loop update},\ }\href
  {https://doi.org/10.1103/PhysRevB.59.R14157} {\bibfield  {journal} {\bibinfo
  {journal} {Phys. Rev. B}\ }\textbf {\bibinfo {volume} {59}},\ \bibinfo
  {pages} {R14157} (\bibinfo {year} {1999})}\BibitemShut {NoStop}%
\bibitem [{\citenamefont {Weber}(2021)}]{PhysRevB.103.L041105}%
  \BibitemOpen
  \bibfield  {author} {\bibinfo {author} {\bibfnamefont {M.}~\bibnamefont
  {Weber}},\ }\bibfield  {title} {\bibinfo {title} {{Valence bond order in a
  honeycomb antiferromagnet coupled to quantum phonons}},\ }\href
  {https://doi.org/10.1103/PhysRevB.103.L041105} {\bibfield  {journal}
  {\bibinfo  {journal} {Phys. Rev. B}\ }\textbf {\bibinfo {volume} {103}},\
  \bibinfo {pages} {L041105} (\bibinfo {year} {2021})}\BibitemShut {NoStop}%
\bibitem [{\citenamefont {Sandvik}\ \emph {et~al.}(1997)\citenamefont
  {Sandvik}, \citenamefont {Singh},\ and\ \citenamefont
  {Campbell}}]{PhysRevB.56.14510}%
  \BibitemOpen
  \bibfield  {author} {\bibinfo {author} {\bibfnamefont {A.~W.}\ \bibnamefont
  {Sandvik}}, \bibinfo {author} {\bibfnamefont {R.~R.~P.}\ \bibnamefont
  {Singh}},\ and\ \bibinfo {author} {\bibfnamefont {D.~K.}\ \bibnamefont
  {Campbell}},\ }\bibfield  {title} {\bibinfo {title} {{Quantum Monte Carlo in
  the interaction representation: Application to a spin-Peierls model}},\
  }\href {https://doi.org/10.1103/PhysRevB.56.14510} {\bibfield  {journal}
  {\bibinfo  {journal} {Phys. Rev. B}\ }\textbf {\bibinfo {volume} {56}},\
  \bibinfo {pages} {14510} (\bibinfo {year} {1997})}\BibitemShut {NoStop}%
\bibitem [{\citenamefont {Weber}\ \emph {et~al.}(2016)\citenamefont {Weber},
  \citenamefont {Assaad},\ and\ \citenamefont
  {Hohenadler}}]{PhysRevB.94.245138}%
  \BibitemOpen
  \bibfield  {author} {\bibinfo {author} {\bibfnamefont {M.}~\bibnamefont
  {Weber}}, \bibinfo {author} {\bibfnamefont {F.~F.}\ \bibnamefont {Assaad}},\
  and\ \bibinfo {author} {\bibfnamefont {M.}~\bibnamefont {Hohenadler}},\
  }\bibfield  {title} {\bibinfo {title} {{Continuous-time quantum Monte Carlo
  for fermion-boson lattice models: Improved bosonic estimators and application
  to the Holstein model}},\ }\href {https://doi.org/10.1103/PhysRevB.94.245138}
  {\bibfield  {journal} {\bibinfo  {journal} {Phys. Rev. B}\ }\textbf {\bibinfo
  {volume} {94}},\ \bibinfo {pages} {245138} (\bibinfo {year}
  {2016})}\BibitemShut {NoStop}%
\bibitem [{\citenamefont
  {Sandvik}(2010{\natexlab{b}})}]{doi:10.1063/1.3518900}%
  \BibitemOpen
  \bibfield  {author} {\bibinfo {author} {\bibfnamefont {A.~W.}\ \bibnamefont
  {Sandvik}},\ }\bibfield  {title} {\bibinfo {title} {{Computational Studies of
  Quantum Spin Systems}},\ }\href {https://doi.org/10.1063/1.3518900}
  {\bibfield  {journal} {\bibinfo  {journal} {AIP Conference Proceedings}\
  }\textbf {\bibinfo {volume} {1297}},\ \bibinfo {pages} {135} (\bibinfo {year}
  {2010}{\natexlab{b}})}\BibitemShut {NoStop}%
\bibitem [{\citenamefont {Weber}\ \emph {et~al.}(2020)\citenamefont {Weber},
  \citenamefont {Parisen~Toldin},\ and\ \citenamefont
  {Hohenadler}}]{PhysRevResearch.2.023013}%
  \BibitemOpen
  \bibfield  {author} {\bibinfo {author} {\bibfnamefont {M.}~\bibnamefont
  {Weber}}, \bibinfo {author} {\bibfnamefont {F.}~\bibnamefont
  {Parisen~Toldin}},\ and\ \bibinfo {author} {\bibfnamefont {M.}~\bibnamefont
  {Hohenadler}},\ }\bibfield  {title} {\bibinfo {title} {{Competing orders and
  unconventional criticality in the Su-Schrieffer-Heeger model}},\ }\href
  {https://doi.org/10.1103/PhysRevResearch.2.023013} {\bibfield  {journal}
  {\bibinfo  {journal} {Phys. Rev. Research}\ }\textbf {\bibinfo {volume}
  {2}},\ \bibinfo {pages} {023013} (\bibinfo {year} {2020})}\BibitemShut
  {NoStop}%
\bibitem [{\citenamefont {Okamoto}\ and\ \citenamefont
  {Nomura}(1992)}]{OKAMOTO1992433}%
  \BibitemOpen
  \bibfield  {author} {\bibinfo {author} {\bibfnamefont {K.}~\bibnamefont
  {Okamoto}}\ and\ \bibinfo {author} {\bibfnamefont {K.}~\bibnamefont
  {Nomura}},\ }\bibfield  {title} {\bibinfo {title} {{Fluid-dimer critical
  point in S = 12 antiferromagnetic Heisenberg chain with next nearest neighbor
  interactions}},\ }\href
  {https://doi.org/https://doi.org/10.1016/0375-9601(92)90823-5} {\bibfield
  {journal} {\bibinfo  {journal} {Physics Letters A}\ }\textbf {\bibinfo
  {volume} {169}},\ \bibinfo {pages} {433} (\bibinfo {year}
  {1992})}\BibitemShut {NoStop}%
\bibitem [{\citenamefont {Eggert}(1996)}]{PhysRevB.54.R9612}%
  \BibitemOpen
  \bibfield  {author} {\bibinfo {author} {\bibfnamefont {S.}~\bibnamefont
  {Eggert}},\ }\bibfield  {title} {\bibinfo {title} {Numerical evidence for
  multiplicative logarithmic corrections from marginal operators},\ }\href
  {https://doi.org/10.1103/PhysRevB.54.R9612} {\bibfield  {journal} {\bibinfo
  {journal} {Phys. Rev. B}\ }\textbf {\bibinfo {volume} {54}},\ \bibinfo
  {pages} {R9612} (\bibinfo {year} {1996})}\BibitemShut {NoStop}%
\bibitem [{\citenamefont {Weber}\ \emph
  {et~al.}(2015{\natexlab{a}})\citenamefont {Weber}, \citenamefont {Assaad},\
  and\ \citenamefont {Hohenadler}}]{PhysRevB.91.235150}%
  \BibitemOpen
  \bibfield  {author} {\bibinfo {author} {\bibfnamefont {M.}~\bibnamefont
  {Weber}}, \bibinfo {author} {\bibfnamefont {F.~F.}\ \bibnamefont {Assaad}},\
  and\ \bibinfo {author} {\bibfnamefont {M.}~\bibnamefont {Hohenadler}},\
  }\bibfield  {title} {\bibinfo {title} {{Phonon spectral function of the
  one-dimensional Holstein-Hubbard model}},\ }\href
  {https://doi.org/10.1103/PhysRevB.91.235150} {\bibfield  {journal} {\bibinfo
  {journal} {Phys. Rev. B}\ }\textbf {\bibinfo {volume} {91}},\ \bibinfo
  {pages} {235150} (\bibinfo {year} {2015}{\natexlab{a}})}\BibitemShut
  {NoStop}%
\bibitem [{\citenamefont {Hohenadler}\ \emph {et~al.}(2011)\citenamefont
  {Hohenadler}, \citenamefont {Fehske},\ and\ \citenamefont
  {Assaad}}]{PhysRevB.83.115105}%
  \BibitemOpen
  \bibfield  {author} {\bibinfo {author} {\bibfnamefont {M.}~\bibnamefont
  {Hohenadler}}, \bibinfo {author} {\bibfnamefont {H.}~\bibnamefont {Fehske}},\
  and\ \bibinfo {author} {\bibfnamefont {F.~F.}\ \bibnamefont {Assaad}},\
  }\bibfield  {title} {\bibinfo {title} {{Dynamic charge correlations near the
  Peierls transition}},\ }\href {https://doi.org/10.1103/PhysRevB.83.115105}
  {\bibfield  {journal} {\bibinfo  {journal} {Phys. Rev. B}\ }\textbf {\bibinfo
  {volume} {83}},\ \bibinfo {pages} {115105} (\bibinfo {year}
  {2011})}\BibitemShut {NoStop}%
\bibitem [{\citenamefont {Weber}\ \emph
  {et~al.}(2015{\natexlab{b}})\citenamefont {Weber}, \citenamefont {Assaad},\
  and\ \citenamefont {Hohenadler}}]{PhysRevB.91.245147}%
  \BibitemOpen
  \bibfield  {author} {\bibinfo {author} {\bibfnamefont {M.}~\bibnamefont
  {Weber}}, \bibinfo {author} {\bibfnamefont {F.~F.}\ \bibnamefont {Assaad}},\
  and\ \bibinfo {author} {\bibfnamefont {M.}~\bibnamefont {Hohenadler}},\
  }\bibfield  {title} {\bibinfo {title} {{Excitation spectra and correlation
  functions of quantum Su-Schrieffer-Heeger models}},\ }\href
  {https://doi.org/10.1103/PhysRevB.91.245147} {\bibfield  {journal} {\bibinfo
  {journal} {Phys. Rev. B}\ }\textbf {\bibinfo {volume} {91}},\ \bibinfo
  {pages} {245147} (\bibinfo {year} {2015}{\natexlab{b}})}\BibitemShut
  {NoStop}%
\bibitem [{\citenamefont {Jullien}\ and\ \citenamefont
  {Haldane}(1983)}]{JulienHaldane83}%
  \BibitemOpen
  \bibfield  {author} {\bibinfo {author} {\bibfnamefont {R.}~\bibnamefont
  {Jullien}}\ and\ \bibinfo {author} {\bibfnamefont {F.~D.~M.}\ \bibnamefont
  {Haldane}},\ }\href@noop {} {\bibfield  {journal} {\bibinfo  {journal} {Bull.
  Am. Phys. Soc}\ }\textbf {\bibinfo {volume} {28}},\ \bibinfo {pages} {34}
  (\bibinfo {year} {1983})}\BibitemShut {NoStop}%
\bibitem [{\citenamefont {{Laflorencie}}\ \emph {et~al.}(2001)\citenamefont
  {{Laflorencie}}, \citenamefont {{Capponi}},\ and\ \citenamefont
  {{S{\o}rensen}}}]{2001EPJB...24...77L}%
  \BibitemOpen
  \bibfield  {author} {\bibinfo {author} {\bibfnamefont {N.}~\bibnamefont
  {{Laflorencie}}}, \bibinfo {author} {\bibfnamefont {S.}~\bibnamefont
  {{Capponi}}},\ and\ \bibinfo {author} {\bibfnamefont {E.~S.}\ \bibnamefont
  {{S{\o}rensen}}},\ }\bibfield  {title} {\bibinfo {title} {{Finite size
  scaling of the spin stiffness of the antiferromagnetic S ={\textonehalf} XXZ
  chain}},\ }\href {https://doi.org/10.1007/s100510170024} {\bibfield
  {journal} {\bibinfo  {journal} {European Physical Journal B}\ }\textbf
  {\bibinfo {volume} {24}},\ \bibinfo {pages} {77} (\bibinfo {year}
  {2001})}\BibitemShut {NoStop}%
\bibitem [{\citenamefont
  {Giamarchi}(2003)}]{10.1093/acprof:oso/9780198525004.001.0001}%
  \BibitemOpen
  \bibfield  {author} {\bibinfo {author} {\bibfnamefont {T.}~\bibnamefont
  {Giamarchi}},\ }\href
  {https://doi.org/10.1093/acprof:oso/9780198525004.001.0001} {\emph {\bibinfo
  {title} {{Quantum Physics in One Dimension}}}}\ (\bibinfo  {publisher}
  {Oxford University Press},\ \bibinfo {year} {2003})\BibitemShut {NoStop}%
\bibitem [{\citenamefont {Bl\"ote}\ \emph {et~al.}(1986)\citenamefont
  {Bl\"ote}, \citenamefont {Cardy},\ and\ \citenamefont
  {Nightingale}}]{PhysRevLett.56.742}%
  \BibitemOpen
  \bibfield  {author} {\bibinfo {author} {\bibfnamefont {H.~W.~J.}\
  \bibnamefont {Bl\"ote}}, \bibinfo {author} {\bibfnamefont {J.~L.}\
  \bibnamefont {Cardy}},\ and\ \bibinfo {author} {\bibfnamefont {M.~P.}\
  \bibnamefont {Nightingale}},\ }\bibfield  {title} {\bibinfo {title}
  {Conformal invariance, the central charge, and universal finite-size
  amplitudes at criticality},\ }\href
  {https://doi.org/10.1103/PhysRevLett.56.742} {\bibfield  {journal} {\bibinfo
  {journal} {Phys. Rev. Lett.}\ }\textbf {\bibinfo {volume} {56}},\ \bibinfo
  {pages} {742} (\bibinfo {year} {1986})}\BibitemShut {NoStop}%
\bibitem [{\citenamefont {Affleck}(1986)}]{PhysRevLett.56.746}%
  \BibitemOpen
  \bibfield  {author} {\bibinfo {author} {\bibfnamefont {I.}~\bibnamefont
  {Affleck}},\ }\bibfield  {title} {\bibinfo {title} {Universal term in the
  free energy at a critical point and the conformal anomaly},\ }\href
  {https://doi.org/10.1103/PhysRevLett.56.746} {\bibfield  {journal} {\bibinfo
  {journal} {Phys. Rev. Lett.}\ }\textbf {\bibinfo {volume} {56}},\ \bibinfo
  {pages} {746} (\bibinfo {year} {1986})}\BibitemShut {NoStop}%
\end{thebibliography}

\end{document}